\documentclass{article}

\usepackage{arxiv}

\usepackage[utf8]{inputenc} 
\usepackage[T1]{fontenc}    
\usepackage{hyperref}       
\usepackage{url}            
\usepackage{booktabs}       
\usepackage{amsmath}        
\usepackage{amssymb}        
\usepackage{nicefrac}       
\usepackage{microtype}      
\usepackage{graphicx}       
\usepackage{natbib}         
\usepackage{doi}            
\usepackage{xcolor}         
\usepackage{CJK}            
\usepackage{enumerate}      
\usepackage{bm}             
\usepackage{indentfirst} 

\hypersetup{
  colorlinks=true,
  linkcolor=red,
  citecolor=orange,
  urlcolor=blue,
}

\graphicspath{{../figure}}

\usepackage{multicol} 

\usepackage{framed}
\usepackage{multirow}

\def\bu{\bm{u}}

\def\dt{\mbox{${\Delta t}$}}

\makeatletter
\@addtoreset{equation}{section}
\makeatother

\title{
Energy-conserving finite difference scheme for compressible magnetohydrodynamic flow 
at low Mach numbers using nonconservative Lorentz force
}

\author{\href{https://orcid.org/0000-0002-4875-8174}{\includegraphics[scale=0.06]{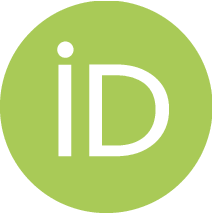}\hspace{1mm}Hideki Yanaoka}
\thanks{Email address for correspondence: yanaoka@iwate-u.ac.jp} \\
    Faculty of Science and Engineering, Iwate University, \\
    4-3-5 Ueda, Morioka, Iwate 020-8551, Japan \\
}

\renewcommand{\shorttitle}{Energy-conserving finite difference scheme for compressible magnetohydrodynamic flow}

\makeatletter
\hypersetup{
pdfusetitle,
bookmarksnumbered,
pdftitle={\@title},
pdfsubject={},
pdfauthor={Hideki Yanaoka},
pdfkeywords={Energy-conserving finite difference scheme for compressible magnetohydrodynamic flow at low Mach numbers using nonconservative Lorentz force}
}
\makeatother

\begin{document}

\maketitle

\begin{abstract}
In magnetohydrodynamic (MHD) flows, incompressibility is assumed for low Mach numbers. 
However, even at low Mach numbers, the Mach number influences flow and magnetic fields. 
Therefore, it is necessary to develop a method that can stably analyze 
low Mach number compressible MHD flows without using the incompressible assumption. 
This study constructs an energy-conserving finite difference method 
to analyze compressible MHD flows at low Mach numbers with the nonconservative Lorentz force. 
This analysis method discretizes the Lorentz force 
so that the transformation between conservative and nonconservative forms holds. 
This scheme simultaneously relaxes velocity, pressure, density, and internal energy, 
and stable convergence solutions can be obtained. 
In this study, we analyze four types of models 
and verify the accuracy and convergence of this numerical method. 
In the analyses of two- and three-dimensional ideal periodic inviscid MHD flows, 
it is clarified that momentum, magnetic flux density, 
and total energy are conserved discretely. 
The total energy is conserved even in a nonuniform grid. 
Even without correction for the magnetic flux density, 
the divergence-free condition of the magnetic flux density is satisfied discretely. 
Analysis of a Taylor decaying vortex under a magnetic field clarifies 
that the present numerical method can be applied to incompressible flows 
and can accurately predict the trend of energy attenuation. 
In the Orszag--Tang vortex analysis, 
an increase in Mach number reduces the magnitude of vorticity and current density. 
In addition, compression work increases more than expansion work, 
and the influence of compressibility appears. 
An increase in Mach number slightly delays the transition to turbulent flow. 
This numerical method has excellent energy conservation properties 
and can accurately predict energy conversion. 
Therefore, we believe that this method can contribute to predicting energy conversion 
in complex and unsteady compressible MHD flows.
\end{abstract}



\section{Introduction}

The numerical analyses of magnetohydrodynamics (MHD) flows have been performed 
for incompressible and compressible fluids 
\citep{Toth_2000,Munz_et_al_2000,Liu&Wang_2001,Dedner_et_al_2002,Gawlik_et_al_2011,Ni&Li_2012,Kraus_et_al_2016,Hu_et_al_2017,Hiptmair_et_al_2018}. 
As shock waves occur in compressible flows at high Mach numbers, 
high Mach number flows have been investigated using shock wave capture schemes. 
Incompressibility is assumed for computational simplicity when analyzing low Mach number flows. 
However, when the temperature difference between a heat source and fluid increases, 
the temperature dependence of density increases, 
and the incompressibility assumption does not hold. 
Compressible flows at low Mach numbers occur in flow fields with combustion or high-temperature heat sources 
and near walls in supersonic boundary layer flows. 
In addition, compressibility must be considered in sound wave analysis. 
When controlling flows with density changes using magnetic fields, 
it is necessary to develop a computational method that can analyze low Mach number flows. 
However, problems with calculation stability and convergence arise. 
Furthermore, another significant problem in the analysis of MHD flow is 
to satisfy the constraint subjected to the Gauss law for magnetism, 
that is, the divergence-free condition of magnetic flux density \citep{Toth_2000}.

In the analysis of compressible fluids, 
it is necessary to capture convective velocity and sound wave propagation. 
The Courant number is given by $\nu = |V \pm c_0|\Delta t/\Delta x$, 
where $\Delta x$ is a grid width, $\Delta t$ is a time step, $c_0$ is a sound speed, 
and $V$ is a convective velocity. 
If the stability condition for calculation is $\nu < 1$, 
the time step must satisfy the condition $\Delta t < \Delta x/|V \pm c|$. 
As the sound wave is faster than convective velocity, 
the Courant number is severely restricted 
when solving the fundamental equations of compressible fluids. 
Therefore, compared with the analysis using the governing equations of incompressible fluids, 
we cannot set a large time increment, 
and a long timescale is required until the flow field reaches a statistically steady state.

There are two numerical methods for solving the momentum conservation equation governing flow: 
density- and pressure-based methods. 
Density-based schemes solve the density from the mass conservation equation 
and find the pressure from the equation of state. 
This numerical method is generally used for compressible flow analyses 
but is not suitable for incompressible flow analyses with low Mach numbers. 
At low Mach numbers, the problem of stiffness in the convective term arises 
because the speed of sound is high compared to convective velocity. 
This problem worsens convergence. 
Using the quasi-compressible method \citep{Chorin_1967} based on the density-based method, 
incompressible fluid flows can also be analyzed. 
However, it is necessary to improve this method for applications to unsteady flows 
and the flow field with mixed incompressible and compressible characteristics. 
As another computational method, 
a preconditioning method that modifies the eigenvalues of the equation has been proposed \citep{Turkel_1987, Choi&Merkle_1993}.

The pressure-based method has been used for incompressible flow analyses 
and has been extended compressible flows \citep{Patnaik_et_al_1987, Rhie_1989, Karki&Patankar_1989, Chen&Pletcher_1991, Demirdzic_et_al_1993}. 
In this scheme, the Poisson equation for pressure or pressure correction value is derived 
from the mass and momentum conservation equations. 
After obtaining the pressure, the density is calculated from the equation of state.

In the analysis for incompressible flows at high Reynolds numbers, 
computational instability can be suppressed 
if transport quantities such as mass, momentum, and kinetic energy are discretely preserved in inviscid flows. 
However, when analyzing compressible flows, 
even if discrete conservation of transport quantities is established, 
the calculation becomes unstable at high Reynolds numbers. 
Therefore, instead of solving the energy conservation equation, 
a method to solve the entropy conservation equation was proposed \citep{Harten_1983}. 
In addition, as using the total energy equation is known to lead to unstable computations, 
Honein and Moin \citep{Honein&Moin_2004} used the internal energy equation. 
For the analysis of high Reynolds number flows, 
the setting of the dependent variable in the energy equation affects the stability of the calculation.

When analyzing compressible flows using the density-based method, 
the pressure in the momentum conservation equation is determined by the equation of state. 
At low Mach numbers, dynamic pressure is lower than the thermodynamic pressure. 
As the dynamic pressure significantly affects the momentum conservation equation, 
high accuracy is required when calculating the pressure using the equation of state. 
When analyzing low Mach number flows, 
the low Mach number approximation \citep{Rehm&Baum_1978,Quere_et_al_1992} has been used. 
This method replaces the pressure in the equation of state with a constant thermodynamic pressure, 
decoupling the pressure link between the equation of state and the conservation of momentum equation. 
Therefore, density is affected by temperature only and does not change with pressure.

There are also methods of analyzing low Mach flows without using the low Mach approximation 
\citep{Patnaik_et_al_1987, Wall_et_al_2002, Morinishi_2009, Morinishi_2010, Hou&Mahesh_2005, Bijl&Wesseling_1998, Kwatra_et_al_2009, Hennink_et_al_2021}. 
Patnaik et al. \citep{Patnaik_et_al_1987} proposed a density-based method (barely implicit correction: BIC) 
that transforms the total energy equation into an elliptic equation for the pressure correction value 
and removes the time-step limitation due to the speed of sound. 
The BIC method modifies the velocity and total energy after solving the pressure correction value. 
Wall et al. \citep{Wall_et_al_2002} proposed a method to convert the Poisson equation for pressure correction 
into the Helmholtz equation using the pressure-based method 
to avoid the Courant number limitation. 
The fully conservative finite difference scheme 
proposed by Morinishi \citep{Morinishi_2009, Morinishi_2010} is an improved version of the scheme of Wall et al. \citep{Wall_et_al_2002}, 
using the implicit midpoint rule, 
transport quantities and square quantities are preserved discretely in time and space directions. 
The method proposed by Hou and Mahesh \citep{Hou&Mahesh_2005} uses a collocated grid 
and, like Wall et al. \citep{Wall_et_al_2002}, the time levels of density, pressure, and temperature are staggered from that of velocity. 
Similarly to Patnaik et al. \citep{Patnaik_et_al_1987}, 
the density is solved from the mass conservation equation, 
and the Poisson equation for the pressure correction value is derived from the total energy equation. 
They also used a nondimensionalization, similar to Bijl and Wesseling \citep{Bijl&Wesseling_1998}, 
so that for low Mach numbers, the energy equation reduces to the form of the divergence-free condition for the velocity. 
Kwatra et al. \citep{Kwatra_et_al_2009} proposed a method to solve the pressure evolution equation. 
As the Mach number approaches zero, 
the Poisson equation for pressure reduces to the Poisson equation for incompressible flow. 
In this method, the density is solved from the mass conservation equation, 
and after solving the pressure, the velocity and total energy are corrected. 
Dumbser et al. \citep{Dumbser_et_al_2019} and Fambri \citep{Fambri_2021} proposed 
a semi-implicit finite volume solver for compressible MHD flows. 
Similarly to existing studies \citep{Patnaik_et_al_1987, Kwatra_et_al_2009}, 
the Poisson equation for pressure is derived from the total energy equation, 
and after solving the pressure, velocity and total energy are corrected. 
Hennink et al. \citep{Hennink_et_al_2021} used the discontinuous Galerkin method 
to construct a pressure-based method for low Mach number flows. 
They adopted the method of solving mass flux instead of velocity. 
Boscheri and Tavelli \citep{Boscheri&Tavelli_2022} proposed a method to calculate low Mach number flows 
using a semi-implicit method. 
In this method, the density is solved from the mass conservation equation. 
Also, by substituting the momentum equation into the total energy equation, 
the Poisson equation for pressure (pressure wave equation) is derived, 
and the pressure is solved. 
As described above, various methods have been proposed to analyze low Mach number flows, 
but energy conservation properties have not been investigated in the previous studies 
\citep{Hou&Mahesh_2005, Kwatra_et_al_2009, Hennink_et_al_2021}.

The Lorentz force occurs in MHD flows and significantly affects the flow and magnetic fields. 
As it is included as a body force in the momentum conservation equation, 
the equation is in a nonconservative form. 
As the Lorentz force is nonconservative, 
Toth \citep{Toth_2000}, Munz et al. \citep{Munz_et_al_2000}, and Dedner et al. \citep{Dedner_et_al_2002} 
transformed the Lorentz force into a conservative form using the Ampere law 
and solved the conservative fundamental equation. 
Ni and Li \citep{Ni&Li_2012} have proposed a method to convert the Lorentz force 
into a divergent form using a distance-vector. 
Some studies \citep{Liu&Wang_2001, Gawlik_et_al_2011, Kraus_et_al_2016, Hu_et_al_2017, Hiptmair_et_al_2018}, 
have proposed a structure-preserving numerical method for MHD flows. 
Using this method, excellent energy and helicity conservation properties have been demonstrated 
\citep{Gawlik_et_al_2011, Kraus_et_al_2016}. 
The author \citep{Yanaoka_2023} clarified the influence of the transformation 
of the nonconservative and conservative forms of the Lorentz force 
on the energy conservation property. 
Dumbser et al. \citep{Dumbser_et_al_2019} and Fambri \citep{Fambri_2021} proposed 
a semi-implicit finite volume solver for compressible MHD flows 
and also analyzed low Mach number flows. 
Until now, MHD flows at low Mach numbers where density changes occur have received little attention, 
and, in addition, the energy conservation properties in such flow fields have not been investigated in detail.

Herein, we construct a numerical analysis method that can analyze low Mach number flows 
using a pressure-based solution method, 
intending to control flows accompanied by density changes using magnetic fields. 
Using this numerical method, we analyze flows in a wide range of Mach numbers, 
from incompressible to low Mach number flows, 
and verify the validity of this computational method. 

The remainder of this paper is organized as follows: 
Section \ref{fundamental_equation} presents the fundamental equations 
and also derives the energy transport equations. 
In section \ref{discretization}, we derive the method for discretizing the Lorentz force 
and the discretization equation for the governing equation. 
We also derive the total energy conservation equation discretely. 
In section \ref{numerical_method}, we propose a simultaneous relaxation method to solve the governing equations. 
In Section \ref{verification}, we verify the computational accuracy, convergence, 
and conservation properties of the present numerical method. 
Finally, Section \ref{summary} summarizes the results.

\section{Modeling and fundamental equations}
\label{fundamental_equation}

\subsection{Fundamental equation}

In this study, we deal with compressible MHD flows at low Mach numbers 
and analyze flows without shock waves. 
We also assume that the fluid is an ideal gas. 
The fundamental equations governing compressible flows are 
the transport equations for mass, momentum, and internal energy, 
the solenoidal constraint imposed by Gauss's law for magnetism, 
and the Faraday equation. 
These equations are given as follows:
\begin{equation}
  \frac{Wo^2}{Re} \frac{\partial \rho}{\partial t} 
  + \frac{\partial \rho u_j}{\partial x_j} = 0,
  \label{mass}
\end{equation}
\begin{equation}
  \frac{Wo^2}{Re} \frac{\partial \rho u_i}{\partial t} 
  + \frac{\partial \rho u_j u_i}{\partial x_j} 
  = - \frac{\partial p}{\partial x_i} 
  + \frac{1}{Re} \frac{\partial \tau_{ij}}{\partial x_j} 
  + \frac{1}{Al^2} \epsilon_{ijk} j_j B_k,
  \label{momentum}
\end{equation}
\begin{align}
  \frac{Wo^2}{Re} \frac{\partial \rho e}{\partial t} 
  + \frac{\partial \rho u_j e}{\partial x_j} 
  &= - \frac{\kappa}{Re Pr} \frac{\partial q_j}{\partial x_j} 
  - (\kappa - 1) (\kappa Ma^2 p + 1) \frac{\partial u_i}{\partial x_i} 
  \nonumber \\
  &+ \frac{\kappa (\kappa - 1) Ma^2}{Re} \tau_{ij} \frac{\partial u_i}{\partial x_j} 
   + \frac{\kappa (\kappa - 1) Ma^2}{Al^2 Re_m} j_i^2,
  \label{energy_e}
\end{align}
\begin{equation}
  \frac{\partial B_i}{\partial x_i} = 0,
  \label{divergence_magnetic}
\end{equation}
\begin{equation}
  \frac{Wo^2}{Re} \frac{\partial B_i}{\partial t} 
  + \epsilon_{ijk} \frac{\partial E_k}{\partial x_j} = 0,
  \label{Faraday}
\end{equation}
where $t$, $\rho$, $u_i$, $p$, $e$, $j_i$, $B_i$, and $E_i$ represent 
the time, density of fluid, 
velocity at the coordinate $x_i$, pressure, internal energy, 
current density, magnetic flux density, and electric field, respectively. 
The symbol $\epsilon_{ijk}$ represents an alternation symbol. 
The term $\epsilon_{ijk} j_j B_k/Al^2$ in Eq. (\ref{momentum}) 
expresses the Lorentz force. 
The second term on the right side of the internal energy equation (\ref{energy_e}) 
represents work due to volume change, 
and the third term represents the viscous dissipation term. 
We assume that the flow is a Newtonian fluid. 
The viscous stress tensor $\tau_{ij}$ and the heat flux vector $q_j$ are defined as
\begin{equation}
  \tau_{ij} = \mu \left( \frac{\partial u_i}{\partial x_j} 
                       + \frac{\partial u_j}{\partial x_i} 
  - \frac{2}{3} \frac{\partial u_k}{\partial x_k} \delta_{ij} \right),
\end{equation}
\begin{equation}
  q_j = - k \frac{\partial T}{\partial x_j},
\end{equation}
where $\mu$, $k$, and $T$ represent the viscosity coefficient of fluid, 
thermal conductivity, and temperature, respectively. 
$\delta_{ij}$ represents the Kronecker delta function. 
As for the reference values used for nondimensionalization, 
the length is $l_\mathrm{ref}$, velocity is $u_\mathrm{ref}$, 
time is $t_\mathrm{ref}$, internal energy is $e_\mathrm{ref}$, 
temperature is $T$, magnetic flux density is $B_\mathrm{ref}$, 
viscosity coefficient is $\mu_\mathrm{ref}$, 
and thermal conductivity is $k_\mathrm{ref}$. 
Using these reference values, 
the variables in the fundamental equations were nondimensionalized as follows:
\begin{subequations}
\begin{equation}
  x_i^* = \frac{x_i}{l_\mathrm{ref}}, \quad
  u_i^* = \frac{u_i}{u_\mathrm{ref}}, \quad
  \rho^* = \frac{\rho}{\rho_\mathrm{ref}}, \quad
  p^* = \frac{p - p_\mathrm{ref}}{\rho_\mathrm{ref} u_\mathrm{ref}^2}, \quad
  e^* = \frac{e}{e_\mathrm{ref}}, \quad
  E^* = \frac{E}{e_\mathrm{ref}}, \quad
\end{equation}
\begin{equation}
  T^* = \frac{T}{T_\mathrm{ref}}, \quad
  E_i^* = \frac{E_i}{u_\mathrm{ref} B_\mathrm{ref}}, \quad 
  \psi^* = \frac{\psi}{l_\mathrm{ref} u_\mathrm{ref} B_\mathrm{ref}}, \quad
  j_i^* = \frac{j_i}{B_\mathrm{ref}/(\mu_{m,f} l_\mathrm{ref})},
\end{equation}
\begin{equation}
  B_i^* = \frac{B_i}{B_\mathrm{ref}}, \quad
  A_i^* = \frac{A_i}{l_\mathrm{ref} B_\mathrm{ref}}, \quad
  t^* = \frac{t}{t_\mathrm{ref}}, \quad
  \mu^* = \frac{\mu}{\mu_\mathrm{ref}}, \quad
  k^* = \frac{k}{k_\mathrm{ref}}.
  \label{non-dimensionalize}
\end{equation}
\end{subequations}
where $*$ represents the nondimensional variable, 
and is omitted in the fundamental equations. 
In addition, $e_\mathrm{ref} = c_v T_\mathrm{ref}$, 
$c_v$ is the constant volume specific heat, and $p_\mathrm{ref}$ is the reference pressure. 
This study sets $\mu^* = 1$ and $k^* = 1$ 
without considering the temperature dependence of physical property values. 
The nondimensional parameters in these fundamental equations are defined as follows: 
$Re$, $Wo$, $Pr$, $Ma$, $Al$, and $Re_m$ represent the Reynolds, Womersley, 
Prandtl, Mach, Alfv\'{e}n, and magnetic Reynolds numbers, respectively:
\begin{subequations}
\begin{equation}
  Re = \frac{u_\mathrm{ref} l_\mathrm{ref}}{\nu_\mathrm{ref}}, \quad 
  Wo = l_\mathrm{ref} \sqrt{\frac{1}{\nu_\mathrm{ref} t_\mathrm{ref}}}, \quad 
  Al = \frac{\sqrt{\rho \mu_m} u_\mathrm{ref}}{B_\mathrm{ref}}, \quad  
  Pr = \frac{\nu_\mathrm{ref}}{\alpha_\mathrm{ref}}, \quad 
\end{equation}
\begin{equation}
  Ma = \frac{u_\mathrm{ref}}{c_\mathrm{ref}}, \quad 
  Re_m = \frac{u_\mathrm{ref} l_\mathrm{ref}}{\nu_{m,f}}.
\end{equation}
\end{subequations}
where $\rho$ and $\nu$ represent the density and kinematic viscosity of the fluid, respectively, 
and $\mu_m$ represents the magnetic permeability 
related to the magnetic diffusivity $\nu_m$ and the electrical conductivity $\sigma$, 
as $\nu_m = 1/(\sigma \mu_m)$.

The equation of state for an ideal gas is given as follows:
\begin{equation}
  \kappa Ma^2 p + 1 = \rho e,
  \label{eqn_of_state}
\end{equation}

The current density defined by Ampere's law and Ohm's law, respectively, is as follows:
\begin{equation}
  j_i = \epsilon_{ijk} \frac{\partial B_k}{\partial x_j} 
  \label{Ampere}
\end{equation}
\begin{equation}
  j_i = Re_m \left( E_i + \epsilon_{ijk} u_j B_k \right).
  \label{Ohm}
\end{equation}

The magnetic flux density equation is obtained by revising 
Eq. (\ref{Faraday}) using Ohm's law (\ref{Ohm}) as follows:
\begin{equation}
  \frac{Wo^2}{Re} \frac{\partial B_i}{\partial t} 
  + \frac{\partial (u_j B_i - B_j u_i)}{\partial x_j} 
  = - \frac{1}{Re_m} \epsilon_{ijk} \frac{\partial j_k}{\partial x_j},
  \label{magnetic_field}
\end{equation}
When $Re_m = \infty$, Eq. (\ref{magnetic_field}) is in a conservative form, 
which is the form used by \citep{Toth_2000,Dedner_et_al_2002}. 

The induced electric field is expressed using the electric potential $\psi$ as
\begin{equation}
  E_i = - \frac{\partial \psi}{\partial x_i} 
  - \frac{Wo^2}{Re} \frac{\partial A_i}{\partial t},
  \label{electric_potential}
\end{equation}
where $A_i$ is the magnetic vector potential satisfying $B_i = \epsilon_{ijk} \partial_j A_k$. 
The conservation law of electric charge is given as
\begin{equation}
  \frac{\partial j_i}{\partial x_i} = 0.
  \label{electric_charge_law}
\end{equation}
The Poisson equation for the electric potential can be obtained 
using Eqs. (\ref{Ohm}) and (\ref{electric_potential_poisson}) as follows:
\begin{equation}
  \frac{\partial}{\partial x_i} 
  Re_m \left( - \frac{\partial \psi}{\partial x_i} 
  - \frac{Wo^2}{Re} \frac{\partial A_i}{\partial t} 
  + \frac{\partial \epsilon_{ijk} u_j B_k}{\partial x_i} \right) = 0.
  \label{electric_potential_poisson}
\end{equation}
Applying the Coulomb gauge $\partial_i A_i = 0$ yields the following Poisson's equation:
\begin{equation}
  \frac{\partial^2 \psi}{\partial x_i^2} 
  = \frac{\partial \epsilon_{ijk} u_j B_k}{\partial x_i}.
\end{equation}
Applying Ampere's (\ref{Ampere}) and Ohm's laws (\ref{Ohm}) 
to Eq. (\ref{electric_potential}), 
the equation for the magnetic vector potential is obtained as follows:
\begin{equation}
  \frac{Wo^2}{Re} \frac{\partial A_i}{\partial t} 
  + \epsilon_{ijk} B_j u_k = - \frac{\partial \psi}{\partial x_i} 
  - \frac{1}{Re_m} j_i,
  \label{vector_potential}
\end{equation}

When the magnetic permeability is constant, 
the Lorentz force in Eq. (\ref{momentum}) 
can be revised from a nonconservative to a conservative form as follows:
\begin{align}
  F_i &= \frac{1}{Al^2} \epsilon_{ijk} j_j B_k \nonumber \\ 
  &= \frac{1}{Al^2} \left[ 
  \frac{\partial B_j B_i}{\partial x_j} 
  - \frac{1}{2} \frac{\partial B_j^2}{\partial x_i} 
  - \frac{\partial B_j}{\partial x_j} B_i \right],
  \label{Lorentz}
\end{align}
where $B_i^2/(2Al^2)$ is the magnetic pressure. 
The last term is zero according to the solenoidal constraint. 
The momentum equation is transformed using Eq. (\ref{Lorentz}), 
and the terms of the equation, other than the final term, are expressed 
in the conservative form as follows:
\begin{equation}
  \frac{Wo^2}{Re} \frac{\partial \rho u_i}{\partial t} 
  + \frac{\partial}{\partial x_j} 
  \left( \rho u_j u_i - \frac{1}{Al^2} B_j B_i \right) 
  = - \frac{\partial P}{\partial x_i} 
  + \frac{1}{Re} \frac{\partial \tau_{ij}}{\partial x_j} 
  - \frac{1}{Al^2} \epsilon_{ijk} B_i \frac{\partial B_j}{\partial x_j},
  \label{momentum2}
\end{equation}
where $P$ is a component obtained by adding a magnetic pressure component to pressure 
as follows:
\begin{equation}
  P = p + \frac{1}{2} \frac{1}{Al^2} B_j^2.
\end{equation}
The final term of Eq. (\ref{momentum2}) becomes zero 
when the solenoidal constraint $\partial_j B_j = 0$ is satisfied. 
Equation (\ref{Lorentz}) indicates that the direction of Lorentz force is 
perpendicular to the magnetic field 
if the solenoidal constraint is satisfied. 
If $\partial_j B_j$ is not zero, a nonphysical Lorentz force proportional to 
$\partial_j B_j$ also occurs in the direction parallel to the magnetic field. 
Therefore, the time should be advanced while satisfying the solenoidal constraint.

Equation (\ref{Lorentz}) represents the transformation of the Lorentz force. 
The Lorentz forces in (\ref{momentum}) and (\ref{momentum2}) are 
in the nonconservative and conservative forms, respectively. 
If the conversion of the Lorentz force in Eq. (\ref{Lorentz}) holds discretely, 
the nonconservative Lorentz force can be converted to a conservative form. 
Therefore, Eq. (\ref{momentum}) can be transformed 
into Eq. (\ref{momentum2}) using discretized Eq. (\ref{Lorentz}). 
Thus, if the constraint condition of the magnetic flux density is satisfied, 
Eq. (\ref{momentum}) becomes conservative, 
and momentum is conserved for $Re = Re_m = \infty$. 
Moreover, the work done by the Lorentz force changes 
the kinetic and magnetic energies; this change affects energy conservation properties. 
If the Lorentz force cannot be transformed discretely as in Eq. (\ref{Lorentz}), 
the discrete forms of the Lorentz forces in Eqs. (\ref{momentum}) 
and (\ref{momentum2}) are different. 
The form of the Lorentz force can affect the conservation of energy and momentum 
and the conversion of energy. 
A previous study \citep{Yanaoka_2023} clarified 
that even when the Lorentz force in Eq. (\ref{momentum}) is discretized, 
the nonconservative Lorentz force is converted to the conservative form, 
and the transformation between the nonconservative and conservative forms is established. 
In Section \ref{discretization}, 
the discretization method of the Lorentz force is described 
such that this transformation holds. 
Analyzing various models confimed 
that the calculation stability in both nonconservative 
and conservative forms is the same in uniform grids. 
However, as the transformation is not established on nonuniform grids, 
the calculation using a conservative form that does not discretely satisfy 
the preservation of total energy becomes unstable. 
Therefore, this study adopts a method that uses a nonconservative form of the Lorentz force.

\subsection{Energy equations}

Each energy is made dimensionless as follows:
\begin{equation}
  e^* = \frac{e}{e_\mathrm{ref}}, \quad 
  K^* = \frac{K}{e_\mathrm{ref}}, \quad 
  E^* = \frac{E}{e_\mathrm{ref}}, \quad 
  M^* = \frac{M}{e_\mathrm{ref}}, \quad 
  E_t^* = \frac{E_t}{e_\mathrm{ref}}, 
\end{equation}
where $c_\mathrm{ref}^2 = \kappa(\kappa-1) e_\mathrm{ref}$. 
By calculating the inner product of the velocity $u_i$ 
and Eq. (\ref{momentum}), 
the transport equation for the kinetic energy $K$ is obtained. 
The total energy $E$, not including magnetic energy, and its equation are given as
\begin{equation}
  \rho E = \rho e + \frac{1}{2} \kappa (\kappa - 1) Ma^2 \rho u_i^2,
  \label{total_energy}
\end{equation}
\begin{align}
  & 
  \frac{Wo^2}{Re} \frac{\partial \rho E}{\partial t} 
  + \frac{\partial \rho u_j E}{\partial x_j} 
  = - \frac{\kappa}{Re Pr} \frac{\partial q_j}{\partial x_j} 
  - (\kappa - 1) \frac{\partial u_i (\kappa Ma^2 p + 1)}{\partial x_i} 
  \nonumber \\
  & \quad 
  + \frac{\kappa (\kappa - 1) Ma^2}{Re} \frac{\partial \tau_{ij} u_i}{\partial x_j} 
  + \frac{1}{Fr^2} \rho g n_i u_i 
  + \frac{\kappa (\kappa - 1) Ma^2}{Al^2} 
  \left[ u_i \epsilon_{ijk} j_j B_k + \frac{1}{Re_m} j_i^2 \right], 
  \label{energy_E}
\end{align}
Discontinuities such as shock waves occur at compressible flows. 
Therefore, in numerical calculations, 
discontinuous variables such as density are accepted 
as solutions of partial differential equations, 
so solutions are generally expressed in the weak form \citep{Morinishi_2009, Morinishi_2010}. 
Therefore, Eq. (\ref{energy_E}) is used instead of Eq. (\ref{energy_e}) 
for compressible flow analysis. 
For compressible flows at low Mach numbers, 
using the total energy equation is known to be computationally unstable \citep{Honein&Moin_2004}. 
Therefore, suppression of nonlinear instability is more significant than capturing discontinuities. 
This study uses the internal energy equation (\ref{energy_e}), 
similar to the previous studies \citep{Honein&Moin_2004, Morinishi_2009, Morinishi_2010}.

Subsequently, by calculating the inner product of the Faraday equation (\ref{Faraday}) 
and magnetic flux density $B_i$, 
the transport equation for the magnetic energy $M$ is obtained. 
The magnetic energy $M$ and its equation are given as
\begin{equation}
  \rho M = \frac{1}{2 Al^2} \kappa (\kappa - 1) Ma^2 B_i^2,
  \label{magnetic_energy}
\end{equation}
\begin{equation}
  \frac{Wo^2}{Re} \frac{\partial (\rho M)}{\partial t} 
  = - \frac{\kappa (\kappa - 1) Ma^2}{Al^2} 
  \left( \frac{\partial \epsilon_{ijk} E_j B_k}{\partial x_i} 
  + u_i \epsilon_{ijk} j_j B_k + \frac{1}{Re_m} j_i^2 \right).
  \label{energy_M}
\end{equation}
By summing Eqs. (\ref{energy_E}) and (\ref{energy_M}), 
the equation for the total energy $E_t = E + M$, 
which is the sum of kinetic energy, internal energy, and magnetic energy, is obtained. 
The total energy $E_t$ and its equation are given as
\begin{equation}
  \rho E_t = \rho e + \frac{1}{2} \kappa (\kappa - 1) Ma^2 
  \left( \rho u_i^2 + \frac{1}{Al^2} B_i^2 \right),
  \label{total_energy.mag}
\end{equation}
\begin{align}
  & 
  \frac{Wo^2}{Re} \frac{\partial (\rho E_t)}{\partial t} 
  + \frac{\partial \rho u_j E}{\partial x_j} 
  = - \frac{\kappa}{Re Pr} \frac{\partial q_j}{\partial x_j} 
  - (\kappa - 1) \frac{\partial u_i (\kappa Ma^2 p + 1)}{\partial x_i} 
  \nonumber \\
  & \quad 
  + \frac{\kappa (\kappa - 1) Ma^2}{Re} \frac{\partial \tau_{ij} u_i}{\partial x_j} 
  - \frac{\kappa (\kappa - 1) Ma^2}{Al^2} 
  \frac{\partial \epsilon_{ijk} E_j B_k}{\partial x_i},
  \label{energy_Et}
\end{align}
where the superscript $*$ in the above formula is omitted. 
In Eq. (\ref{energy_Et}), 
the work terms $u_i \epsilon_{ijk} j_j B_k$ done by the Lorentz force 
appearing in Eqs. (\ref{energy_E}) and (\ref{energy_M}) cancel each other. 
However, energy is exchanged between the velocity and magnetic fields 
through this term. 
The Lorentz forces appearing in Eqs. (\ref{energy_E}) and (\ref{energy_M}) 
should be calculated by the same discretization and interpolation. 
If the calculation method of the Lorentz force is inconsistent, 
the energy conversion cannot be captured correctly.

When $Re = Re_m = \infty$, Eq. (\ref{energy_Et}) is expressed as follows:
\begin{equation}
  \frac{\partial (\rho E_t)}{\partial t} 
  + \frac{\partial \rho u_j E}{\partial x_j} 
  = - (\kappa - 1) \frac{\partial u_i (\kappa Ma^2 p + 1)}{\partial x_i} 
  - \frac{\kappa (\kappa - 1) Ma^2}{Al^2} 
  \frac{\partial \epsilon_{ijk} E_j B_k}{\partial x_i},
  \label{energy_inviscid_Et}
\end{equation}
where $Wo = \sqrt{Re}$ is set to remove $Re$. 
From the above equation, the transport equation for the total energy $E_t$, 
which is the sum of the kinetic, internal, and magnetic energies, is conservative; 
that is, Eq. (\ref{energy_inviscid_Et}) states that the total energy is conserved.

\subsection{Entropy equation}

Using the thermodynamic relations, 
the following equation for entropy $s$ is obtained:
\begin{equation}
  \frac{Wo^2}{Re} \frac{\partial \rho s}{\partial t} 
  + \frac{\partial \rho u_j s}{\partial x_j} 
  = \frac{1}{T} \left( - \frac{\kappa}{Re Pr} \frac{\partial q_j}{\partial x_j} 
  + \frac{Ec}{Re} \tau_{ij} \frac{\partial u_i}{\partial x_j} \right),
 \label{entropy_eq}
\end{equation}
where the entropy is dimensionless using $c_v$ and given as
\begin{equation}
  s = \ln \frac{\kappa Ma^2 p + 1}{\rho^{\kappa}}.
  \label{entropy}
\end{equation}
$Ec$ is the Eckert number, defined as $Ec = u_\mathrm{ref}^2/(c_v T_\mathrm{ref})$. 
In inviscid flows, the total amount of entropy is conserved. 
Therefore, if the entropy changes with time 
when investigating the time variation of the total amount, 
we can find the occurrence of a nonphysical phenomenon.

\subsection{Magnetic helicity equation}

For an ideal inviscid incompressible MHD flow, 
the total energy $E_t$ is a conserved quantity. 
The magnetic helicity $H_m = B_i A_i/(\rho_0 \mu_m)$ is also preserved, 
where $\rho_0$ is a reference density, 
and the magnetic permeability $\mu_m$ is set to be constant. 
Magnetic helicity is made dimensionless as $H_m^* = H_m/(L_\mathrm{ref}e_\mathrm{ref})$. 
Using Eqs. (\ref{Faraday}) and (\ref{vector_potential}) for $Re_m = \infty$, 
the magnetic helicity and its equation can be obtained as follows:
\begin{equation}
  H_m = \frac{1}{Al^2} \kappa (\kappa - 1) Ma^2 B_i A_i,
  \label{magnetic_helicity}
\end{equation}
\begin{equation}
  \frac{\partial H_m}{\partial t} 
  = \frac{\kappa (\kappa - 1) Ma^2}{Al^2} \left[ 
  - \frac{\partial}{\partial x_j} (\epsilon_{jki} E_k A_i) 
  + 2 B_i (\epsilon_{ijk} u_j B_k) 
  - \frac{\partial B_i \psi}{\partial x_i} 
  + \psi \frac{\partial B_i}{\partial x_i} \right],
  \label{magnetic_helicity_eq}
\end{equation}
where $B_i$ and $\epsilon_{ijk} u_j B_k$ are orthogonal; 
threfore, the inner product is zero. 
Additionally, using the divergence-free condition (\ref{divergence_magnetic}), 
the fourth term on the right side is zero. 
The magnetic helicity is expressed as a conservative equation, 
and the total amount of magnetic helicity is conserved 
under the assumption of periodic flow:
\begin{equation}
  \frac{\partial H_m}{\partial t} 
  = - \frac{\kappa (\kappa - 1) Ma^2}{Al^2} \left[ 
    \frac{\partial}{\partial x_j} (\epsilon_{jki} E_k A_i) 
  + \frac{\partial B_i \psi}{\partial x_i} \right].
  \label{magnetic_helicity_ideal_eq}
\end{equation}

\subsection{Incompressible limit}

In the incompressible limit where the Mach number is $Ma \rightarrow 0$, 
the pressure work and viscous dissipation terms in the internal energy equation 
(\ref{energy_e}) are eliminated, 
and the internal energy equation is transformed into the energy equation 
for the temperature $T$. 
Furthermore, from the state equation (\ref{eqn_of_state}), $\rho e = 1$ 
and the flow field is isothermal. 
Therefore, the above fundamental equations reduce to equations for incompressible fluids 
in the limit $Ma \rightarrow 0$. 
This is because $p^* = (p-p_\mathrm{ref})/(\rho u_\mathrm{ref}^2)$ is used 
for nondimensionalization of the pressure. 
Here, the reference pressure $p_\mathrm{ref}$ is obtained 
as $p_\mathrm{ref} = (\kappa-1)\rho e$ using the mainstream or initial densities and temperatures. 
A similar nondimensionalization is used in existing work \citep{Bijl&Wesseling_1998}.

\section{Discretization of transport equation}
\label{discretization}

For periodic inviscid incompressible flows without applied magnetic fields, 
the transport quantity, such as the kinetic energy, 
must be discretely conserved \citep{Morinishi_1996a,Morinishi_1998}. 
The generation of nonphysical kinetic energy leads to computational instability. 
Additionally, the transformation between the conservative and nonconservative forms 
of convection terms must be discretely satisfied 
\citep{Morinishi_1996a,Morinishi_1998}. 
In a flow field without an applied magnetic field, 
a fully conservative finite difference method, 
in which the transport quantity is discretely conserved in the spatiotemporal direction, 
has been proposed. 
The transformation between conservative and nonconservative forms 
of the advection term has been established \citep{Ham_et_al_2002,Morinishi_2009,Morinishi_2010}. 
In this study, the fully conservative finite difference method is applied to analyze MHD flows at low Mach numbers, 
as in \citep{Ham_et_al_2002,Morinishi_2009,Morinishi_2010}. 
In addition, we adopt a spatiotemporal staggered grid, 
similar to the existing studies \citep{Wall_et_al_2002, Morinishi_2009, Morinishi_2010}. 
The governing equation is discretized so that the transport quantity is discretely conserved 
in the spatiotemporal direction 
while maintaining second-order accuracy and compatibility between conservative and nonconservative forms. 
Velocity, internal energy, and magnetic flux density are placed on the same time level, 
and density and pressure are placed on a time level that is half the time offset from the velocity. 
The implicit midpoint rule for the time derivative 
and the second-order central difference for the spatial derivative are applied. 
Below, the discretization of each transport quantity equation is described in detail.

\subsection{Definitions of finite difference and interpolation operations}

The Cartesian coordinates $x_m$ in the physical space are transformed into 
the computational space $\xi_m$ for discretization in a nonuniform grid. 
The relationship $x_m = x_m(\xi_m)$ is assumed between both spaces. 
By letting a dependent variable such as the velocity, pressure, 
and magnetic flux density be $\Phi$, 
the first derivative can be converted as follows:
\begin{equation}
  \frac{\partial \Phi}{\partial x_1} 
    = \frac{1}{J} \frac{\partial (J \xi_{1,1} \Phi)}{\partial \xi_1}, \quad 
  \frac{\partial \Phi}{\partial x_2} 
    = \frac{1}{J} \frac{\partial (J \xi_{2,2} \Phi)}{\partial \xi_2}, \quad 
  \frac{\partial \Phi}{\partial x_3} 
    = \frac{1}{J} \frac{\partial (J \xi_{3,3} \Phi)}{\partial \xi_3},
\end{equation}
where $J$ is the Jacobian defined as $J = x_{1,1} x_{2,2} x_{3,3}$. 
$\xi_{i,i}$ is given as
\begin{equation}
  \xi_{1,1} = \frac{1}{J} x_{2,2} x_{3,3}, \quad 
  \xi_{2,2} = \frac{1}{J} x_{3,3} x_{1,1}, \quad 
  \xi_{3,3} = \frac{1}{J} x_{1,1} x_{2,2}.
\end{equation}

The variable at a cell center $(i, j, k)$ is defined as $\Phi_{i,j,k}$ and $\Psi_{i,j,k}$. 
For the $x$ ($\xi_1$)-direction, 
the second-order central difference equation and interpolation for the variable $\Phi$ 
and the permanent product for two variables are given, respectively, as follows: 
\citep{Morinishi_1996a,Morinishi_1998}:
\begin{equation}
  \left. \frac{\partial \Phi}{\partial x_1} \right|_{i,j,k} 
  = \delta_{\xi_1} \Phi 
  = \frac{1}{J} \frac{(J \xi_{1,1} \bar{\Phi}^{\xi_1})_{i+1/2,j,k} 
                    - (J \xi_{1.1} \bar{\Phi}^{\xi_1})_{i-1/2,j,k}}{\Delta \xi_1},
\end{equation}
\begin{equation}
  \left. \bar{\Phi}^{\xi_1} \right|_{i+1/2,j,k} 
  = \frac{\Phi_{i,j,k} + \Phi_{i+1,j,k}}{2},
\end{equation}
\begin{equation}
  \left. \widetilde{\Phi \Psi}^{\xi_1} \right|_{i+1/2,j,k} 
  = \frac{\Phi_{i,j,k} \Psi_{i+1,j,k} 
        + \Phi_{i+1,j,k} \Psi_{i,j,k}}{2},
\end{equation}
where $\Delta \xi_1$ is the grid spacing in the computational space. 
The definitions of the $x_2$ ($\xi_2$) and $x_3$ ($\xi_3$)-directions are identical. 
The Jacobian is defined at a cell center. 
The index $j$ representing the direction of the difference $\delta_{\xi_j}$ 
is considered a tensor component and follows the summation convention. 
The indices $j$ of the interpolation $\bar{\Phi}^{\xi_j}$ 
and permanent product $\tilde{\Phi \Psi}^{\xi_j}$ do not follow the convention. 
The indices $j$ change simultaneously with the indices of the tensor components 
in the same term.
Derivative terms that are not directly related to conservation properties, 
such as momentum and total energy, are discretized 
without coordinate transformation, as follows:
\begin{equation}
  \left. \frac{\partial \Phi}{\partial x_1} \right|_{i,j,k} 
  = \delta_{x_1} \Phi 
  = \frac{\bar{\Phi}^{x_1}_{i+1/2,j,k} - \bar{\Phi}^{x_1}_{i-1/2,j,k}}{\Delta x_1},
\end{equation}
where $\Delta x_{i} = x_{i+1/2}-x_{i-1/2}$ is the grid spacing. 
If a variable at time level $n$ is defined as $\Phi^n$, 
the derivative and interpolation of the variable for time are similarly expressed as follows:
\begin{equation}
  \left. \frac{\partial \Phi}{\partial t} \right|^{n+1/2} 
  = \delta_t \Phi 
  = \frac{\Phi^{n+1} - \Phi^{n}}{\dt},
\end{equation}
\begin{equation}
  \Phi^{n+1/2} = \bar{\Phi}^t = \frac{\Phi^{n+1} + \Phi^{n}}{2},
\end{equation}
where $\Delta t$ is a time increment. 
For derivations in the subsequent subsections, 
the following discrete relational formula is used \citep{Morinishi_1996a,Morinishi_1998}:
\begin{equation}
  \delta_{\xi_i} \Psi \bar{\Phi}^{\xi_i} 
  = \overline{\Psi \delta_{\xi_i} \bar{\Phi}^{\xi_i}}^{\xi_i} 
  + \Phi \delta_{\xi_i} \Psi,
\end{equation}
\begin{equation}
  \bar{\Psi}^t \delta_t \Phi + \bar{\Phi}^t \delta_t \Psi = \delta_t \Psi \Phi,
\end{equation}
\begin{equation}
  \frac{1}{2} \delta_t \Phi^2 
  = \bar{\Phi}^t \delta_t \Phi.
\end{equation}

\subsection{Discretization of the Lorentz force}

This study uses a staggered grid. 
The velocities, $u_1$, $u_2$, and $u_3$, are defined 
at the cell interfaces, $(i+1/2, j, k)$, $(i, j+1/2, k)$, and $(i, j, k+1/2)$, respectively. 
As with the velocity field, 
the magnetic flux densities, $B_1$, $B_2$, and $B_3$, are defined 
at the cell interfaces, $(i+1/2, j, k)$, $(i, j+1/2, k)$, and $(i, j, k+1/2)$, respectively. 
The definition point of the electric field is different from that of the magnetic field. 
The current densities, $j_1$, $j_2$, and $j_3$, are defined at the midpoints of the cell edge, 
$(i, j+1/2, k+1/2)$, $(i+1/2, j, k+1/2)$, and $(i+1/2, j+1/2, k)$, respectively. 
The electric field $E_i$ is similar. 
The method of spatially shifting the definition points of the electric 
and magnetic fields is similar to that described in \citep{Yee_1966}. 
However, when the electric potential is obtained from the charge conservation law 
(\ref{electric_charge_law}) using Ohm's law, 
the current densities, $j_1$, $j_2$, and $j_3$, are defined 
at the cell interfaces in the same manner as the velocity. 
Scalar quantities such as pressure and internal energy are defined 
at the cell center $(i, j, k)$.

In this study, the nonconservative Lorentz force is obtained 
through the weighted interpolation of magnetic flux and current densities 
using the Jacobian \citep{Yanaoka_2023}. 
The nonconservative Lorentz force is expressed discretely as follows:
\begin{equation}
  F_i = \frac{1}{Al^2} \frac{1}{\bar{J}^{\xi_i}} 
  \overline{\epsilon_{ijk} \overline{\bar{J}^{\xi_i}}^{\xi_k} 
  j_j \bar{B_k}^{\xi_i}}^{\xi_k}.
  \label{compact_interpolation_F^nc}
\end{equation}
Current density is used in Eq. (\ref{compact_interpolation_F^nc}), 
and therefore, the current density that satisfies the charge conservation law should be used. 
In this study, the Lorentz force is calculated using the current density 
defined at the midpoint of the cell edge. 
The current density (\ref{Ampere}) is discretized as follows:
\begin{equation}
  j_i = \epsilon_{ijk} \delta_{x_j} B_k.
  \label{compact_interpolation_J}
\end{equation}
The charge conservation law $\partial_i j_i = 0$ is satisfied 
at the grid point $(i+1/2, j+1/2, k+1/2)$ as follows.
\begin{equation}
  \frac{\partial j_i}{\partial x_i} 
  = \delta_{x_1} (\delta_{x_2} B_3 - \delta_{x_3} B_2) 
  + \delta_{x_2} (\delta_{x_3} B_1 - \delta_{x_1} B_3) 
  + \delta_{x_3} (\delta_{x_1} B_2 - \delta_{x_2} B_1) = 0.
\end{equation}
In this study, compact interpolation refers to calculating the Lorentz force 
via the interpolation defined by Eq. (\ref{compact_interpolation_F^nc}) 
using Eq. (\ref{compact_interpolation_J}) \citep{Yanaoka_2023}.

\subsection{Discretization of mass, momentum, and internal energy equations}

This study uses the fully conservative finite difference method 
proposed in \citep{Morinishi_2009,Morinishi_2010} 
for discretization of the mass and momentum conservation equations. 
Using the same discretization method, 
Eqs. (\ref{mass}) and (\ref{momentum}) are discretized as
\begin{equation}
  \frac{Wo^2}{Re} 
  \delta_t \rho + \frac{1}{J} \delta_{\xi_j} U_j = 0,
\end{equation}
\begin{align}
  \frac{Wo^2}{Re} \frac{1}{\bar{J}^{\xi_i}} 
  \delta_t \overline{J \bar{\rho}^t}^{\xi_i} u_i 
  + \frac{1}{\bar{J}^{\xi_i}} 
    \delta_{\xi_j} \overline{\bar{U_j}^t}^{\xi_i} \overline{\hat{u_i}}^{\xi_j} 
  &= - \frac{1}{\bar{J}^{\xi_i}} J \xi_{i,i} \delta_{\xi_i} \overline{\bar{p}^t}^t 
  + \frac{1}{Re} \frac{1}{\bar{J}^{\xi_i}} \delta_{\xi_j} J \xi_{j,j} \bar{\tau_{ij}}^t 
  \nonumber \\
  &
  + \frac{1}{Al^2} \frac{1}{\bar{J}^{\xi_i}} 
  \overline{\epsilon_{ijk} \overline{\bar{J}^{\xi_i}}^{\xi_k} 
  \bar{j_j}^t \overline{\bar{B_k}^t}^{\xi_i}}^{\xi_k}, 
  \label{momentum_fdm}
\end{align}
\begin{align}
  \frac{Wo^2}{Re} \frac{1}{J} 
  \delta_t J \bar{\rho}^t e 
  + \frac{1}{J} \delta_{\xi_j} \bar{U_j}^t \overline{\hat{e}}^{\xi_j} 
  &= - \frac{\kappa}{Re Pr} \frac{1}{J} \delta_{\xi_j} J \xi_{j,j} \bar{q_j}^t 
  - (\kappa - 1) (\kappa Ma^2 \overline{\bar{p}^t}^t + 1) \frac{1}{J} \delta_{\xi_i} J \xi_{i,i} \hat{u_i} 
  \nonumber \\
  &
  + \frac{\kappa (\kappa - 1) Ma^2}{Re} \frac{1}{J} \bar{\tau_{ij}}^t J \xi_{j,j} \delta_{\xi_j} \hat{u_j} 
  + \frac{\kappa (\kappa - 1) Ma^2}{Al^2 Re_m} \frac{1}{J} 
    \overline{\overline{J \bar{j_i}^t \bar{j_i}^t}^{\xi_j}}^{\xi_k},
  \label{energy_e_fdm}
\end{align}
respectively, where $U_j$ is the mass flux defined as
\begin{equation}
  U_j = J \xi_{j,j} \overline{\bar{\rho}^t}^{\xi_j} u_j.
\end{equation}
The compatibility of the convective term \citep{Morinishi_1998} is maintained 
by calculating the interpolated value $\overline{U_i}^{\xi_i}$ 
using the contravariant velocity $\xi_{i,i}u_i$ 
and discretizing the convection term. 
Additionally, $\hat{u_i}$ and $\hat{e}$ in the above equation are the square-root density weighted interpolation, 
defined as follows \citep{Morinishi_2009, Morinishi_2010}:
\begin{equation}
  \hat{u_i} 
  = \frac{\overline{\sqrt{\overline{J \bar{\rho}^t}^{\xi_i}} u_i}^t}
         {\overline{\sqrt{\overline{J \bar{\rho}^t}^{\xi_i}}}^t}, \quad 
  \hat{e} 
  = \frac{\overline{\sqrt{\bar{\rho}^t} e}^t}
         {\overline{\sqrt{\bar{\rho}^t}}^t}, \quad 
  \bar{\rho}^t = \frac{\rho^{n+3/2} + \rho^{n+1/2}}{2}.
\end{equation}
This interpolated value was introduced to construct a fully conservative finite difference scheme 
\citep{Morinishi_2009, Morinishi_2010}. 
Double time interpolation of pressure $\overline{\bar{p}^t}^t$ was introduced 
to treat pressure implicitly \citep{Wall_et_al_2002}. 
We will discuss this in detail in the later section \ref{numerical_method}.

In the energy equations (\ref{energy_E}) and (\ref{energy_M}), 
there is a term $j_i^2/Re_m$ that represents Joule heat. 
The two terms cancel each other out in the total energy equation (\ref{energy_Et}), 
and the term $j_i^2/Re_m$ does not appear. 
To predict energy conversion correctly, 
the term $j_i^2/Re_m$ in the two equations (\ref{energy_E}) and (\ref{energy_M}) 
should be calculated by the same interpolation form. 
In this study, the heat generation term is found in Eq. (\ref{energy_e_fdm}) 
using the same current density used to calculate the Lorentz force. 
The same current density is also used in the Faraday equation, 
as explained in the following subsection \ref{Faraday_fdm}. 
Similarly to the term $j_i^2/Re_m$ that appears in the magnetic energy discretization equation, 
the heat generation term in Eq. (\ref{energy_e_fdm}) is calculated 
by weighted interpolation using the Jacobian. 
The method of calculating this heating term does not affect the energy conservation properties 
in an ideal inviscid MHD flow. 
The derivation of the magnetic energy equation is explained in Subsection \ref{subsection_magnetic_energy_eq}.

\subsection{Discretization of Faraday's equation}
\label{Faraday_fdm}

Here, I describe the discretization of Faraday's equation (\ref{Faraday}) 
and velify that the magnetic flux density equation (\ref{magnetic_field}) 
can be discretely derived from Eq. (\ref{Faraday}) 
using compact interpolation \citep{Yanaoka_2023}. 
Equation (\ref{Faraday}) is discretized as follows:
\begin{align}
  \frac{Wo^2}{Re} \delta_t B_i 
  &= - \epsilon_{ijk} \delta_{x_j} \bar{E_k}^t 
  = - \epsilon_{ijk} \frac{1}{\bar{J}^{\xi_i}} \delta_{\xi_j} J \xi_{j,j} \bar{E_k}^t 
  \nonumber \\
  &= 
  - \epsilon_{ijk} \frac{1}{\bar{J}^{\xi_i}} \delta_{\xi_j} J \xi_{j,j} 
    \left( \frac{1}{Re_m} \bar{j_k}^t - \epsilon_{klm} 
    \overline{\hat{u_l}}^{\xi_m} \overline{\bar{B_m}^t}^{\xi_l} \right) 
  \nonumber \\
  &= 
    \frac{1}{\bar{J}^{\xi_i}} \delta_{\xi_j} J \xi_{j,j} 
    \left( \epsilon_{ijk} \epsilon_{klm} 
    \overline{\hat{u_l}}^{\xi_m} \overline{\bar{B_m}}^{\xi_l} \right) 
  - \frac{1}{Re_m} \frac{1}{\bar{J}^{\xi_i}} 
    \epsilon_{ijk} \delta_{\xi_j} J \xi_{j,j} \bar{j_k}^t 
  \nonumber \\
  &= 
  - \frac{1}{\bar{J}^{\xi_i}} \delta_{\xi_j} \left( 
    J \xi_{j,j} \overline{\hat{u_j}}^{\xi_i} \overline{\bar{B_i}^t}^{\xi_j} 
  - J \xi_{j,j} \overline{\bar{B_j}^t}^{\xi_i} \overline{\hat{u_i}}^{\xi_j} \right) 
  - \frac{1}{Re_m} \frac{1}{\bar{J}^{\xi_i}} 
    \epsilon_{ijk} \delta_{\xi_j} J \xi_{j,j} \bar{j_k}^t, 
  \label{magnetic_flux_density_fdm}
\end{align}
where $\epsilon_{ijk}\epsilon_{klm} = \delta_{il}\delta_{jm}-\delta_{im}\delta_{jl}$ is used. 
Evidently, Eqs. (\ref{Faraday}) and (\ref{magnetic_field}) can be discretely transformed into each other. 
Furthermore, when $Re_m = \infty$, 
the discretization equation of Eq. (\ref{magnetic_field}) also has a conservative form. 
$\hat{u_j}$ in the above equation is the square-root density weighted interpolation. 
The work $u_i \epsilon_{ijk} j_j B_k$ due to the Lorentz force 
appearing in Eqs. (\ref{energy_E}) and (\ref{energy_M}) cancel each other out in Eq. (\ref{energy_Et}). 
Therefore, the work caused by the Lorentz force appearing in Eqs. (\ref{energy_E}) and (\ref{energy_M}) 
must be approximated using the same form. 
Therefore, it is necessary to use the velocity $\hat{u_j}$ used in the momentum equation. 
The use of this velocity is very significant to conserve the total amount of energy.

The discretization method for the convection terms, 
$\partial_{\xi_j} J \xi_{j,j} u_j B_i$ and $- \partial_{\xi_j} J \xi_{j,j} B_j u_i$, 
in this equation is different from that for the convection terms in the momentum equation (\ref{momentum}). 
When discretizing the convection term $\partial_{\xi_j} J \xi_{j,j} u_j u_i$, 
the interpolated value of $\overline{\bar{U_j}^t}^{\xi_i} \overline{\hat{u_i}}^{\xi_j}$ is used 
in Eq. (\ref{momentum_fdm}) to satisfy the transformation of the convection terms \citep{Morinishi_1998}. 
In the conservation form, $\partial_{\xi_j} B_j B_i$, of Lorentz force, 
the interpolated value $\overline{J \xi_{j,j}  B_j}^{\xi_i} \bar{B_i}^{\xi_j}$ 
is used in Eq. (\label{compact_interpolation_F^c}). 
In Eq. (\ref{magnetic_flux_density_fdm}), the interpolated value 
$(J \xi_{j,j}) \overline{\hat{u_j}}^{\xi_i} \overline{\bar{B_i}^t}^{\xi_j}$ is used. 
As with the momentum equation, 
Eq. (\ref{Faraday}) can be discretized 
using each contravariant component of velocity and magnetic flux density. 
However, the magnetic energy equation (\ref{energy_M}) cannot be derived 
discretely from Faraday's equation (\ref{Faraday}).

Calculating the divergence of the formula (\ref{magnetic_flux_density_fdm}) 
at the cell center reveals that the time variation of $\partial_i B_i$ 
is discretely zero, as follows:
\begin{align}
  \frac{Wo^2}{Re} \delta_t \delta_{x_i} B_i 
  &= - \delta_{x_i} \epsilon_{ijk} \delta_{x_j} \bar{E_k}^t 
  \nonumber \\
  &= 
  - \delta_{x_1} \left( \delta_{x_2} \bar{E_3}^t - \delta_{x_3} \bar{E_2}^t \right) 
  - \delta_{x_2} \left( \delta_{x_3} \bar{E_1}^t - \delta_{x_1} \bar{E_3}^t \right) 
  - \delta_{x_3} \left( \delta_{x_1} \bar{E_2}^t - \delta_{x_2} \bar{E_1}^t \right) 
  = 0.
  \label{time_variation_constraint}
\end{align}

\subsection{Discretization of the magnetic vector potential equation}

As the magnetic vector potential $A_i$ is defined 
as $\check{B_i} = \epsilon_{ijk}\partial_j A_k$, 
it must satisfy the constraint (\ref{divergence_magnetic}) of magnetic flux density. 
Similarly to the current density $j_i$, 
the magnetic flux densities, $\check{B_1}$, $\check{B_2}$, and $\check{B_3}$, 
associated with the magnetic vector potential $A_i$ are defined 
at the midpoint of the cell edge, $(i, j+1/2, k+1/2)$, $(i+1/2, j, k+1/2)$, 
and $(i+1/2, j+1/2, k)$ , respectively. 

Equation (\ref{vector_potential}) is discretized as follows:
\begin{equation}
  \frac{Wo^2}{Re} \delta_t A_i 
  + \epsilon_{ijk} \overline{\check{\bar{B_j}^t} \overline{\hat{u_k}}^{\xi_i}}^{\xi_k} 
  = - \frac{1}{\bar{J}^{\xi_i}} \delta_{\xi_i} J \xi_{i,i} \bar{\psi}^t 
  - \frac{1}{Re_m} \bar{j_i}^t,
  \label{vector_potential_fdm}
\end{equation}
\begin{equation}
  j_i = \epsilon_{ijk} \delta_{x_j} \check{B_k},
\end{equation}
\begin{equation}
  \check{B_i} = \epsilon_{ijk} \delta_{x_j} A_k.
\end{equation}
In Eq. (\ref{vector_potential_fdm}), 
the same velocity $\hat{u_k}$ used in Eq. (\ref{magnetic_flux_density_fdm}), 
which is a discretized version of the Faraday equation, is used. 
We used the velocity $\hat{u_k}$ to discretely derive the magnetic helicity equation. 
The magnetic helicity equation (\ref{magnetic_helicity_eq}) is derived from 
the Faraday equation (\ref{Faraday}) and the magnetic vector potential equation (\ref{vector_potential}). 
At this time, the term $B_i (\epsilon_{ijk} u_j B_k)$ appears in Eq. (\ref{magnetic_helicity_eq}). 
This term arises from two equations, (\ref{Faraday}) and (\ref{vector_potential}). 
Therefore, in Eq. (\ref{vector_potential_fdm}), 
it is necessary to use the same velocity $\hat{u_k}$ that was used in Eq. (\ref{magnetic_flux_density_fdm}).

The divergence-free condition $\partial_i B_i = 0$ of the magnetic flux density 
of Eq. (\ref{divergence_magnetic}) is satisfied 
at the grid point $(i+1/2, j+1/2, k+1/2)$ as follows:
\begin{equation}
  \frac{\partial \check{B_i}}{\partial x_i} 
  = \delta_{x_1} (\delta_{x_2} A_3 - \delta_{x_3} A_2) 
  + \delta_{x_2} (\delta_{x_3} A_1 - \delta_{x_1} A_3) 
  + \delta_{x_3} (\delta_{x_1} A_2 - \delta_{x_2} A_1) = 0.
\end{equation}
As described above, 
when Eq. (\ref{vector_potential}) is discretized as Eq.  (\ref{vector_potential_fdm}), 
the magnetic vector potential that satisfies the constraint condition 
of magnetic flux density can be obtained. 

Subsequently, the magnetic flux density equation (\ref{magnetic_field}) can be obtained 
by rotating the magnetic vector potential equation (\ref{vector_potential}). 
Coordinate transformations are not required; 
hence, the following discretized magnetic vector potential equation is used:
\begin{equation}
  \frac{Wo^2}{Re} \delta_t A_k 
  = - \epsilon_{klm} \overline{\check{B_l} \overline{\hat{u_m}}^{x_k}}^{x_m} 
  - \delta_{x_k} \bar{\psi}^t - \frac{1}{Re_m} \bar{j_k}^t.
\end{equation}
Calculating the rotation of the above equation gives 
the discretized equation for the magnetic flux density as follows:
\begin{align}
  \frac{Wo^2}{Re} \delta_t \epsilon_{ijk} \delta_{x_j} A_k 
  = \frac{Wo^2}{Re} \delta_t \check{B_i} 
  &= 
  - \epsilon_{ijk} \delta_{x_j} \left( 
  \epsilon_{klm} \overline{\check{\bar{B_l}^t} \overline{\hat{u_m}}^{x_k}}^{x_m} 
  - \delta_{x_k} \bar{\psi}^t - \frac{1}{Re_m} \bar{j_k}^t \right) 
  \nonumber \\
  &= 
  - \epsilon_{ijk} \epsilon_{klm} \delta_{x_j} \overline{\check{\bar{B_l}^t} \overline{\hat{u_m}}^{x_k}}^{x_m} 
  - \frac{1}{Re_m} \epsilon_{ijk} \delta_{x_j} \bar{j_k}^t,
\end{align}
where $\epsilon_{ijk} \epsilon_{klm} = \delta_{il}\delta_{jm}-\delta_{im}\delta_{jl}$. 
The equations for the magnetic flux densities, 
$\check{B_1}$, $\check{B_2}$, and $\check{B_3}$, can be discretely derived 
at the midpoints of the cell edge, $(i, j+1/2, k+1/ 2) $, $(i+1/2, j, k+1/2)$, 
and $(i+1/2, j+1/2, k)$, respectively.

The Lorentz force can also be obtained using the magnetic vector potential. 
The nonconservative Lorentz force is discretely expressed as follows:
\begin{equation}
  F_i = \frac{1}{Al^2} \frac{1}{\bar{J}^{\xi_i}} 
  \overline{\epsilon_{ijk} 
  \overline{J \bar{j_j}^{\xi_j}}^{\xi_i} \overline{\check{B_k}}^{\xi_j}}^{\xi_k}.
  \label{vector_potential_F}
\end{equation}
As the Lorentz force (\ref{vector_potential_F}) is determined by the second-order differential 
of the magnetic vector potential, the above-discretized formula may decrease accuracy. 
Additionally, the conservation of momentum and total energy deteriorate 
in an ideal periodic inviscid MHD flow 
because Eq. (\ref{vector_potential_F}) cannot be discretely transformed into a conservative form of the Lorentz force.

Furthermore, the magnetic energy equation can be derived 
using the magnetic flux density $\check{B_i}$ calculated from the magnetic vector potential $A_i$. 
However, as many transformations of the dependent variable occur, 
numerous interpolations are required. 
Therefore, the magnetic energy equation is derived 
from the discretized Faraday's equation, 
namely, the discretized magnetic flux density equation (\ref{magnetic_flux_density_fdm}), 
as described in Subsection \ref{subsection_magnetic_energy_eq}.

\subsection{Derivation of the magnetic energy equation}
\label{subsection_magnetic_energy_eq}

If the magnetic energy equation (\ref{energy_M}) can be derived discretely 
from Faraday's equation (\ref{Faraday}), 
the discrete total energy conservation equation can be derived. 
The magnetic energy $\rho M$ is defined at the cell center $(i, j, k)$ as follows:
\begin{equation}
  \bar{\rho}^t M = \frac{\kappa (\kappa - 1) Ma^2}{2 Al^2} \frac{1}{J} 
  \overline{\bar{J}^{\xi_i} B_i B_i}^{\xi_i} 
\end{equation}
Calculating the inner product of the discretized Faraday's equation (\ref{magnetic_flux_density_fdm}) 
with the magnetic flux density $B_i^{n+1/2}$ yields 
the discretized equation of the magnetic energy 
as described in \citep{Yanaoka_2023}
\begin{align}
  \frac{Wo^2}{Re} \delta_t \bar{\rho}^t M 
  &= - \frac{\kappa (\kappa - 1) Ma^2}{Al^2} \frac{1}{J} 
  \overline{\bar{J}^{x_i} \bar{B_i}^t \epsilon_{ijk} \delta_{x_j} \bar{E_k}^t}^{x_i} 
  \nonumber \\
  &= - \frac{\kappa (\kappa - 1) Ma^2}{Al^2} \frac{1}{J} 
  \overline{\bar{B_i}^t \epsilon_{ijk} \delta_{\xi_j} J \xi_{j,j} \bar{E_k}^t}^{\xi_i} 
  \nonumber \\
  &= - \frac{\kappa (\kappa - 1) Ma^2}{Al^2} \frac{1}{J} \left( 
  \overline{\delta_{\xi_j} \epsilon_{ijk} J \xi_{j,j} \bar{E_k}^t \overline{\bar{B_i}^t}^{\xi_j}}^{\xi_i} 
  - \overline{\overline{\epsilon_{ijk} J \xi_{j,j} \bar{E_k}^t \delta_{\xi_j} \bar{B_i}^t}^{\xi_j}}^{\xi_i} 
  \right) 
  \nonumber \\
  &= - \frac{\kappa (\kappa - 1) Ma^2}{Al^2} \frac{1}{J} \left( 
  \overline{\delta_{\xi_j} \epsilon_{ijk} J \xi_{j,j} \bar{E_k}^t \overline{\bar{B_i}^t}^{\xi_j}}^{\xi_i} 
  + \overline{\overline{\bar{E_k}^t J \epsilon_{ijk} \delta_{x_i} \bar{B_j}^t}^{\xi_i}}^{\xi_j} 
  \right) 
  \nonumber \\
  &= - \frac{\kappa (\kappa - 1) Ma^2}{Al^2} \frac{1}{J} \left( 
  \overline{\delta_{\xi_j} \epsilon_{ijk} J \xi_{j,j} \bar{E_k}^t \overline{\bar{B_i}^t}^{\xi_j}}^{\xi_i} 
  + \overline{\overline{\bar{E_k}^t J \bar{j_k}^t}^{\xi_i}}^{\xi_j} 
  \right).
\end{align}
As the electric field is defined at the same point as the current density, 
it is given by $E_k = j_k/Re_m-\epsilon_{kij} \overline{\hat{u_i}}^{\xi_j} \bar{B_j}^{\xi_i}$. 
The above equation can be transformed using $E_k$ as follows:
\begin{align}
  \frac{Wo^2}{Re} \delta_t \bar{\rho}^t M 
  &= - \frac{\kappa (\kappa - 1) Ma^2}{Al^2} \frac{1}{J} \left[ 
  \overline{\delta_{\xi_j} \epsilon_{ijk} J \xi_{j,j} \bar{E_k}^t \overline{\bar{B_i}^t}^{\xi_j}}^{\xi_i} 
  + \overline{\overline{J \left( 
  \frac{1}{Re_m} \bar{j_k}^t - \epsilon_{kij} \overline{\hat{u_j}}^{\xi_j} \overline{\bar{B_j}^t}^{\xi_i} 
  \right) \bar{j_k}^t}^{\xi_i}}^{\xi_j} \right] 
  \nonumber \\
  &= - \frac{\kappa (\kappa - 1) Ma^2}{Al^2} \frac{1}{J} \left[ 
  \overline{\delta_{\xi_j} \epsilon_{ijk} J \xi_{j,j} \bar{E_k}^t \overline{\bar{B_i}^t}^{\xi_j}}^{\xi_i} 
  + \overline{\overline{J \frac{1}{Re_m} \bar{j_k}^t \bar{j_k}^t}^{\xi_i}}^{\xi_j} 
  - \overline{\overline{(J \epsilon_{kij} \overline{\hat{u_i}}^{\xi_j} \overline{\bar{B_j}^t}^{\xi_i}) \bar{j_k}^t}^{\xi_i}}^{\xi_j} 
  \right] 
  \nonumber \\
  &= - \frac{\kappa (\kappa - 1) Ma^2}{Al^2} \frac{1}{J} \left[ 
  \overline{\delta_{\xi_j} \epsilon_{ijk} J \xi_{j,j} \bar{E_k}^t \overline{\bar{B_i}^t}^{\xi_j}}^{\xi_i} 
  + \overline{\overline{\frac{1}{Re_m} J \bar{j_k}^t \bar{j_k}^t}^{\xi_i}}^{\xi_j} 
  + \overline{\overline{\overline{\hat{u_i}}^{\xi_k} (J \epsilon_{ijk} \bar{j_j}^t \overline{\bar{B_k}^t}^{\xi_i})}^{\xi_i}}^{\xi_k} 
  \right]. \quad
  \label{energy_M_fdm}
\end{align}
The third term on the right side is the work done by the Lorentz force. 
Additionally, the third term is interpolated using the Jacobian $J$. 
The interpolation form of the Lorentz force is consistent with 
Eq. (\ref{compact_interpolation_F^nc}). 
In discretizing Eq. (\ref{magnetic_flux_density_fdm}), 
the velocity $\hat{u_i}$ was used. 
Therefore, in the discretized magnetic energy equation (\ref{energy_M_fdm}), 
the work done by the Lorentz force appears as 
$\overline{\overline{\overline{\hat{u_i}}^{\xi_k} (J \epsilon_{ijk} \bar{j_j}^t \overline{\bar{B_k}^t}^{\xi_i})}^{\xi_i}}^{\xi_k}$ 
using the velocity $\hat{u_i}$.

Further, the time derivative term of Eq. (\ref{energy_M}) is considered. 
Applying the implicit midpoint rule to the time derivative affords 
the time derivative of the magnetic energy as follows:
\begin{align}
  \frac{Wo^2}{Re} \frac{\kappa (\kappa - 1) Ma^2}{Al^2} \bar{B_i}^t \delta_t B_i 
  &= \frac{Wo^2}{Re} \frac{\kappa (\kappa - 1) Ma^2}{Al^2} 
    \frac{1}{J} \overline{\bar{J}^{\xi_i} \bar{B_i}^t \delta_t B_i}^{\xi_i} 
  \nonumber \\
  &= \frac{Wo^2}{Re} \frac{\kappa (\kappa - 1) Ma^2}{Al^2} 
    \frac{1}{J} \overline{\bar{J}^{\xi_i} \delta_t B_i^2/2}^{\xi_i} 
  = \frac{Wo^2}{Re} \delta_t \bar{\rho}^t M.
  \label{energy_M_time}
\end{align}
The magnetic energy equation (\ref{energy_M}) can be derived discretely 
in both time and space directions.

\subsection{Derivation of the total energy equation}

The total energy $\rho E$, which is the sum of internal energy and kinetic energy, 
is defined at the cell center $(i, j, k)$ as follows:
\begin{equation}
  \bar{\rho}^t E = \bar{\rho}^t e 
  + \frac{\kappa (\kappa - 1) Ma^2}{2} \frac{1}{J} 
    \overline{\overline{J \bar{\rho}^t}^{\xi_i} u_i u_i}^{\xi_i}.
\end{equation}
When no magnetic field is applied, 
the total energy equation can be derived discretely 
from the conservation equations of momentum and internal energy \citep{Morinishi_2009, Morinishi_2010}. 
By calculating the inner product of the discretized momentum equation (\ref{momentum}) 
and the velocity $\hat{u_i}^{n+1/2}$, 
the discretized kinetic energy equation can be derived \citep{Morinishi_2009, Morinishi_2010}. 
For $Re = \infty$, the total energy equation (\ref{energy_Et}) is derived discretely as follows:
\begin{align}
  \delta_t \bar{\rho}^t E 
  &= - \frac{1}{J} \delta_{\xi_j} \left[ 
    \bar{U_j}^t \overline{\hat{e}}^{\xi_j} 
  + \frac{\kappa (\kappa - 1) Ma^2}{2} 
    \overline{\overline{\bar{U_j}^t}^{\xi_i} 
    \widetilde{\hat{u_i} \hat{u_i}}^{\xi_j}}^{\xi_i} \right] 
  - (\kappa - 1) \frac{1}{J} \delta_{\xi_i} J \xi_{i,i} 
  \hat{u_i} \overline{(\kappa Ma^2 \overline{\bar{p}^t}^t + 1)}^{\xi_i} 
  \nonumber \\
  &
  + \frac{\kappa (\kappa - 1) Ma^2}{Al^2} 
  \frac{1}{J} \overline{
  \hat{u_i} \overline{\epsilon_{ijk} \overline{\bar{J}^{\xi_i}}^{\xi_k} 
  \bar{j_j}^t \overline{\bar{B_k}^t}^{\xi_i}}^{\xi_k}}^{\xi_i}, 
  \label{inviscid_energy_E_fdm}
\end{align}
where $Wo = \sqrt{Re}$ is set to remove $Re$. 
In Eq. (\ref{inviscid_energy_E_fdm}), 
the work from the Lorentz force appears in the last term on the right side.

For $Re_m = \infty$, 
the discretized magnetic energy equation (\ref{energy_M_fdm}) is expressed as follows:
\begin{equation}
  \delta_t \bar{\rho}^t M 
  = - \frac{\kappa (\kappa - 1) Ma^2}{Al^2} \frac{1}{J} \left[ 
  \overline{\delta_{\xi_j} \epsilon_{ijk} J \xi_{j,j} \bar{E_k}^t \overline{\bar{B_i}^t}^{\xi_j}}^{\xi_i} 
  + \overline{\overline{\overline{\hat{u_i}}^{\xi_k} (J \epsilon_{ijk} \bar{j_j}^t \overline{\bar{B_k}^t}^{\xi_i})}^{\xi_i}}^{\xi_k} 
  \right].
  \label{ideal_energy_M_fdm}
\end{equation}

The total energy $\rho E_t = \rho E + \rho M$ is defined at the cell center $(i, j, k)$ as follows:
\begin{equation}
  \bar{\rho}^t E_t = \bar{\rho}^t e 
  + \frac{\kappa (\kappa - 1) Ma^2}{2} \frac{1}{J} \left( 
    \overline{\overline{J \bar{\rho}^t}^{\xi_i} u_i u_i}^{\xi_i}
  + \frac{1}{Al^2} \overline{\bar{J}^{\xi_i} B_i B_i}^{\xi_i} \right).
\end{equation}
Taking the sum of Eqs. (\ref{inviscid_energy_E_fdm}) and (\ref{ideal_energy_M_fdm}) 
yields the equation for the total energy $\rho E_t = \rho E + \rho M$ as follows:
\begin{align}
  \delta_t \bar{\rho}^t E_t 
  &= 
  - \frac{1}{J} \delta_{\xi_j} \left[ 
    \bar{U_j}^t \overline{\hat{e}}^{\xi_j} 
  + \frac{\kappa (\kappa - 1) Ma^2}{2} 
    \overline{\overline{\bar{U_j}^t}^{\xi_i} 
    \widetilde{\hat{u_i} \hat{u_i}}^{\xi_j}}^{\xi_i} \right] 
  - (\kappa - 1) \frac{1}{J} \delta_{\xi_i} J \xi_{i,i} 
  \hat{u_i} \overline{(\kappa Ma^2 \overline{\bar{p}^t}^t + 1)}^{\xi_i} 
  \nonumber \\
  & 
  + \frac{\kappa (\kappa - 1) Ma^2}{Al^2} \frac{1}{J} \left[ 
  \overline{
  \hat{u_i} \overline{\epsilon_{ijk} \overline{\bar{J}^{\xi_i}}^{\xi_k} 
  \bar{j_j}^t \overline{\bar{B_k}^t}^{\xi_i}}^{\xi_k}}^{\xi_i} 
  - \overline{\delta_{\xi_j} J \xi_j \epsilon_{ijk} \bar{E_k}^t \overline{\bar{B_i}^t}^{\xi_j}}^{\xi_i} 
  - \overline{\overline{\overline{\hat{u_i}}^{\xi_k} (J \epsilon_{ijk} \bar{j_j}^t \overline{\bar{B_k}^t}^{\xi_i})}^{\xi_i}}^{\xi_k} \right].
  \label{total_energy_eq1_fdm}
\end{align}
By applying the implicit midpoint rule to the time derivative, 
the total energy equation (\ref{energy_inviscid_Et}) can be derived discretely 
in both time and space directions \citep{Yanaoka_2023}. 
The work done by the Lorentz force appears in the third and fifth terms on the right side. 
Two terms $\overline{\hat{u_i} \overline{\epsilon_{ijk} \overline{\bar{J}^{\xi_i}}^{\xi_k} \bar{j_j}^t \overline{\bar{B_k}^t}^{\xi_i}}^{\xi_k}}^{\xi_i}$ 
and $\overline{\overline{\overline{\hat{u_i}}^{\xi_k} (J \epsilon_{ijk} \bar{j_j}^t \overline{\bar{B_k}^t}^{\xi_i})}^{\xi_i}}^{\xi_k}$ 
have the same form of weighted interpolation by the Jacobian 
but with a different interpolation form. 
If these terms approximately cancel each other, 
the total energy is preserved even discretely.

\subsection{Derivation of the magnetic helicity equation}

The magnetic helicity $H_m$ shown by the formula (\ref{magnetic_helicity}) is defined 
at the cell center $(i, j, k)$ as follows:
\begin{equation}
  H_m = \frac{1}{Al^2} \kappa (\kappa - 1) Ma^2 
  \frac{1}{J} \overline{\bar{J}^{\xi_i} B_i A_i}^{\xi_i}.
\end{equation}
The time derivative of the magnetic helicity is expressed discretely as follows:
\begin{align}
  \delta_t H_m 
  &= \frac{1}{Al^2} \kappa (\kappa - 1) Ma^2 
    \frac{1}{J} \overline{\delta_t \bar{J}^{\xi_i} (B_i A_i)}^{\xi_i} 
  \nonumber \\
  &= \frac{1}{Al^2} \kappa (\kappa - 1) Ma^2 \frac{1}{J} 
  \left( \overline{\bar{J}^{\xi_i} \bar{A_i}^t \delta_t B_i}^{\xi_i} 
       + \overline{\bar{J}^{\xi_i} \bar{B_i}^t \delta_t A_i}^{\xi_i} \right).
  \label{magnetic_helicity_time_fdm}
\end{align}
By calculating the inner product of the discretized equation (\ref{magnetic_flux_density_fdm}) 
at $Re_m = \infty$ and the magnetic vector potential $A_i^{n+1/2}$, 
the first term of Eq. (\ref{magnetic_helicity_time_fdm}) is obtained as follows:
\begin{align}
  \overline{\bar{J}^{\xi_i} \bar{A_i}^t \delta_t B_i}^{\xi_i} 
  &= - \overline{\bar{J}^{\xi_i} \bar{A_i}^t \epsilon_{ijk} \frac{1}{\bar{J}^{\xi_j}} \delta_{\xi_j} J \xi_{j,j} \bar{E_k}^t}^{\xi_i} 
  = - \epsilon_{ijk} \left( \delta_{\xi_j} J \xi_{j,j} \overline{\bar{A_i}^t}^{\xi_j} \bar{E_k}^t 
  - \overline{\bar{E_k}^t \delta_{\xi_j} J \xi_{j,j} \bar{A_i}^t}^{\xi_j} \right) 
  \nonumber \\
  &= - \delta_{\xi_j} \epsilon_{jki} J \xi_{j,j} \overline{\bar{A_i}^t}^{\xi_j} \bar{E_k}^t 
  + \overline{\epsilon_{kij} \bar{E_k}^t \delta_{\xi_i} J \xi_{i,i} \bar{A_j}^t}^{\xi_j} 
  \nonumber \\
  &= - \delta_{\xi_j} \epsilon_{jki} J \xi_{j,j} \overline{\bar{A_i}^t}^{\xi_j} \bar{E_k}^t 
  + \overline{(\epsilon_{kij} \overline{\hat{u_i}}^{\xi_j} \overline{\bar{B_j}^t}^{\xi_i}) (J \check{\bar{B_k}^t})}^{\xi_j}.
\end{align}
By calculating the inner product of the discretized equation (\ref{vector_potential_fdm}) 
at $Re_m = \infty$ and the magnetic flux density $B_i^{n+1/2}$, 
the second term of Eq. (\ref{magnetic_helicity_time_fdm}) is obtained as follows:
\begin{align}
  \overline{\bar{J}^{\xi_i} \bar{B_i}^t \delta_t A_i}^{\xi_i} 
  &= - \overline{
  \epsilon_{ijk} \bar{J}^{\xi_i} \bar{B_i}^t \overline{\check{\bar{B_j}^t} \overline{\hat{u_k}}^{\xi_i}}^{\xi_k} 
  }^{\xi_i} 
  - \overline{\bar{B_i}^t \delta_{\xi_i} J \xi_{i,i} \bar{\psi}^t}^{\xi_i} 
  \nonumber \\
  &= \overline{ 
  \epsilon_{ijk} \bar{J}^{\xi_i} \bar{B_i}^t \overline{\check{\bar{B_j}^t} \overline{\hat{u_k}}^{\xi_i}}^{\xi_k} 
  }^{\xi_i} 
  - \delta_{\xi_i} J \xi_{i,i} \bar{B_i}^t \overline{\bar{\psi}^t}^{\xi_i} 
  + \bar{\psi}^t \delta_{\xi_i} J \xi_{i,i} \bar{B_i}^t. 
  \label{magnetic_helocity_eq_fdm}
\end{align}
Therefore, 
the time derivative (\ref{magnetic_helicity_time_fdm}) of the magnetic helicity 
is expressed as follows:
\begin{align}
  \delta_t H_m 
  &= \frac{\kappa (\kappa - 1) Ma^2}{Al^2} \frac{1}{J} \left[ 
  - \delta_{\xi_j} \epsilon_{jki} J \xi_{j,j} \overline{\bar{A_i}^t}^{\xi_j} \bar{E_k}^t 
  + \overline{(\epsilon_{kij} \overline{\hat{u_i}}^{\xi_j} \overline{\bar{B_j}^t}^{\xi_i}) (J \check{\bar{B_k}^t})}^{\xi_j} 
  \right. \nonumber \\
  & \left. 
  + \overline{ 
  \epsilon_{ijk} \bar{J}^{\xi_i} \bar{B_i}^t \overline{\check{\bar{B_j}^t} \overline{\hat{u_k}}^{\xi_i}}^{\xi_k} 
  }^{\xi_i} 
  - \delta_{\xi_i} J \xi_{i,i} \bar{B_i}^t \overline{\bar{\psi}^t}^{\xi_i} 
  + \bar{\psi}^t \delta_{\xi_i} J \xi_{i,i} \bar{B_i}^t \right]. 
\end{align}
If $\partial_i B_i = 0$ is satisfied discretely, 
the last term on the right side of the above equation approaches zero asymptotically. 
Additionally, because the two vectors, $B_i$ and $\epsilon_{ijk} B_j u_k$, are orthogonal, 
their inner product is zero. 
However, the inner product of two vectors is not strictly zero 
in the discretized equation. 
Therefore, if $\overline{(\epsilon_{kij} \overline{\hat{u_i}}^{\xi_j} \overline{\bar{B_j}^t}^{\xi_i}) (J \check{\bar{B_k}^t})}^{\xi_j}$ 
and $\overline{\epsilon_{ijk} \bar{J}^{\xi_i} \bar{B_i}^t \overline{\check{\bar{B_j}^t} \overline{\hat{u_k}}^{\xi_i}}^{\xi_k}}^{\xi_i}$ 
discretely approach zeros, then magnetic helicity is preserved discretely.

\section{Numerical method}
\label{numerical_method}

This study adopts a spatiotemporal staggered grid, 
similar to the existing studies \citep{Wall_et_al_2002, Morinishi_2009, Morinishi_2010}. 
This method allows the transport quantity to be conserved discretely in the spatiotemporal direction. 
The Newton method is used to solve the unsteady solution. 
Using the implicit midpoint rule, Eqs. (\ref{mass}), (\ref{momentum}), (\ref{energy_e}), 
(\ref{Faraday}), and (\ref{vector_potential}) are given as
\begin{subequations}
\begin{equation}
  \frac{Wo^2}{Re} \frac{\rho^{n+3/2,m+1} - \rho^{n+1/2}}{\dt} 
  = H_{\rho}^{n+1,m+1},
  \label{implicit.mass}
\end{equation}
\begin{equation}
  H_{\rho}^{n+1,m+1} = - \frac{\partial (\rho u_j)^{n+1,m+1}}{\partial x_j},
  \label{implicit.Hrho}
\end{equation}
\end{subequations}
\begin{subequations}
\begin{equation}
  \frac{Wo^2}{Re} \frac{(\rho u_i)^{n+1,m+1} - (\rho u_i)^n}{\dt} 
  = H_{u_i}^{n+1/2,m+1} - \frac{\partial p^{n+1/2,m+1}}{\partial x_i},
  \label{implicit.u}
\end{equation}
\begin{equation}
  H_{u_i}^{n+1/2,m+1} 
  = - \frac{\partial (\rho u_j)^{n+1/2,m+1} \hat{u_i}^{n+1/2,m+1}}{\partial x_j} 
  + \frac{1}{Re} \frac{\partial \tau_{ij}^{n+1/2,m+1}}{\partial x_j} 
  + \frac{1}{Al^2} \epsilon_{ijk} j_j^{n+1/2,m+1} B_k^{n+1/2,m+1},
  \label{implicit.Hui}
\end{equation}
\end{subequations}
\begin{subequations}
\begin{equation}
  \frac{Wo^2}{Re} \frac{(\rho e)^{n+1,m+1} - (\rho e)^n}{\dt} 
  = H_e^{n+1/2,m+1} 
  - (\kappa - 1) (\kappa Ma^2 p^{n+1/2,m+1} + 1) 
  \frac{\partial \hat{u_i}^{n+1/2,m+1}}{\partial x_i},
  \label{implicit.e}
\end{equation}
\begin{align}
  H_e^{n+1/2,m+1} 
  &= - \frac{\partial (\rho u_j)^{n+1/2,m+1} e^{n+1/2,m+1}}{\partial x_j} 
  - \frac{\kappa}{Re Pr} \frac{\partial q_j^{n+1/2,m+1}}{\partial x_j} 
  \nonumber \\
  &
  + \frac{\kappa (\kappa - 1) Ma^2}{Re} 
  \tau_{ij}^{n+1/2,m+1} \frac{\partial \hat{u_i}^{n+1/2,m+1}}{\partial x_j} 
  + \frac{\kappa (\kappa - 1) Ma^2}{Al^2 Re_m} |j_i^{n+1/2,m+1}|^2,
  \label{implicit.He}
\end{align}
\end{subequations}
\begin{subequations}
\begin{equation}
  \frac{Wo^2}{Re} \frac{B_i^{n+1,m+1} - B_i^n}{\dt} = H_{B_i}^{n+1/2,m+1},
  \label{implicit.B}
\end{equation}
\begin{equation}
  H_{B_i}^{n+1/2} 
  = - \epsilon_{ijk} \frac{\partial E_k^{n+1/2,m+1}}{\partial x_j},
  \label{implicit.HBi}
\end{equation}
\end{subequations}
\begin{subequations}
\begin{equation}
  \frac{Wo^2}{Re} \frac{A_i^{n+1,m+1} - A_i^n}{\dt} = H_{A_i}^{n+1/2,m+1} 
  - \frac{\partial \psi^{n+1/2,m+1}}{\partial x_i},
  \label{implicit.A}
\end{equation}
\begin{equation}
  H_{A_i}^{n+1/2,m+1} 
  = - \epsilon_{ijk} B_j^{n+1/2,m+1} u_k^{n+1/2,m+1} 
  - \frac{1}{Re_m} j_i^{n+1/2,m+1},
  \label{implicit.Hai}
\end{equation}
\end{subequations}
where $n$ and $m$ indicate the time and Newton iterative levels, respectively. 
The temporal levels are defined as the $n$ level for the velocity, 
internal energy, and magnetic flux density 
and the $n+1/2$ level for the density and pressure. 
In addition, $\hat{u_i}$ and $\hat{e}$ in the above equation are 
the square-root density weighted interpolation, 
defined as follows \citep{Morinishi_2009, Morinishi_2010}:
\begin{equation}
  \hat{u_i}^{n+1/2,m+1} 
  = \frac{\overline{\sqrt{\overline{J \bar{\rho}^t}^{\xi_i}} u_i}^t}
         {\overline{\sqrt{\overline{J \bar{\rho}^t}^{\xi_i}}}^t}, \quad 
  \hat{e}^{n+1/2,m+1} 
  = \frac{\overline{\sqrt{\bar{\rho}^t} e}^t}
         {\overline{\sqrt{\bar{\rho}^t}}^t},
\end{equation}
\begin{equation}
  \bar{\rho}^{n+1,m+1} = \frac{\rho^{n+3/2,m+1} + \rho^{n+1/2,m+1}}{2}.
\end{equation}
This interpolated value was introduced 
to construct a fully conservative finite difference scheme \citep{Morinishi_2009, Morinishi_2010}. 
Wall et al. \citep{Wall_et_al_2002} and Ham et al. \citep{Ham_et_al_2002} have not used this interpolation. 
The density is obtained from the following state equation (\ref{eqn_of_state}):
\begin{equation}
  \rho^{n+1,m+1} = \frac{\kappa Ma^2 p^{n+1,m+1} + 1}{e^{n+1,m+1}}.
  \label{implicit.rho}
\end{equation}
To relax the Courant number limitation due to the speed of sound, 
we treat pressure implicitly 
and use the following double-time interpolation of pressure \citep{Wall_et_al_2002}:
\begin{equation}
  \overline{\bar{p}^t}^t = \bar{p}^{n+1/2,m+1} 
  = \left( \frac{1}{4} - \varepsilon \right) p^{n-1/2} 
  + \frac{1}{2}  p^{n+1/2} 
  + \left( \frac{1}{4} + \varepsilon \right) p^{n+3/2,m+1},
  \label{implicit.weighted_average}
\end{equation}
where $\varepsilon$ is a parameter introduced to prevent numerical oscillations 
caused by high wavenumber acoustic modes. 
When nonphysical acoustic modes occur, 
they cannot be dissipated, so we need to prevent such vibrations. 
The previous study \citep{Wall_et_al_2002} uses a value as small as $\varepsilon = 0.005$. 
In this research, we basically set $\varepsilon = 0$.

In this study, the conservation laws of mass and electric charge 
and the constraint of Gauss's law are discretely satisfied at the time $n+1$ level. 
Therefore, Eqs. (\ref{divergence_magnetic}) and (\ref{electric_charge_law}) 
are given as follows, respectively:
\begin{equation}
  \frac{\partial B_i^{n+1,m+1}}{\partial x_i} = 0,
  \label{divergence_magnetic_fdm}
\end{equation}
\begin{equation}
  \frac{\partial j_i^{n+1,m+1}}{\partial x_i} = 0.
  \label{electric_charge_law_fdm}
\end{equation}

The Yee scheme \citep{Yee_1966} is an explicit method, 
wherein the temporal level of a magnetic field is shifted from that of an electric field 
by half a time step. 
When the flow field is also solved, the total energy is not discretely conserved 
for ideal inviscid MHD flows 
unless all dependent variables are defined at the same temporal level. 
Therefore, a total-energy conservative difference scheme cannot be constructed 
using the Yee method \citep{Yee_1966}. 
As explained in Section \ref{discretization}, 
by applying the implicit midpoint rule to Eq. (\ref{Faraday}), 
the total energy equation can be derived discretely. 
Moreover, total energy is discretely conserved in ideal periodic inviscid MHD flows. 
Considering the applications of MHD flow, 
the present implicit method is efficient for applicative calculations. 
The method of spatially shifting the definition points of the electric and magnetic fields, 
like the Yee scheme \citep{Yee_1966}, is the same as the method of this study. 
By adopting such a staggered grid, as explained in Section \ref{discretization}, 
the conservative and nonconservative forms 
of the Lorentz force can be interconverted, and such a transformation is discretely satisfied. 
Furthermore, the magnetic flux density equation (\ref{magnetic_field}) 
can be derived discretely from Faraday's equation (\ref{Faraday}).

By applying the simplified marker and cell (SMAC) method \citep{Amsden&Harlow_1970}, 
Eqs. (\ref{implicit.mass}), (\ref{implicit.u}), and (\ref{implicit.e}) are temporally split as follows:
\begin{subequations}
\begin{equation}
  \frac{Wo^2}{Re} \frac{\rho^{n+1,m+1} \tilde{u_i}^{n+1,m+1} - \rho^n u_i^n}{\dt} 
  = H_{u_i}^{n+1/2,m+1} - \frac{\partial \bar{p}^{n+1/2,m}}{\partial x_i},
  \label{predict.u}
\end{equation}
\begin{equation}
  \frac{Wo^2}{Re} 
  \frac{\rho^{n+1,m+1} u_i^{n+1,m+1} 
      - \rho^{n+1,m+1} \tilde{u_i}^{n+1,m+1}}{\dt} 
  = - \left( \frac{1}{4} + \varepsilon \right) \frac{\partial \Delta p^m}{\partial x_i},
  \label{correct.u.1}
\end{equation}
\begin{equation}
  \frac{Wo^2}{Re} \frac{\rho^{n+1,m} \tilde{e}^{n+1,m+1} - \rho^n e^n}{\dt} 
  = H_e^{n+1/2,m+1} 
  - (\kappa - 1) (\kappa Ma^2 \bar{p}^{n+1/2,m} + 1) 
  \frac{\partial \hat{u_i}^{n+1/2,m}}{\partial x_i},
  \label{predict.e}
\end{equation}
\begin{equation}
  \frac{Wo^2}{Re} \frac{\rho^{n+1,m+1} e^{n+1,m+1} 
                      - \rho^{n+1,m} \tilde{e}^{n+1,m+1}}{\dt} 
  = - \kappa (\kappa - 1) Ma^2 
  \left( \frac{1}{4} + \varepsilon \right) \Delta p^m 
  \frac{\partial \hat{u_i}^{n+1/2,m}}{\partial x_i},
  \label{correct.e.1}
\end{equation}
\begin{equation}
  p^{n+3/2,m+1} = p^{n+3/2,m} + \Delta p^m,
  \label{correct.p.1}
\end{equation}
\end{subequations}
where $\tilde{u_i}^{n+1,m+1}$ and $\tilde{e}^{n+1,m+1}$ are the predicted values of velocity 
and internal energy, respectively, and $\Delta p^m$ is the pressure correction value. 
The velocity in $H_{u_i}^{n+1/2,m+1}$ on the right side of Eq. (\ref{predict.u}) 
is defined as $u_i^{n+1/2,m+1} = (\tilde{u_i}^{n+1,m+1} + u_i^n)/2$. 
The internal energy in $H_e^{n+1/2,m+1}$ on the right side of Eq. (\ref{predict.e}) 
is defined as $e^{n+1/2,m+1} = (\tilde{e}^{n+1,m+1} + e^n)/2$. 
When calculating the velocity $\tilde{u_i}^{n+1,m+1}$, 
the convective term is linearized as 
$\partial_j (\rho u_j)^{n+1/2,m} \hat{u_i}^{n+1/2,m+1})$ using the $m$-level value. 
The magnetic flux density in the Lorentz force is also linearized as $B_i^{n+1/2,m}$. 
In Eqs. (\ref{predict.e}) and (\ref{correct.e.1}), 
the density and velocity are linearized as $\rho^{n+1,m+1} = \rho^{n+1,m}$ 
and $\hat{u_i}^{n+1,m+1} = \hat{u_i}^{n+1,m}$, respectively. 
Once the Newton iteration is completed, 
such a linearized approximation can be ignored, 
and second-order accuracy in the time integration is preserved.

Substituting Eq. (\ref{correct.u.1}) into the mass conservation equation (\ref{implicit.mass}) 
yields Poisson's equation for the pressure correction value $\Delta p$ as follows:
\begin{equation}
  \left( \frac{1}{4} + \varepsilon \right) 
  \frac{\partial}{\partial x_i} \frac{\partial \Delta p^m}{\partial x_i} 
  = \frac{Wo^2}{\dt Re} \left[ 
  \frac{Wo^2}{Re} \frac{\rho^{n+3/2,m+1} - \rho^{n+1/2}}{\dt} 
  + \frac{\partial}{\partial x_i} (\rho^{n+1,m+1} \tilde{u_i}^{n+1,m+1}) \right].
  \label{poisson.1}
\end{equation}
To stabilize calculations, 
we incorporate the effect of the pressure correction value on the density 
in the time derivative of the density of the mass conservation equation (\ref{implicit.mass}) 
as follows \citep{Wall_et_al_2002}:
\begin{equation}
  \left. \frac{\partial \rho}{\partial t} \right|^{n+1,m+1} 
  \approx \frac{\rho^{n+3/2,m+1} - \rho^{n+1/2}}{\dt} 
  + \frac{1}{\dt} \left. \frac{\partial \rho}{\partial p} \right|_e \Delta p.
\end{equation}
$\partial \rho/\partial p|_e$ represents the derivative when the internal energy is constant. 
This derivative term is obtained using the equation of state as follows:
\begin{equation}
  \left. \frac{\partial \rho}{\partial p} \right|_e 
  = \frac{\kappa Ma^2}{e}.
\end{equation}
Using the above equation, 
the Poisson equation (\ref{poisson.1}) can be rewritten into the following Helmholtz equation:
\begin{align}
  &
  \left( \frac{1}{4} + \varepsilon \right) 
  \frac{\partial}{\partial x_i} \frac{\partial \Delta p^m}{\partial x_i} 
  - \frac{Wo^4}{\dt^2 Re^2} 
  \left. \frac{\partial \rho}{\partial p} \right|_e \Delta p^m 
  \nonumber \\
  & \qquad 
  = \frac{Wo^2}{\dt Re} \left[ 
  \frac{Wo^2}{Re} \frac{\rho^{n+3/2,m+1} - \rho^{n+1/2}}{\dt} 
  + \frac{\partial}{\partial x_i} (\rho^{n+1,m+1} \tilde{u_i}^{n+1,m+1}) \right].
  \label{poisson.2}
\end{align}

Bijl and Wesseling \citep{Bijl&Wesseling_1998} and Kwatra et al. \citep{Kwatra_et_al_2009} 
proposed a method to solve the Poisson equation of pressure as a pressure-based method. 
The Poisson equation for pressure used in these existing studies is a complex form. 
On the other hand, the Laplacian operator in the Poisson equation (\ref{poisson.2}) 
for the pressure correction value used in this study is linear and has a simple form.

In the SMAC method \citep{Amsden&Harlow_1970}, 
the right side of Eq. (\ref{poisson.2}) enables self-regulation 
of the velocity divergence error, 
and a stable convergent solution can be obtained using an iterative solver 
such as the successive over-relaxation method. 
However, the iteration of Poisson's equation takes is time consuming. 
To satisfy the continuity condition, 
the velocity and pressure are relaxed simultaneously, 
as in \citep{Hirt_et_al_1975,Takemitsu_1985,Oki&Tanahashi_1993,Yanaoka&Inafune_2023,Yanaoka_2023}. 
The method in this study does not change the form of Helmholtz's equation (\ref{poisson.1}). 
Thus, simultaneous relaxation does not affect the stability 
when solving Helmholtz's equation. 
The simultaneous relaxation of velocity and pressure is performed as follows:
\begin{subequations}
\begin{align}
  &
  \left( \frac{1}{4} + \varepsilon \right) 
  \frac{\partial}{\partial x_i} \frac{\partial \Delta p^{m,l}}{\partial x_i} 
  - \frac{Wo^4}{\dt^2 Re^2} 
  \left. \frac{\partial \rho}{\partial p} \right|_e \Delta p^{m,l} 
  \nonumber \\
  & \qquad 
  = \frac{Wo^2}{\dt Re} \left[ 
  \frac{Wo^2}{Re} \frac{\rho^{n+3/2,m+1,l} - \rho^{n+1/2}}{\dt} 
  + \frac{\partial}{\partial x_i} (\rho^{n+1,m+1.l} \hat{u_i}^{n+1,m+1,l}) \right],
  \label{poisson.3}
\end{align}
\begin{equation}
  \frac{Wo^2}{Re} 
  \frac{\rho^{n+1,m+1,l} u_i^{n+1,m+1,l+1} 
      - \rho^{n+1,m+1,l} u_i^{n+1,m+1,l}}{\dt} 
  = - \left( \frac{1}{4} + \varepsilon \right) \frac{\partial \Delta p^{m,l}}{\partial x_i},
  \label{correct.u.2}
\end{equation}
\begin{equation}
  \frac{Wo^2}{Re} \frac{\rho^{n+1,m+1,l} e^{n+1,m+1,l+1} 
                      - \rho^{n+1,m+1,l} e^{n+1,m+1,l}}{\dt} 
  = - \kappa (\kappa - 1) Ma^2 
  \left( \frac{1}{4} + \varepsilon \right) \Delta p^{m,l} 
  \frac{\partial \hat{u_i}^{n+1/2,m}}{\partial x_i},
  \label{correct.e.2}
\end{equation}
\begin{equation}
  p^{n+3/2,m+1,l+1} = p^{n+3/2,m+1,l} + \Delta p^{m,l},
  \label{correct.p.2}
\end{equation}
\begin{equation}
  \rho^{n+1,m+1,l+1} = \frac{\kappa Ma^2 p^{n+1,m+1,l+1} + 1}{e^{n+1,m+1,l+1}},
  \label{density}
\end{equation}
\end{subequations}
where the superscript $l$ represents the number of iterations. 
When $l = 1$, let $\bu^{n+1,m+1,l} = \tilde{\bu}^{n+1,m+1}$, 
$p^{n+3/2,m+1,l} = p^{n+3/2,m}$, $e^{n+1,m+1,l} = \tilde{e}^{n+1,m+1}$, and $\rho^{n+3/2,m+1,l} = \rho^{n+3/2,m}$. 
In such a scenario, 
the velocity, pressure, internal energy, and density are simultaneously relaxed. 
The calculation is repeated up to a predetermined iteration number. 
After the simultaneous relaxation, 
let $\bu^{n+1,m+1} = \bu^{n+1,m+1,l+1}$, $p^{n+3/2,m+1} = p^{n+3/2,m+1,l+1}$, $e^{n+1,m+1} = e^{n+1,m+1,l+1}$, and $\rho^{n+3/2,m+1} = \rho^{n+3/2,m+1,l+1}$. 
Takemitsu \citep{Takemitsu_1985} proposed a similar method 
that simultaneously iterates the velocity correction equation 
and Poisson equation of the pressure correction. 
However, Poisson's equation for pressure should be solved after correcting the velocity. 
The present numerical method does not require Poisson's equation for obtaining pressure. 

In MHD flow analyses, the magnetic flux density must be calculated 
while satisfying its constraint. 
As in \citep{Evans&Hawley_1988,Dumbser_et_al_2019,Yanaoka_2023}, 
Faraday's equation (\ref{Faraday}) is discretized 
such that its divergence is zero. 
Therefore, the magnetic flux density is not corrected, 
in contrast to existing studies \citep{Dedner_et_al_2002,Brackbill&Barnes_1980}. 
The discretization method is described in Subsection \ref{Faraday_fdm}.

The magnetic vector potential is calculated in the same manner as the velocity. 
The principle of the SMAC method \citep{Amsden&Harlow_1970} is applied 
to calculate Eq. (\ref{implicit.A}) as follows:
\begin{subequations}
\begin{equation}
  \frac{Wo^2}{Re} \frac{\tilde{A_i}^{n+1,m+1} - A_i^n}{\dt} = H_{A_i}^{n+1/2,m+1} 
  - \frac{\partial \psi^{n+1/2,m}}{\partial x_i}, 
  \label{predict.A}
\end{equation}
\begin{equation}
  \frac{Wo^2}{Re} \frac{A_i^{n+1,m+1} - \tilde{A_i}^{n+1,m+1}}{\dt} 
  = - \lambda \frac{\partial \Delta \psi^m}{\partial x_i},
  \label{correct.A.1}
\end{equation}
\begin{equation}
  \psi^{n+1,m+1} = \psi^{n+1,m} + \Delta \psi^m,
  \label{correct.psi.1}
\end{equation}
\end{subequations}
where $\hat{A_i}^{n+1,m+1}$ is the predicted value of the magnetic vector potential, 
and $\Delta \psi$ is the correction for $\psi$. 
The magnetic vector potential in $H_{A_i}^{n+1/2,m+1}$ on the right side of Eq. (\ref{predict.A}) 
is defined as $A_i^{n+1/2,m+1} = (\hat{A_i}^{n+1,m+1} + A_i^n)/2$. 
When calculating the magnetic vector potential $\hat{A_i}^{n+1,m+1}$, 
the convective term is linearized as 
$\epsilon_{ijk} (\epsilon_{jlm} \partial_l \hat{A}_m^{n+1/2,m+1}) u_k^{n+1/2,m}$ 
using the $m$-level value. 
Once the Newton iteration is completed, 
such a linearized approximation can be ignored, 
preserving second-order accuracy in the time integration. 
By applying the Coulomb gauge, 
taking the divergence of Eq. (\ref{correct.A.1}) 
and using the divergence-free condition of the magnetic vector potential at the $n+1$ level, 
Poisson's equation for the correction value $\Delta \psi$ is derived as
\begin{equation}
  \lambda \frac{\partial}{\partial x_i} \frac{\partial \Delta \psi^m}{\partial x_i} 
  = \frac{Wo^2}{Re} \frac{1}{\dt} \frac{\partial \tilde{A_i}^{n+1,m+1}}{\partial x_i}.
  \label{poisson.p_A.1}
\end{equation}
The magnetic vector potential can also be calculated via simultaneous relaxation 
similar to the velocity as follows:
\begin{subequations}
\begin{equation}
  \frac{Wo^2}{Re} \frac{A_i^{n+1,m+1,l+1} - A_i^{n+1,m+1,l}}{\dt} 
  = - \lambda \frac{\partial \Delta \psi^{m,l}}{\partial x_i},
  \label{correct.A.2}
\end{equation}
\begin{equation}
  \psi^{n+1,m+1,l+1} = \psi^{n+1,m,l} + \Delta \psi^{m,l},
  \label{correct.psi.2}
\end{equation}
\begin{equation}
  \lambda \frac{\partial}{\partial x_i} \frac{\partial \Delta \psi^{m,l}}{\partial x_i} 
  = \frac{Wo^2}{Re} \frac{1}{\dt} \frac{\partial A_i^{n+1,m+1,l}}{\partial x_i},
  \label{poisson.psi.2}
\end{equation}
\end{subequations}
where, when $l = 1$, 
let $A_i^{n+1,m+1,l} = \tilde{A_i}^{n+1,m+1}$ and $\psi^{n+1,m+1,l} = \psi^{n+1,m}$. 
The magnetic vector potential $A_i$ and electirc potential $\psi$ 
are then simultaneously relaxed. 
After the simultaneous relaxation, 
let $A_i^{n+1,m+1} = A_i^{n+1,m+1,l+1}$ and $\psi^{n+1,m+1} = \psi^{n+1,m+1,l+1}$. 
Equation (\ref{correct.A.2}) is used as the boundary condition 
to solve Eq. (\ref{poisson.psi.2}).

To analyze steady and unsteady flows, 
the Euler implicit method and implicit midpoint rule are used 
for the time derivative, respectively. 
The biconjugate gradient stabilized method \citep{Vorst_1992} is applied 
to solve simultaneous linear equations. 
These discretized equations are solved by following the subsequent procedure.
\begin{enumerate}[Step 1:]
\setlength{\leftskip}{1.5em}
\setlength{\itemsep}{0em}
\item At $m = 1$, 
      let $\rho^{n+3/2,m} = \rho^{n+1/2}$, $u_i^{n+1,m} = u_i^{n}$, $p^{n+3/2,m} = p^{n+1/2}$, 
      $e^{n+1,m} = e^{n}$, $E^{n+1,m} = E^{n}$, $T^{n+1,m} = T^{n}$, 
      $B_i^{n+1,m} = B_i^n$, $A_i^{n+1,m} = A_i^n$, and $\psi^{n+1,m} = \psi^n$.
\item Solve Eq. (\ref{predict.u}), and predict the velocity $\tilde{u_i}^{n+1,m+1}$.
\item Solve Eq. (\ref{predict.e}), and predict the internal energy $\tilde{e}^{n+1,m+1}$.
\item Calculate the density $\rho^{n+3/2,m+1}$ by Eq. (\ref{implicit.rho}).
\item Solve the pressure correction value $\Delta p^{m,l}$ using Helmholtz's equation (\ref{poisson.2}).
\item Correct the velocity $u_i^{n+1,m+1,l+1}$, pressure $p^{n+3/2,m+1,l+1}$, 
      internal energy $e^{n+1,m+1,l+1}$, and density $\rho^{n+3/2,m+1,l+1}$ 
      using Eqs. (\ref{correct.u.2}), (\ref{correct.p.2}), (\ref{correct.e.2}), 
      and (\ref{density}), respectively. 
      At the end of simultaneous relaxation, 
      set $u_i^{n+1,m+1} = u_i^{n+1,m+1,l+1}$, 
      $p^{n+3/2,m+1} = p^{n+3/2,m+1,l+1}$, 
      $e^{n+1,m+1} = e^{n+1,m+1,l+1}$, 
      and $\rho^{n+3/2,m+1} = \rho^{n+3/2,m+1,l+1}$.
\item Solve the magnetic flux density $B_i^{n+1,m+1}$ using Eq. (\ref{implicit.B}).
      Solve Eq. (\ref{predict.A}) 
      and predict the magnetic vector potential $\tilde{A_i}^{n+1,m+1}$.
\item Solve the correction $\Delta \psi^{m,l}$ using Poisson's equation (\ref{poisson.psi.2}).
      Correct the magnetic vector potential $A_i^{n+1,m+1,l+1}$ 
      and electric potential $\psi^{n+1,m+1,l+1}$ 
      using Eq. (\ref{correct.A.2}) and (\ref{correct.psi.2}), respectively. 
      At the end of simultaneous relaxation, 
      set $A_i^{n+1,m+1} = A_i^{n+1,m+1,l+1}$ and 
      $\psi^{n+1,m+1} = \psi^{n+1,m+1,l+1}$.
\item Repeat from Step 2 to Step 8. 
      After the Newton iteration is completed,
      set $\rho^{n+3/2} = \rho^{n+3/2,m+1}$, $u_i^{n+1} = u_i^{n+1,m+1}$, 
      $p^{n+3/2} = p^{n+3/2,m+1}$, $e^{n+1} = e^{n+1,m+1}$, 
      $T^{n+1} = T^{n+1,m+1}$, $B_i^{n+1} = B_i^{n+1,m+1}$, 
      $A_i^{n+1} = A_i^{n+1,m+1}$, $\psi^{n+1} = \psi^{n+1,m+1}$.
\item Advance the time step and return to Step 1.
\end{enumerate}

\section{Verification of the proposed numerical method}
\label{verification}

For inviscid analysis, 
the energy conservation properties of this numerical method is investigated. 
It is verified that viscous analysis can accurately capture energy conversion. 
Additionally, this study clarifies changes in flow and magnetic fields due to Mach number. 
Below, the coordinate $x_i$, the velocity $u_i$, 
the magnetic flux density $B_i$, and magnetic vector potential $A_i$ are denoted 
as $\bm{x} = (x, y, z)$, $\bm{u} = (u, v, w)$, $\bm{B} = (B_x, B_y, B_z)$, 
and $\bm{A} = (A_x, A_y, A_z)$, respectively.

Subsections \ref{vortex}, \ref{decaying_vortex}, and \ref{OT_vortex} 
deal with two-dimensional models in the $x$-$y$ section. 
However, periodic boundary conditions are applied in the $z$-direction, 
and this study analyzes the computational model three-dimensionally. 
Three components of velocity, magnetic flux density, 
and magnetic vector potential are solved. 
We confirm that no nonphysical component in the $z$-direction occurs. 
In addition, in a two-dimensional problem, magnetic helicity is zero, 
but we also sample the magnetic helicity and confirm that it is zero.

\subsection{Periodic inviscid compressible flow}
\label{inviscid}

We analyze a three-dimensional periodic inviscid compressible flow 
to verify the validity of this numerical method. 
In the flow field, each total amount of momentum and total energy 
inside the computational domain is temporally conserved. 
If an inappropriate difference scheme is used, each total amount is not preserved discretely 
because of the generation of nonphysical momentum and total energy. 
In the inviscid analysis, 
we can verify whether each total amount for the momentum and total energy are conserved in time 
because no energy attenuation due to viscosity occurs.

The calculation area is a cube with sides of $L$. 
As with the previous study \citep{Morinishi_2009}, 
we use the velocity and magnetic flux density that satisfy the divergence-free conditions as initial values. 
The initial velocity and magnetic flux density conditions are three-dimensionally derived 
using vector potential with uniform random numbers. 
The velocity is normalized to satisfy 
that the volume-averaged velocity $\langle \bm{u} \rangle$ becomes zero, 
and the volume-averaged velocity fluctuation $\frac{1}{3} \langle u’^2+v’^2+w’^2 \rangle$ is a constant value $U^2$. 
Similarly, for the initial value of magnetic flux density, 
the magnetic flux density is normalized to satisfy 
that the volume-averaged magnetic flux density $\langle \bm{B} \rangle$ becomes zero, 
and the volume-averaged magnetic flux density fluctuation $\frac{1}{3} \langle B_x’^2+B_y’^2+B_z’^2 \rangle$ is a constant value $B^2$. 
The initial density and temperature values are $\rho_0$ and $T_0$, respectively, 
and are uniform. 
As for boundary conditions, periodic boundary conditions are given to the velocity, 
pressure, internal energy, and magnetic flux density.

The reference values in this calculation are as follows: 
the length is $l_\mathrm{ref} = L$, 
velocity is $u_\mathrm{ref} = U$, 
time is $t_\mathrm{ref} = L/U$, 
density is $\rho_\mathrm{ref} = \rho_0$, 
pressure is  $p_\mathrm{ref} = (\kappa -1) \rho_0 c_v T_0$, 
temperature $T_\mathrm{ref} = T_0$, 
internal energy is $e_\mathrm{ref} = c_v T_0$, 
and magnetic flux density is $B_\mathrm{ref} = B$. 
The specific heat ratio is set to $\kappa = 1.4$, 
and the initial fluctuating Mach number is given as $Ma = U_0/c_0 = 0.2$, 
where $c_0$ is the sound speed at $T_0$. 
The pressure shift parameter $\varepsilon$ is set to $\varepsilon = 0$, 
but for the time interval $\Delta t/(L/U) = 0.05$, to stabilize the calculation, 
the pressure shift parameter $\varepsilon$ was set as $\varepsilon = 0.002$. 
The initial Courant number $\mathrm{CFL} = \Delta t (U_0+c_0)/\Delta x$ 
considering the speed of sound for $\Delta t/(L/U_0) = 0.05$ is CFL = 3.0 
and the local Courant number is CFL = 7.60. 
Herein, $\Delta x$ is the grid width. 
In this analysis, we first use a uniform grid with $11 \times 11 \times 11$ grid points. 
For $\Delta t/(L/U) = 0.01$, we investigated the conservation properties 
for momentum and energy using a nonuniform grid. 
The nonuniform grid was generated using the same method as in the previous study \citep{Yanaoka_2023}.

\begin{figure}[!t]
\begin{minipage}{0.48\linewidth}
\begin{center}
\includegraphics[trim=0mm 0mm 0mm 0mm, clip, width=70mm]{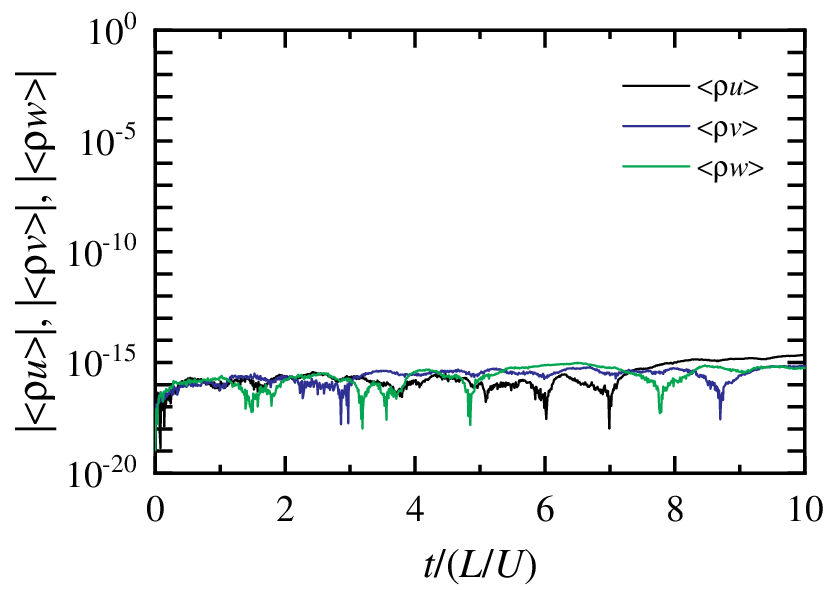} \\
{\small (a) Uniform grid}
\end{center}
\end{minipage}
\hspace{0.02\linewidth}
\begin{minipage}{0.48\linewidth}
\begin{center}
\includegraphics[trim=0mm 0mm 0mm 0mm, clip, width=70mm]{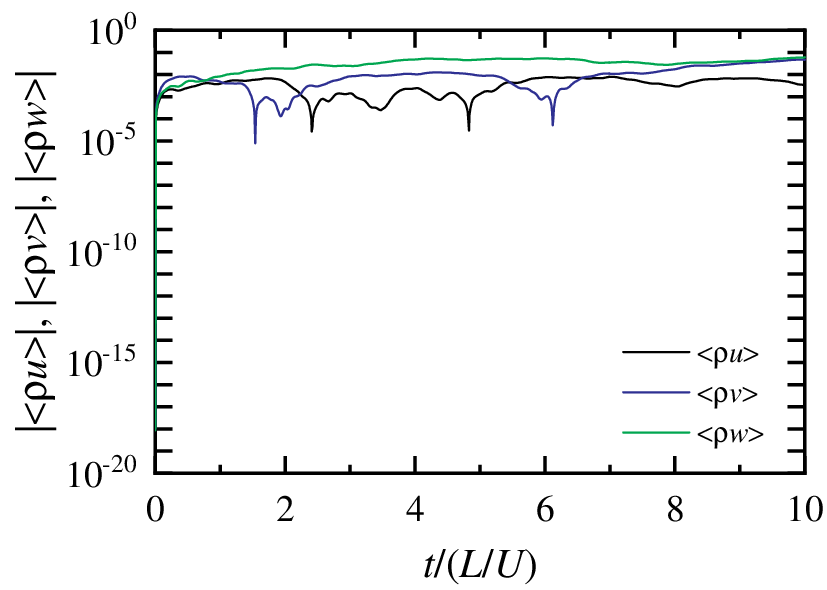} \\
{\small (b) Nonuniform grid}
\end{center}
\end{minipage}
\caption{Time variation of total amount of momentum: 
$\Delta t/(L/U) = 0.01$}
\label{inviscid_momentum}
\end{figure}

\begin{figure}[!t]
\begin{minipage}{0.48\linewidth}
\begin{center}
\includegraphics[trim=0mm 0mm 0mm 0mm, clip, width=70mm]{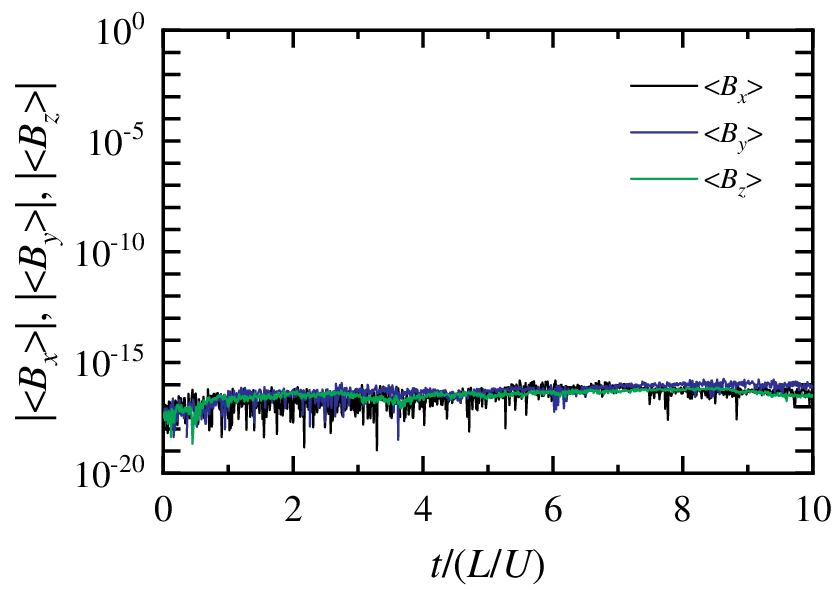} \\
{\small (a) Uniform grid}
\end{center}
\end{minipage}
\hspace{0.02\linewidth}
\begin{minipage}{0.48\linewidth}
\begin{center}
\includegraphics[trim=0mm 0mm 0mm 0mm, clip, width=70mm]{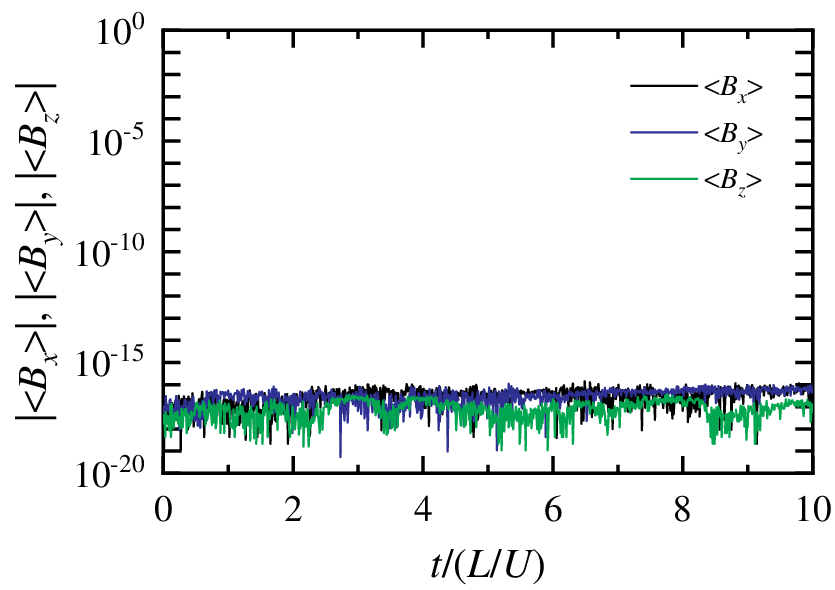} \\
{\small (b) Nonuniform grid}
\end{center}
\end{minipage}
\caption{Time variation of total amount of magnetic flux density: 
$\Delta t/(L/U) = 0.01$}
\label{inviscid_magnetic}
\end{figure}

For the time interval $\Delta t/(L/U_0) = 0.01$, 
Figs. \ref{inviscid_momentum} and \ref{inviscid_magnetic} show 
the total amount for the momentum and magnetic flux density, 
$\langle \rho \bm{u} \rangle$ and $\langle \bm{B} \rangle$, respectively. 
The total amount was determined by volume integration. 
For the uniform grid, 
$|\langle \rho \bm{u} \rangle|$ and $|\langle \bm{B} \rangle|$ 
change on the orders of $10^{-15}$ and $10^{-17}$, respectively, 
and the momentum and magnetic flux density are temporally conserved even at a discrete level. 
In the nonuniform grid, 
the conservation of the total amount for the momentum is worsened. 
The conservation property of magnetic flux density is the same as that for the uniform grid. 
As mentioned in the previous study on incompressible MHD flows \citep{Yanaoka_2023}, 
when using the nonconservative Lorentz force, 
the nonconservative form can be converted to the conservative form on uniform grids. 
However, in the case of nonuniform grids, 
the nonconservative Lorentz force cannot be converted to the conservative form. 
Therefore, the momentum conservation property deteriorates in nonuniform grids.

\begin{figure}[!t]
\begin{minipage}{0.48\linewidth}
\begin{center}
\includegraphics[trim=0mm 0mm 0mm 0mm, clip, width=70mm]{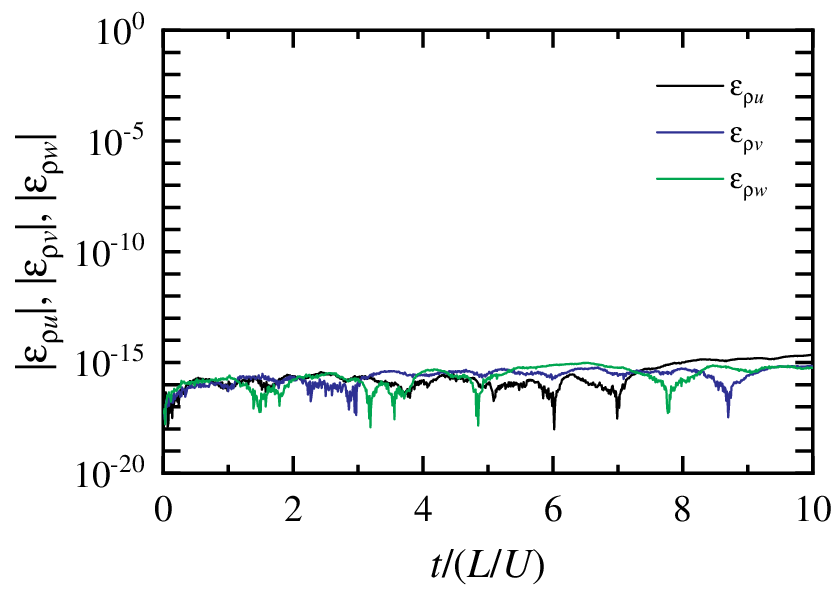} \\
{\small (a) Uniform grid}
\end{center}
\end{minipage}
\hspace{0.02\linewidth}
\begin{minipage}{0.48\linewidth}
\begin{center}
\includegraphics[trim=0mm 0mm 0mm 0mm, clip, width=70mm]{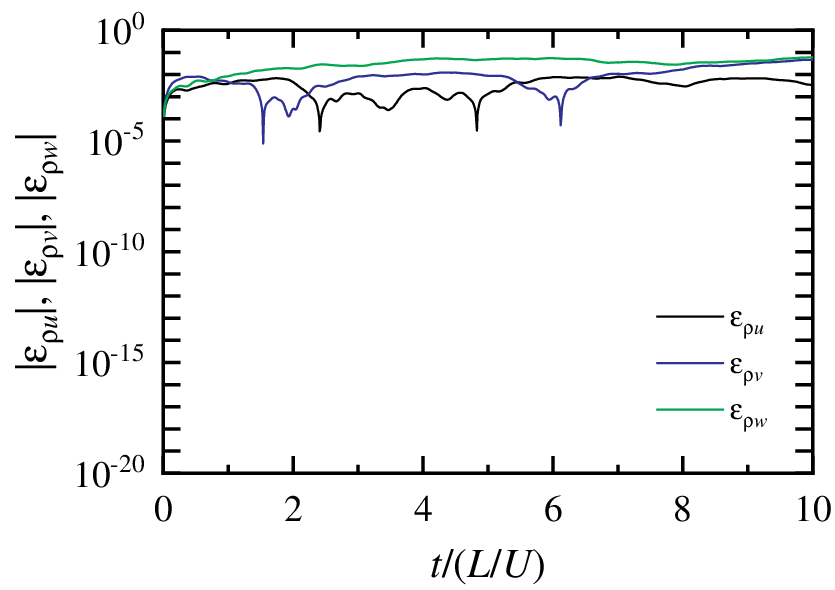} \\
{\small (b) Nonuniform grid}
\end{center}
\end{minipage}
\caption{Time variation of momentum error: $\Delta t/(L/U) = 0.01$}
\label{inviscid_momentum_error}
\end{figure}

\begin{figure}[!t]
\begin{minipage}{0.48\linewidth}
\begin{center}
\includegraphics[trim=0mm 0mm 0mm 0mm, clip, width=70mm]{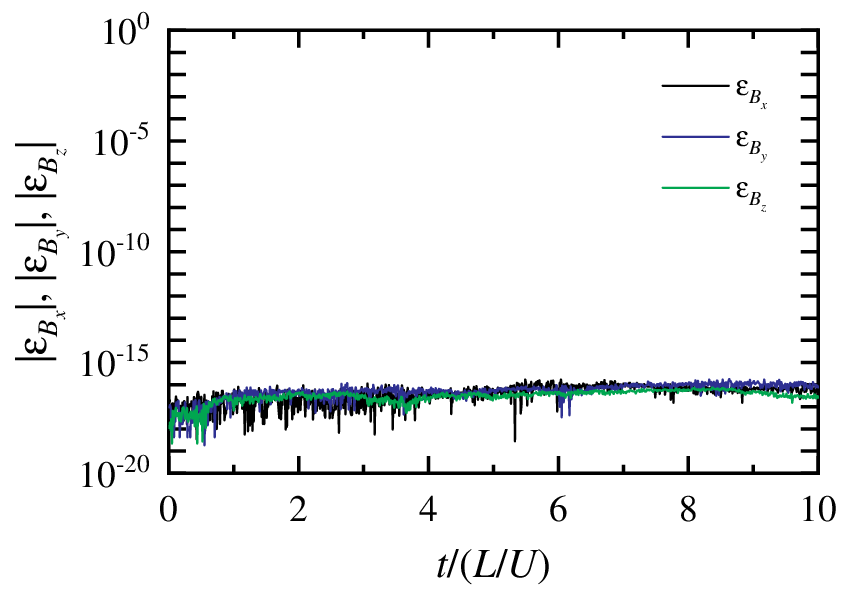} \\
{\small (a) Uniform grid}
\end{center}
\end{minipage}
\hspace{0.02\linewidth}
\begin{minipage}{0.48\linewidth}
\begin{center}
\includegraphics[trim=0mm 0mm 0mm 0mm, clip, width=70mm]{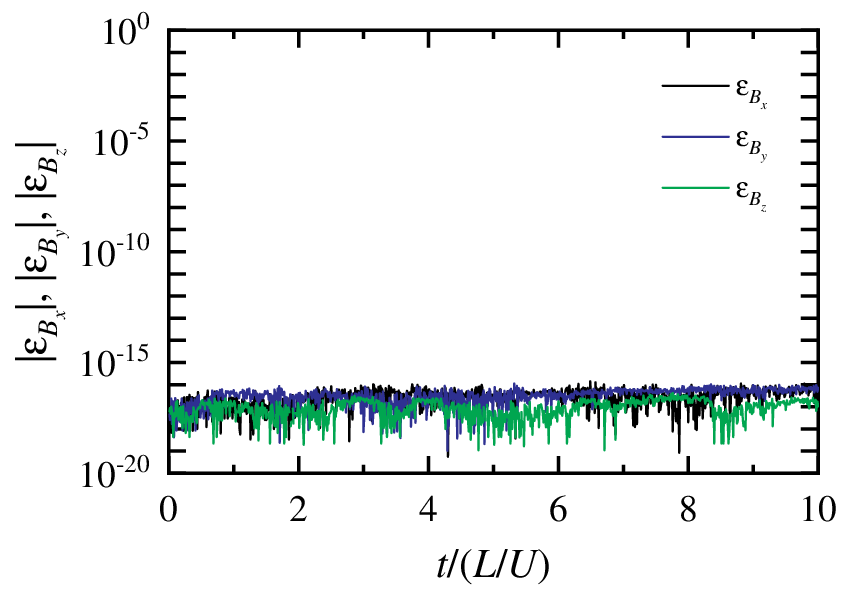} \\
{\small (b) Nonuniform grid}
\end{center}
\end{minipage}
\caption{Time variation of magnetic flux density error: $\Delta t/(L/U) = 0.01$}
\label{inviscid_magnetic_error}
\end{figure}

\begin{figure}[!t]
\begin{minipage}{0.48\linewidth}
\begin{center}
\includegraphics[trim=0mm 0mm 0mm 0mm, clip, width=70mm]{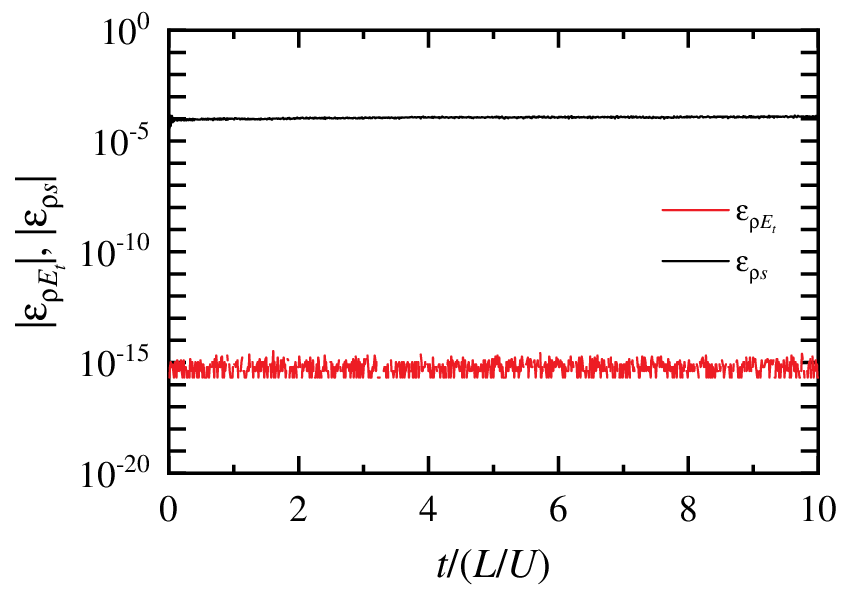} \\
{\small (a) Uniform grid}
\end{center}
\end{minipage}
\hspace{0.02\linewidth}
\begin{minipage}{0.48\linewidth}
\begin{center}
\includegraphics[trim=0mm 0mm 0mm 0mm, clip, width=70mm]{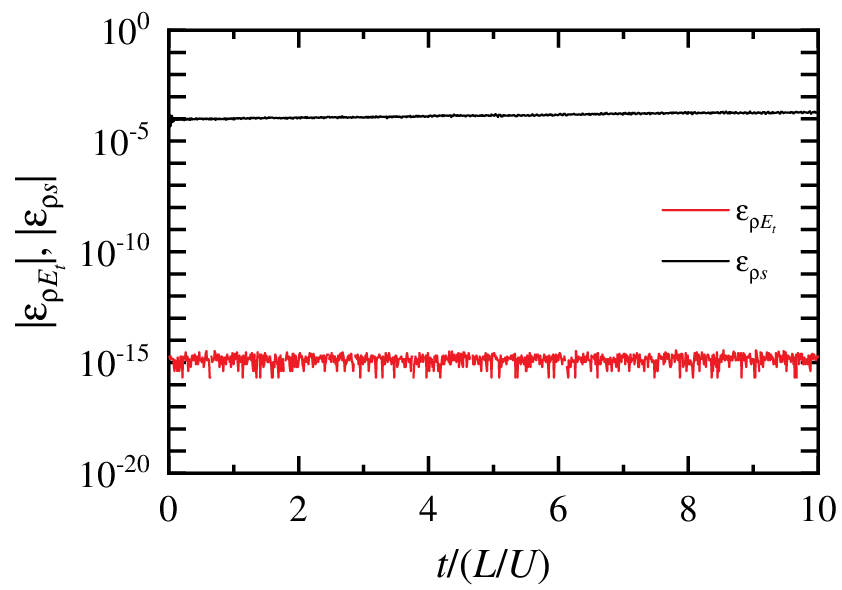} \\
{\small (b) Nonuniform grid}
\end{center}
\end{minipage}
\caption{Time variations of total energy and entropy errors: $\Delta t/(L/U) = 0.01$}
\label{inviscid_energy_error}
\end{figure}

Figures \ref{inviscid_momentum_error} and \ref{inviscid_magnetic_error} 
show the relative errors, 
$\varepsilon_{\rho \bm{u}} = (\langle \rho \bm{u} \rangle - \langle \rho \bm{u} \rangle_0)/\langle (\rho \bm{u})^2 \rangle_0^{1/2}$ 
and $\varepsilon_{\bm{B}} = (\langle \bm{B} \rangle - \langle \bm{B} \rangle_0)/\langle \bm{B}^2 \rangle_0^{1/2}$, 
in the momentum and magnetic flux density for $\Delta t/(L/U) = 0.01$, respectively. 
Figure \ref{inviscid_energy_error} shows the relative error, 
$\varepsilon_{\rho E_t} = (\langle \rho E_t \rangle - \langle \rho E_t \rangle_0)/\langle \rho E_t \rangle_0$, in the total energy 
and absolute error, $\varepsilon_{\rho s} = \langle \rho s \rangle - \langle \rho s \rangle_0$, in the entropy. 
The subscript 0 represents the initial value. 
Herein, as $\langle \rho s \rangle_0$ is zero, 
the error was defined as an absolute error. 
$\langle (\rho \bm{u})^2 \rangle_0^{1/2}$, $\langle \bm{B}^2 \rangle_0^{1/2}$, 
and $\langle \rho E_t \rangle_0$ is in the order of $10^0$. 
Therefore, although the definition of error is different, 
we can compare the magnitude of each error for $\varepsilon_{\rho \bm{u}}$, 
$\varepsilon_{\bm{B}}$, $\varepsilon_{\rho E_t}$, and $\varepsilon_{\rho s}$. 
In the uniform grid, 
$|\varepsilon_{\rho \bm{u}}|$, $|\varepsilon_{\bm{B}}|$, 
and $|\varepsilon_{\rho E_t}|$ change on the orders of $10^{-15}$, $10^{-17}$, and $10^{-16}$, 
and momentum, magnetic flux density, and total energy are conserved in time. 
On the other hand, $|\varepsilon_{\rho s}|$ changes on the order of $10^{-4}$, 
which is a high value compared to the momentum and total energy errors. 
This result is because the entropy conservation equation cannot be derived discretely, 
and the discrete conservation property of entropy deteriorates. 
For the nonuniform grid, 
$|\varepsilon_{\rho \bm{u}}|$, $|\varepsilon_{\bm{B}}|$, 
and $|\varepsilon_{\rho E_t}|$ change around $10^{-2}$, $10^{-17}$, and $10^{-15}$, respectively. 
Although the momentum conservation property deteriorates, 
the magnetic flux density and total energy are well conserved. 
$|\varepsilon_{\rho s}|$ is comparable to the result for the uniform grid. 
The energy conservation properties are similar to those for incompressible MHD flows \citep{Yanaoka_2023}.

\begin{figure}[!t]
\begin{center}
\includegraphics[trim=0mm 0mm 0mm 0mm, clip, width=70mm]{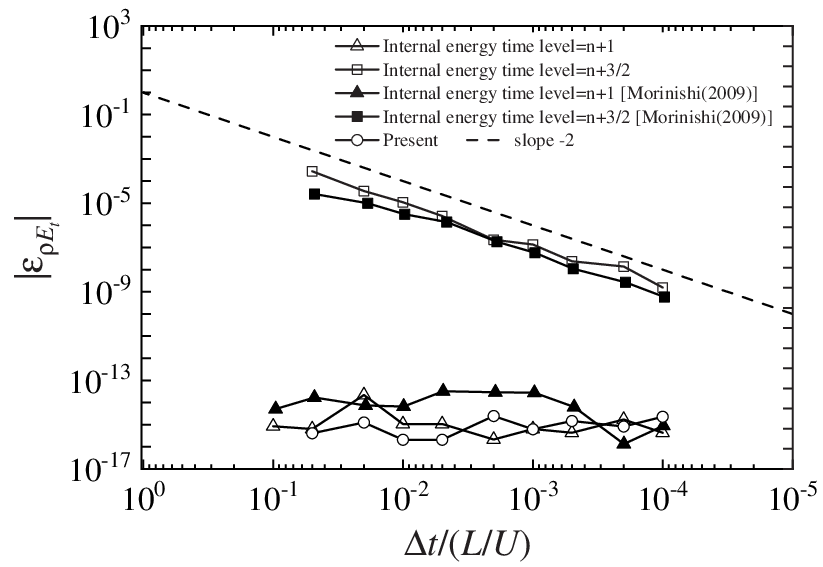}
\end{center}
\caption{Error of total energy at $t/(L/U) = 10$.}
\label{inviscid_error}
\end{figure}

Next, at $t/(L/U_0) = 10$, 
the relative error, $\varepsilon_{\rho E_t}$, in the total energy to $\Delta t/(L/U)$ 
is shown in Fig. \ref{inviscid_error}. 
In the figure, the results are compared 
when the time level of the internal energy is set to $n+3/2$ as in Wall et al. \citep{Wall_et_al_2002} 
and $n+1$ as in Morinishi \citep{Morinishi_2009}. 
The dashed line indicates a straight line with a slope of $-2$. 
Similarly to the result without applying a magnetic field, 
even when a magnetic field is applied, 
the total energy error, $|\varepsilon_{\rho E_t}|$, changes around $10^{-16}-10^{- 15}$ order, 
and the total energy is conserved. 
Furthermore, if we calculate kinetic, internal, and magnetic energies discretely at the same time level, 
we find that the total energy is discretely conserved at the rounding error level.

The maximum error of the mass conservation law 
and the maximum divergence error of the magnetic flux density in this analysis 
are $5.38\times 10^{-12}$ and $1.20\times 10^{-12}$, respectively, for the uniform grid 
and $5.86\times 10^{-14}$ and $1.74\times 10^{-13}$, respectively, for the nonuniform grid.

\subsection{Magnetized advection vortex}
\label{vortex}

We investigated the behavior of an advective vortex and energy conservation properties 
in a flow field with an applied magnetic field as in previous studies \citep{Gawlik_et_al_2011,Dumbser_et_al_2019}. 
The previous study analyzed a magnetized advective vortex in an incompressible flow \citep{Yanaoka_2023}. 
The same vortex model is used for a low Mach number flow. 
The origin is placed at the center of the vortex. 
The $x$- and $y$-axes are the streamwise and vertical directions, respectively, 
and the $z$-axis is vertical to the plane of the paper. 
A uniform magnetic field $B$ is applied in the positive direction of the $x$-axis. 
The stream function and pressure representing the two-dimensional vortex are given as
\begin{align}
  \Psi_z &= \frac{1}{s_v} \omega, \quad 
  \omega = \frac{1}{2} \Gamma_v e^{- s_v \frac{r^2}{2}}, 
  \label{stream_function} \\
  p &= - \frac{1}{2 s_v} \omega^2,
  \label{pressure}
\end{align}
where $r^2 = x^2+y^2$. 
$\Gamma_v$ represents the vortex strength, 
which is the magnitude of the vorticity $\omega_z$ in the $z$-direction 
at the origin. 
The variable $s_v$ is a parameter related to the vortex size. 
The velocities are calculated from $u = \partial \Psi_z/\partial y$ and $v = -\partial \Psi_z/\partial x$. 
Equations (\ref{stream_function}) and (\ref{pressure}) are 
the exact solutions for a steady incompressible inviscid flow with a uniform magnetic field applied. 
The vortex rotates counterclockwise. 
Such vortices line up periodically in the $x$-direction, 
and the interval between vortices is $2L$. 
In this study, we set $\Gamma_v = 4$ and $s_v = 20$ 
so that the induced velocity is low at $r = L$. 
The initial value of the density at the reference temperature $T_0$ is $\rho_0$ and is uniform. 
The initial value of internal energy is found from the equation of state. 
We analyze the flow field in which this vortex advects with a uniform velocity $U$. 
The initial dimensionless streamwise velocity at this time is given as
\begin{equation}
  u = 1 - y \omega.
\end{equation}
For boundary conditions, periodic boundary conditions are given 
to all variables in the $x$-, $y$-, and $z$-directions. 

The reference values used in this calculation are as follows: 
the length is $L_\mathrm{ref} = L$, velocity is $U_\mathrm{ref} = U$, 
time is $t_\mathrm{ref} = L/U$, density is $\rho_\mathrm{ref}$, 
pressure is $p_\mathrm{ref} = (\kappa -1) \rho_0 c_v T_0$, 
temperature is $T_\mathrm{ref} = T_0$, 
internal energy is $e_\mathrm{ref} = c_v T_\mathrm{ref}$, 
and magnetic flux density is $B_\mathrm{ref} = B$. 
The computational domain is $2L$ in the $x$ and $y$-directions, 
and the length in the $z$-direction is a grid width. 
In this calculation, the specific heat ratio is set as $\kappa = 5/3$, 
and the Mach number $Ma = U/c_0$ is varied as $Ma = 0.2$, 0.4, and 0.6. 
Here, $c_0$ is the sound speed at the initial temperature. 
The calculation is performed under the conditions $Re = \infty$ and $Re_m = \infty$. 
The Alfv\'{e}n number is $Al = 1$.

Uniform and nonuniform grids with $N \times N \times 2$ grid points are used. 
$N$ is the number of grid points in the $x$- and $y$-directions, and $N = 41$, 81, and 161. 
We used the same nonuniform grid as in previous research \citep{Yanaoka_2023}. 
Rather than using nonuniform grids to capture the phenomena accurately, 
we use nonuniform grids to investigate changes in energy conservation properties with the grid. 
The Courant number is defined as $\mathrm{CFL} = \Delta t U/\Delta x_\mathrm{min}$ 
using a uniform velocity $U$ and minimum grid width $\Delta x_\mathrm{min}$. 
The time step is set so that the Courant number is CFL = 0.1. 

\begin{figure}[!t]
\begin{minipage}{0.325\linewidth}
\begin{center}
\includegraphics[trim=0mm 0mm 0mm 0mm, clip, width=50mm]{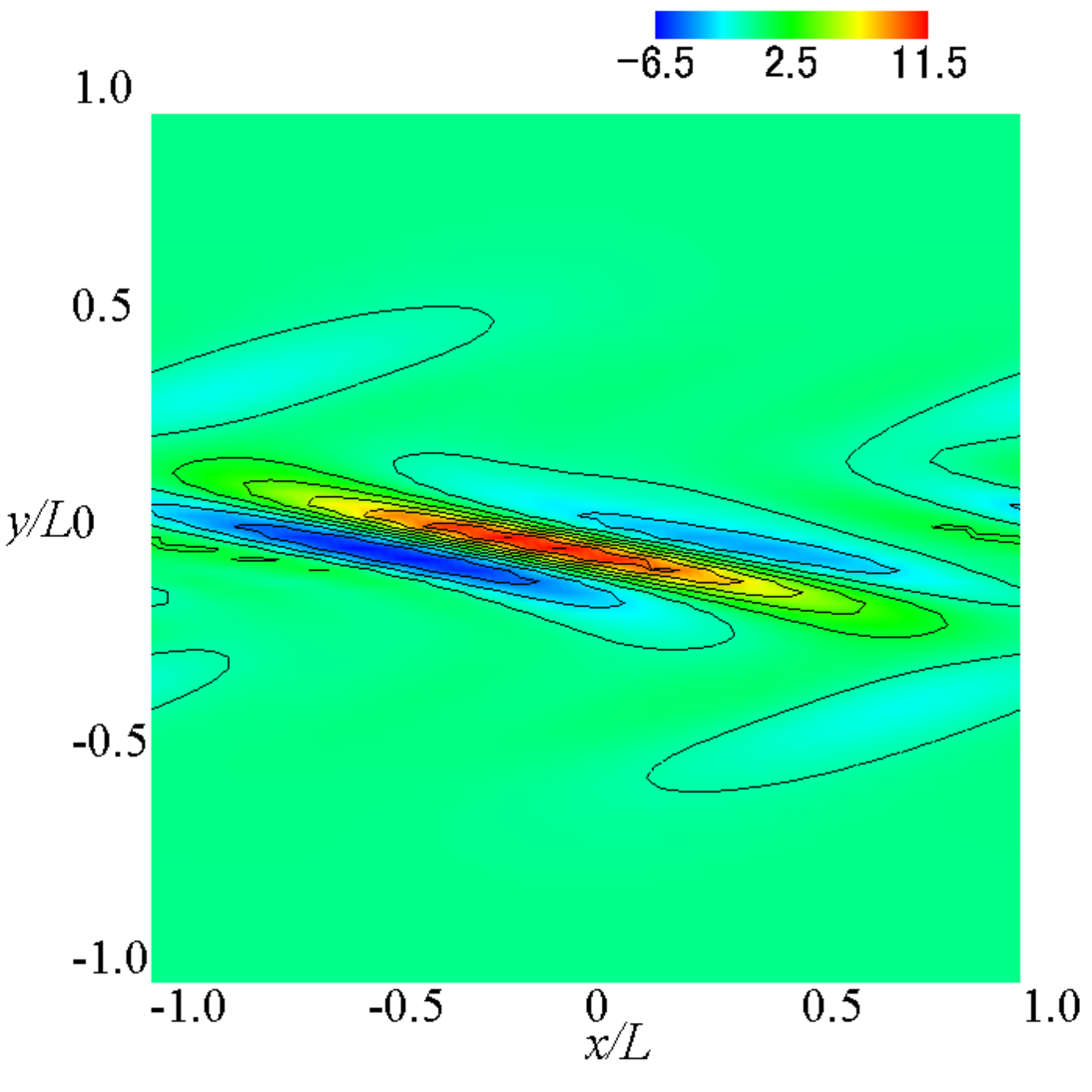} \\
{\small (a) $Ma = 10^{-3}$}
\end{center}
\end{minipage}
\begin{minipage}{0.325\linewidth}
\begin{center}
\includegraphics[trim=0mm 0mm 0mm 0mm, clip, width=50mm]{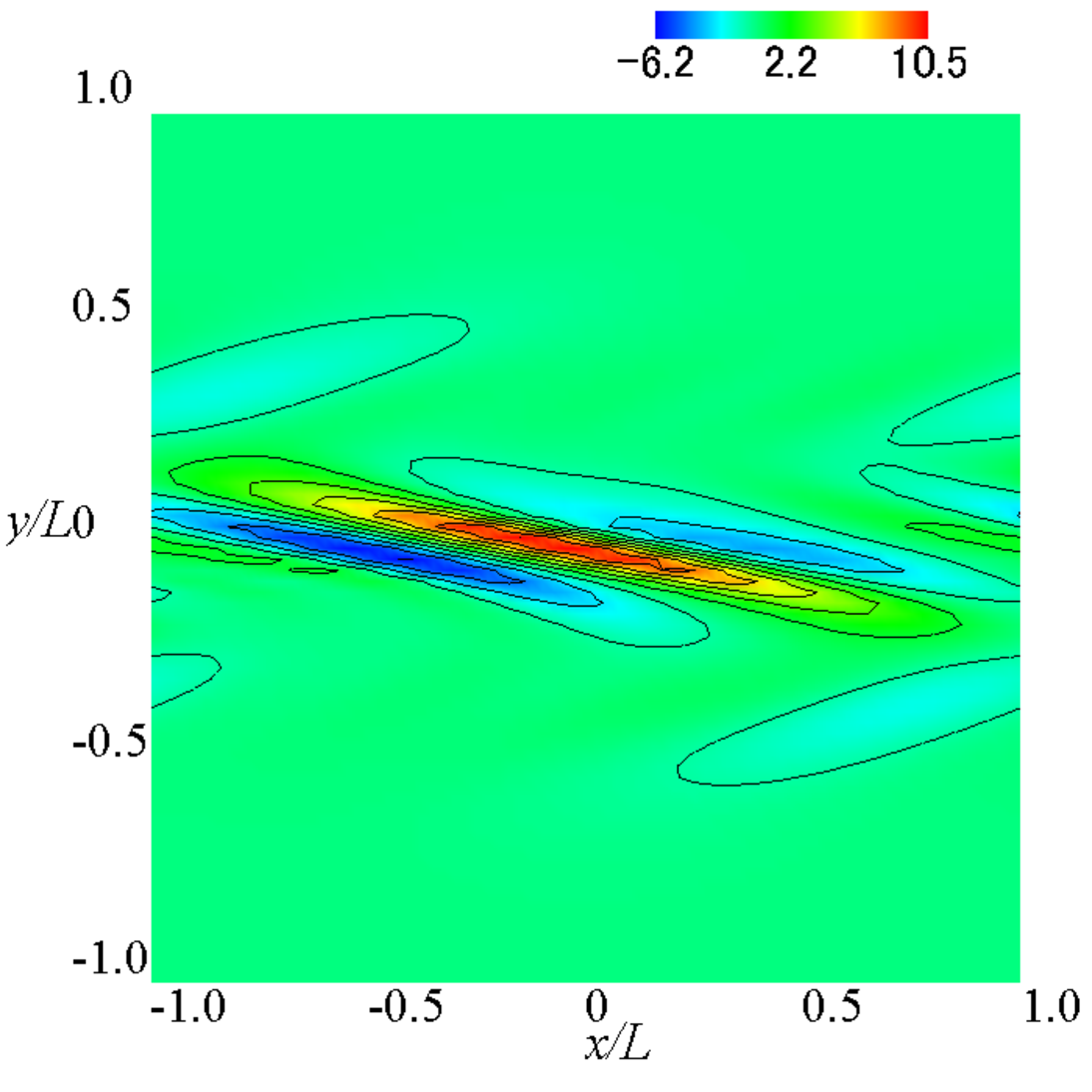} \\
{\small (b) $Ma = 0.1$}
\end{center}
\end{minipage}
\begin{minipage}{0.325\linewidth}
\begin{center}
\includegraphics[trim=0mm 0mm 0mm 0mm, clip, width=50mm]{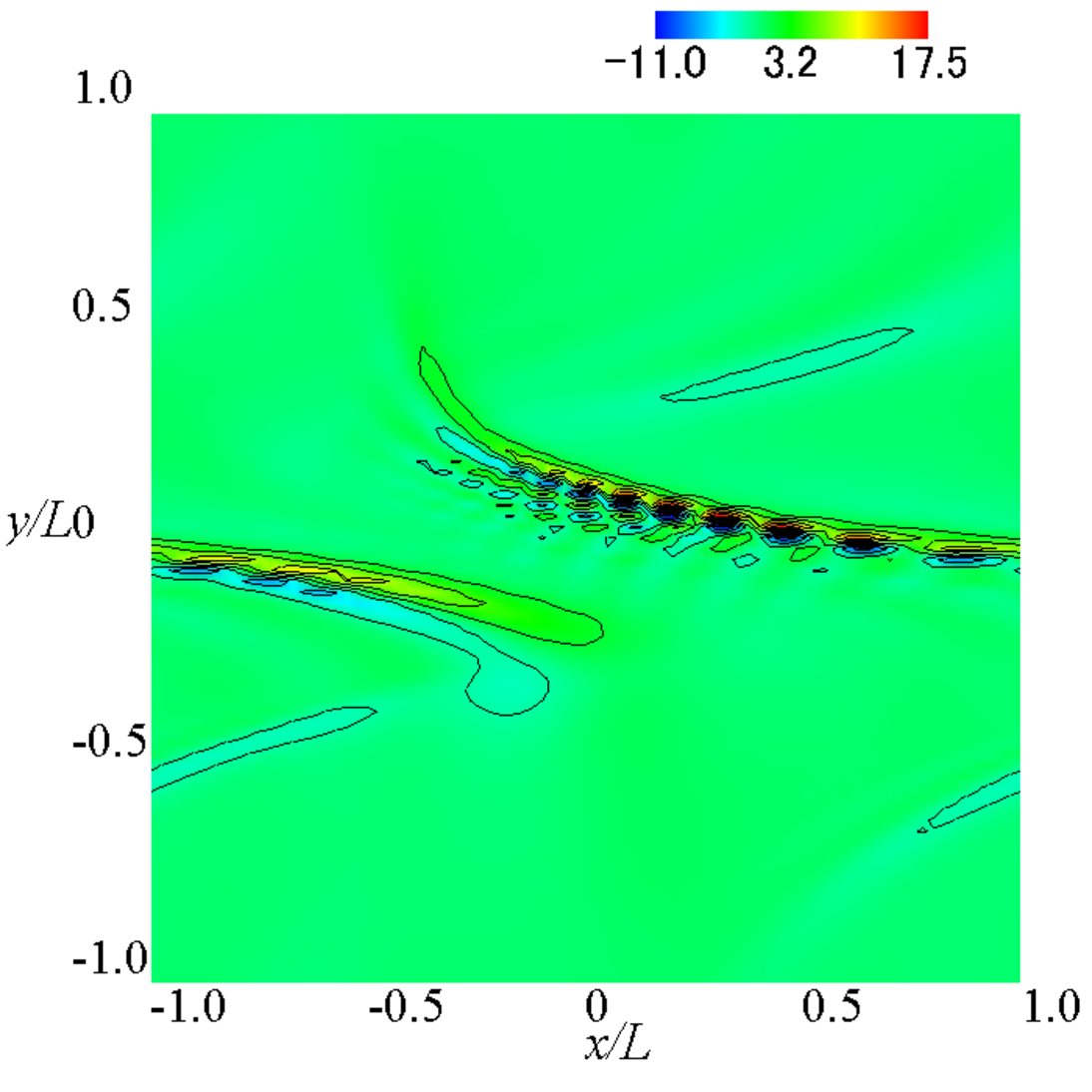} \\
{\small (c) $Ma = 0.5$}
\end{center}
\end{minipage}
\caption{Variation in vorticity contour with Mach number at $t/(L/U) = 10.0$: 
uniform grid ($N = 81$).}
\label{vortex_wz}
\end{figure}

\begin{figure}[!t]
\begin{minipage}{0.325\linewidth}
\begin{center}
\includegraphics[trim=0mm 0mm 0mm 0mm, clip, width=50mm]{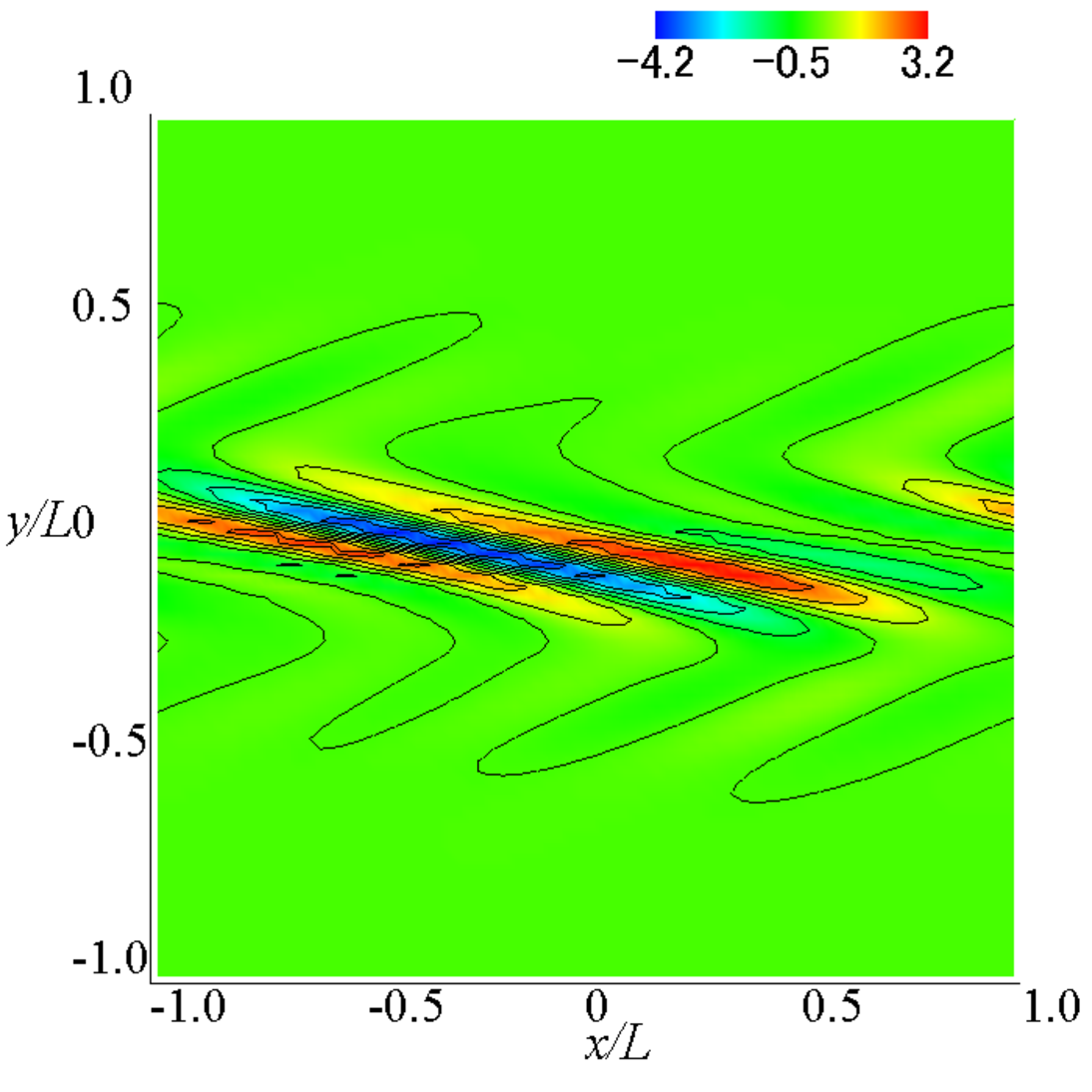} \\
{\small (a) $Ma = 10^{-3}$}
\end{center}
\end{minipage}
\begin{minipage}{0.325\linewidth}
\begin{center}
\includegraphics[trim=0mm 0mm 0mm 0mm, clip, width=50mm]{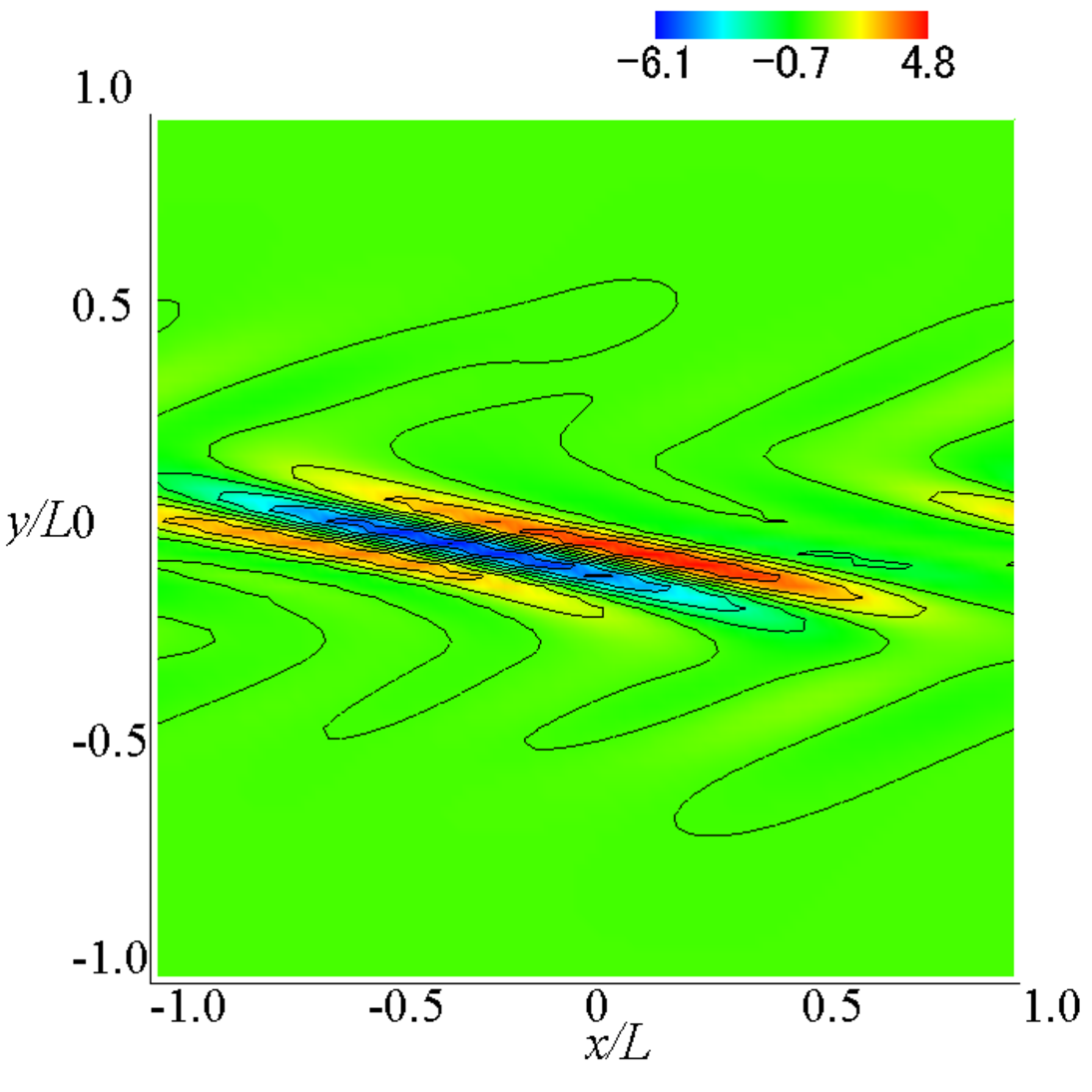} \\
{\small (b) $Ma = 0.1$}
\end{center}
\end{minipage}
\begin{minipage}{0.325\linewidth}
\begin{center}
\includegraphics[trim=0mm 0mm 0mm 0mm, clip, width=50mm]{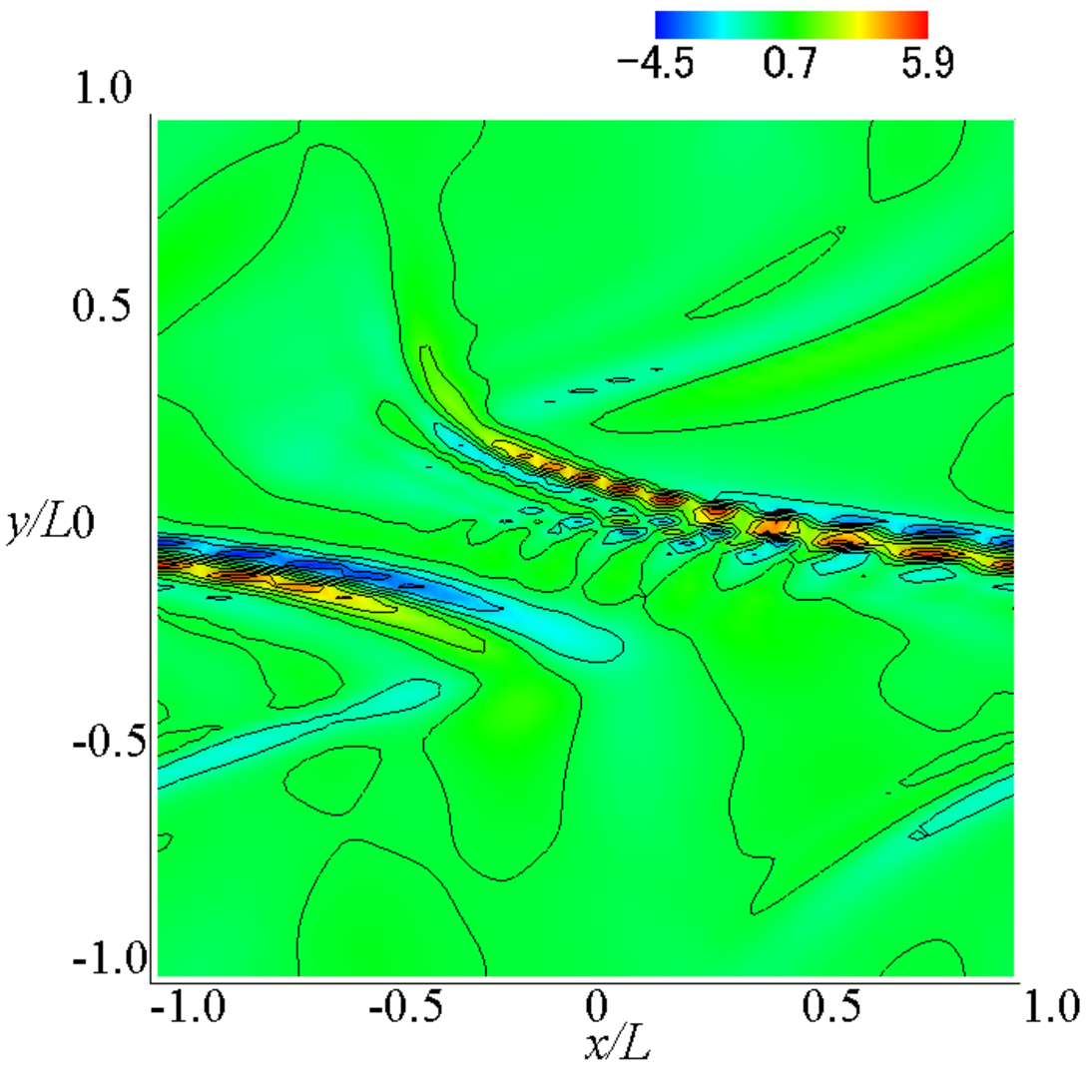} \\
{\small (c) $Ma = 0.5$}
\end{center}
\end{minipage}
\caption{Variation in current density contour with Mach number at $t/(L/U) = 10.0$: 
uniform grid ($N = 81$).}
\label{vortex_jz}
\end{figure}

Figures \ref{vortex_wz} and \ref{vortex_jz} show 
the distribution of vorticity and current density in the $z$-direction at $t/(L/U) = 10$, respectively. 
These results were obtained using the uniform grid with $N = 81$. 
As the vortex rotates counterclockwise, 
the velocity slows down from the uniform velocity for $y > 0$ 
and increases for $y < 0$. 
An induced magnetic field is generated over time, 
and a Lorentz force is formed in the $y$-direction. 
This force deforms the vortex. 
The vortex is significantly deformed from the initial state 
and changes to a thin shear layer. 
A current density layer corresponding to this vortex layer is generated. 
The results for $Ma = 10^{-3}$ and 0.1 agree well with the results of the previous study \citep{Yanaoka_2023}, 
and the influence of Mach number does not appear. 
At $Ma = 0.5$, the vortex and current layers are thinner than the results for low Mach numbers 
and spread in the $y$-direction. 
In addition, it is seen that the absolute values of vorticity and current density increase locally, 
and the vortex and current layers are strengthened by the Mach number.

\begin{figure}[!t]
\begin{minipage}{0.48\linewidth}
\begin{center}
\includegraphics[trim=0mm 0mm 0mm 0mm, clip, width=70mm]{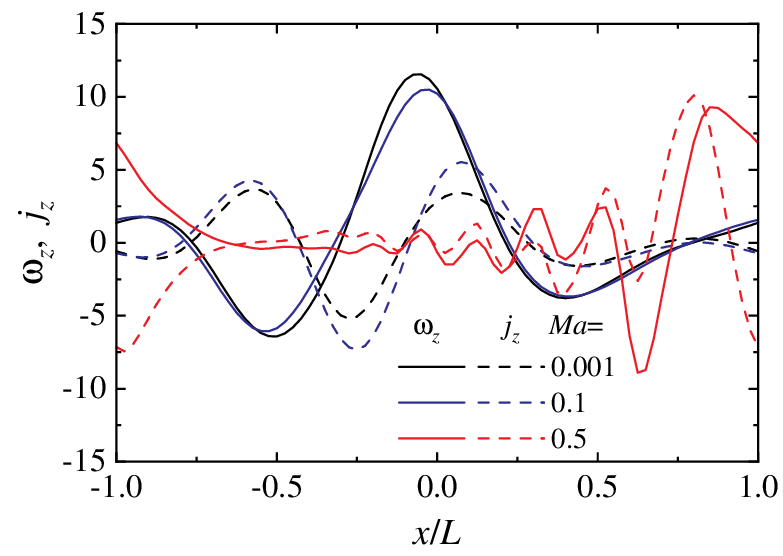} \\
{\small (a) $\omega_z$ and $j_z$}
\end{center}
\end{minipage}
\hspace{0.02\linewidth}
\begin{minipage}{0.48\linewidth}
\begin{center}
\includegraphics[trim=0mm 0mm 0mm 0mm, clip, width=70mm]{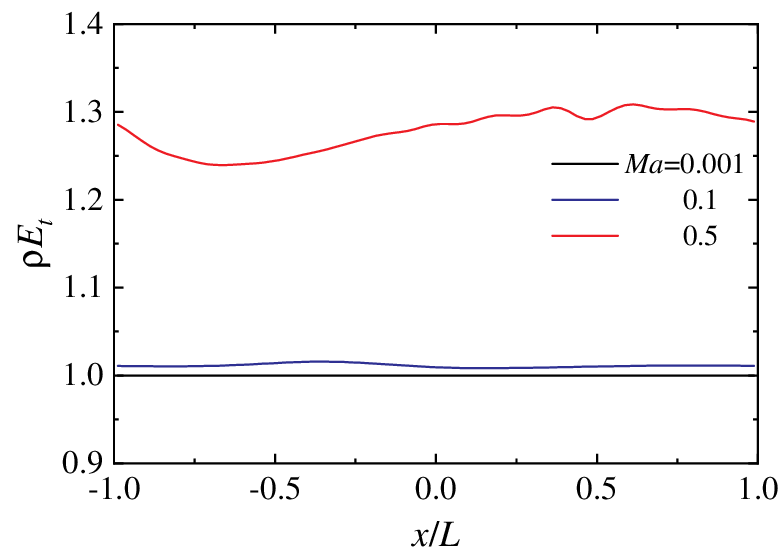} \\
{\small (b) $E_t$}
\end{center}
\end{minipage}
\caption{Distributions of vorticity, current density, and total energy: 
uniform grid ($N = 81$).}
\label{vortex_wz_jz_et}
\end{figure}

The distribution of vorticity, current density, and total energy at $y/L = 0.5$ 
is shown in Fig. \ref{vortex_wz_jz_et}. 
In Figs. \ref{vortex_wz} and \ref{vortex_jz}, 
the results for $Ma = 10^{-3}$ and 0.1 were qualitatively consistent, 
but this figure shows that the slight difference appears quantitatively. 
The distribution for $Ma = 0.5$ is very different from the other results, 
with a significant local increase in total energy.

\begin{figure}[!t]
\begin{minipage}{0.325\linewidth}
\begin{center}
\includegraphics[trim=0mm 0mm 0mm 0mm, clip, width=50mm]{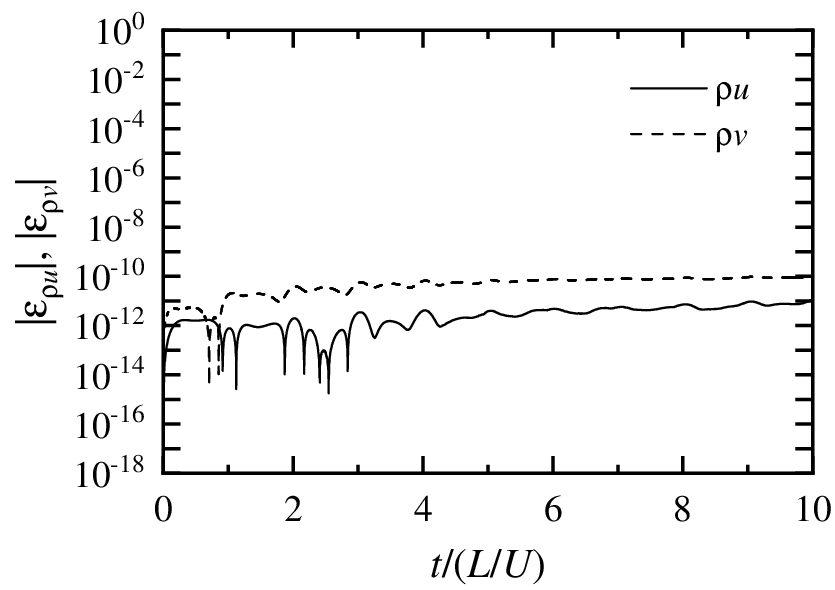} \\
{\small (a) $Ma = 10^{-3}$}
\end{center}
\end{minipage}
\begin{minipage}{0.325\linewidth}
\begin{center}
\includegraphics[trim=0mm 0mm 0mm 0mm, clip, width=50mm]{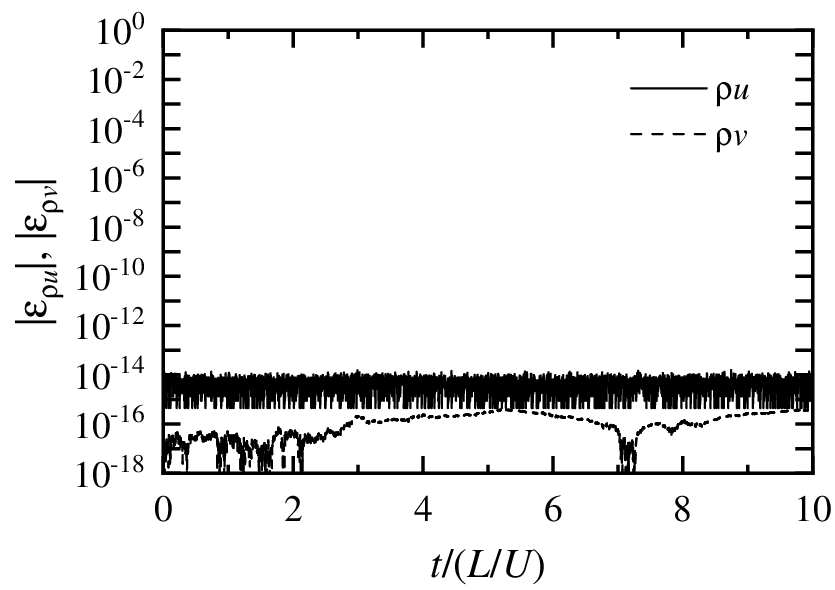} \\
{\small (b) $Ma = 0.1$}
\end{center}
\end{minipage}
\begin{minipage}{0.325\linewidth}
\begin{center}
\includegraphics[trim=0mm 0mm 0mm 0mm, clip, width=50mm]{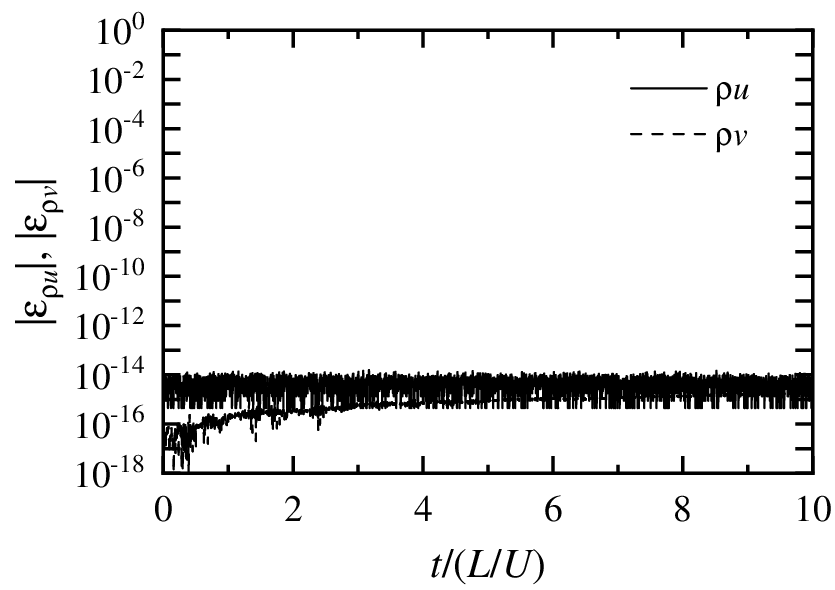} \\
{\small (c) $Ma = 0.5$}
\end{center}
\end{minipage}
\caption{Time variation of momentum error: uniform grid ($N = 81$).}
\label{vortex_rhou}
\end{figure}

\begin{figure}[!t]
\begin{minipage}{0.325\linewidth}
\begin{center}
\includegraphics[trim=0mm 0mm 0mm 0mm, clip, width=50mm]{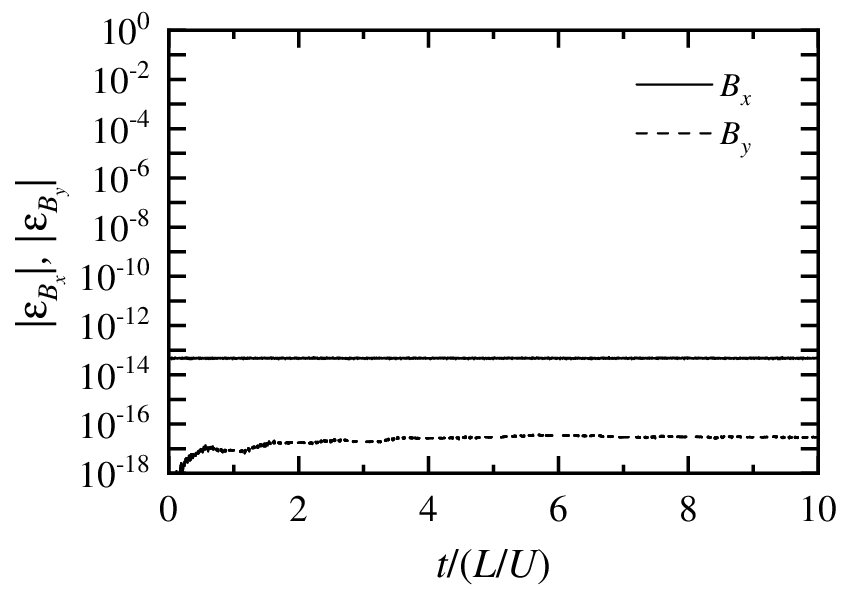} \\
{\small (a) $Ma = 10^{-3}$}
\end{center}
\end{minipage}
\begin{minipage}{0.325\linewidth}
\begin{center}
\includegraphics[trim=0mm 0mm 0mm 0mm, clip, width=50mm]{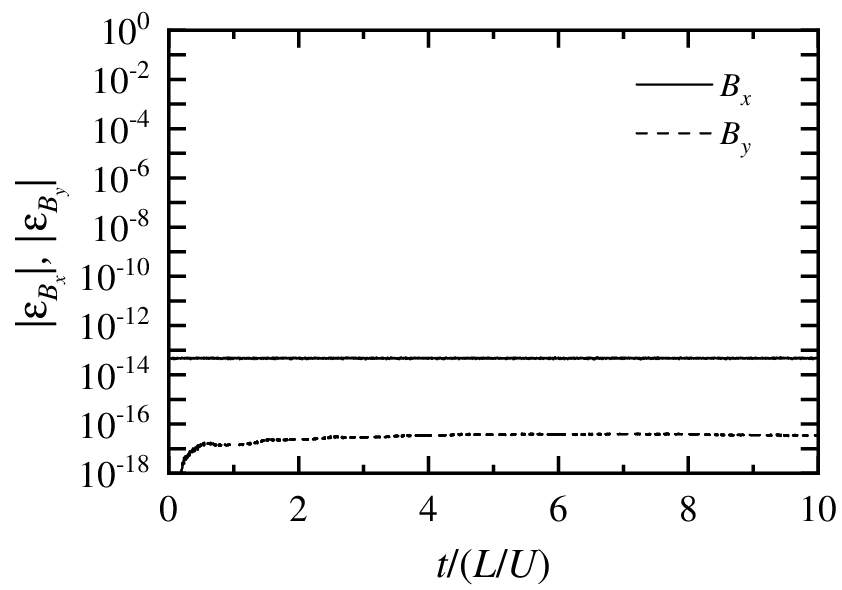} \\
{\small (b) $Ma = 0.1$}
\end{center}
\end{minipage}
\begin{minipage}{0.325\linewidth}
\begin{center}
\includegraphics[trim=0mm 0mm 0mm 0mm, clip, width=50mm]{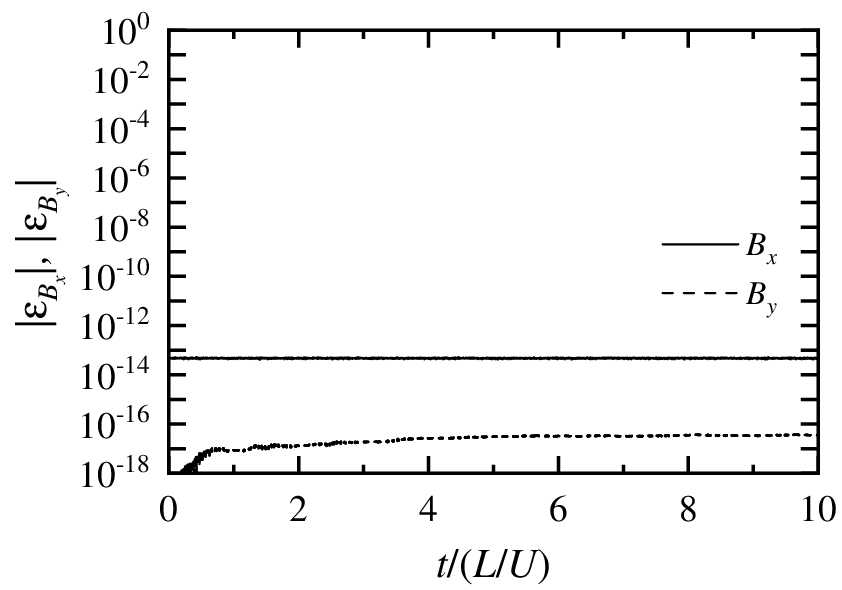} \\
{\small (c) $Ma = 0.5$}
\end{center}
\end{minipage}
\caption{Time variation of magnetic flux density error: uniform grid ($N = 81$).}
\label{vortex_b}
\end{figure}

Figure \ref{vortex_rhou} shows the relative error, 
$\varepsilon_{\rho \bm{u}} = (\langle \rho \bm{u} \rangle - \langle \rho \bm {u} \rangle_0)/\langle (\rho \bm{u})^2 \rangle_0^{1/2}$, 
in the momentum for $N = 81$. 
The subscript 0 represents the initial value. 
The error at $Ma = 10^{-3}$ is around $10^{-11} - 10^{-10}$, 
and the errors at $Ma = 0.1$ and 0.5 are around $10^{-15} - 10^{-14}$. 
When the Mach number is low, the momentum conservation decreases, 
but the error itself is low and not considered a deterioration of the conservation properties. 
Figure \ref{vortex_b} shows the relative error, 
$\varepsilon_{B_x} = (\langle B_x \rangle - \langle B_x \rangle_0)/\langle B_x \rangle_0$, 
in the magnetic flux density $B_x$ and the absolute error, 
$\varepsilon_{B_y} = (\langle B_y \rangle - \langle B_y \rangle_0)$, in $B_y$. 
As $\langle B_y \rangle_0$ is zero, the absolute error is displayed instead of the relative error. 
Regardless of $Ma$, the error is about $10^{-17} - 10^{-14}$, 
and the total amount of magnetic flux density is conserved discretely. 
We also confirmed the total amount for the momentum and magnetic flux density in the $z$-direction. 
The total amount is zero, and no unphysical velocity or magnetic flux density occurs.

\begin{figure}[!t]
\begin{minipage}{0.325\linewidth}
\begin{center}
\includegraphics[trim=0mm 0mm 0mm 0mm, clip, width=50mm]{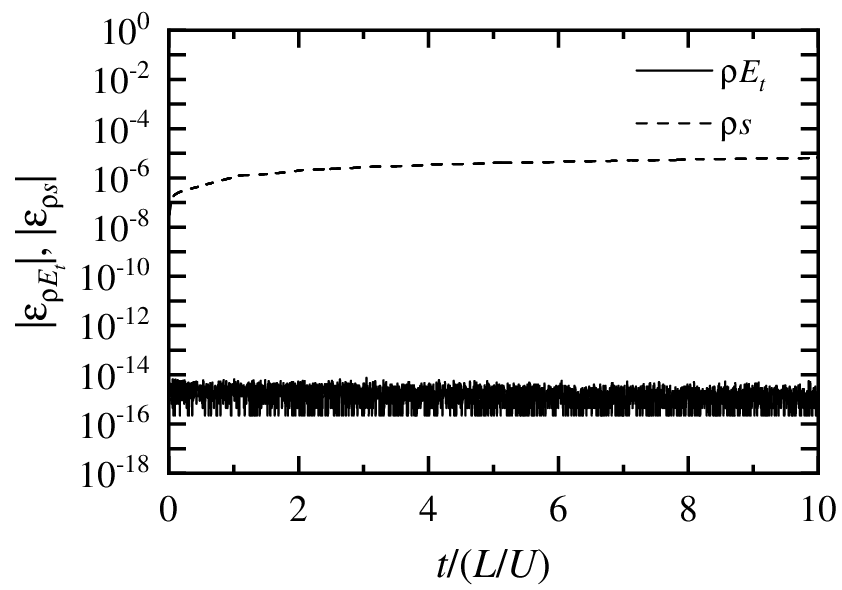} \\
{\small (a) $Ma = 10^{-3}$}
\end{center}
\end{minipage}
\begin{minipage}{0.325\linewidth}
\begin{center}
\includegraphics[trim=0mm 0mm 0mm 0mm, clip, width=50mm]{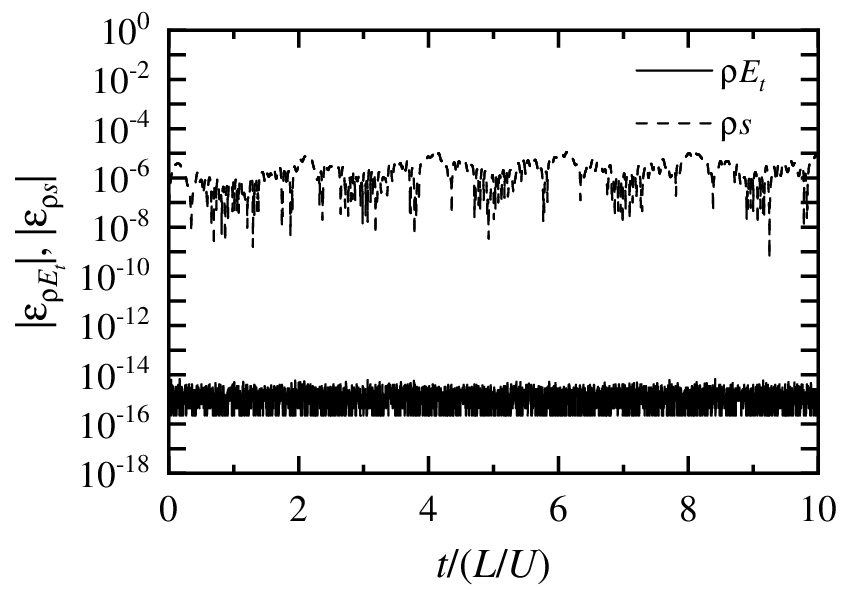} \\
{\small (b) $Ma = 0.1$}
\end{center}
\end{minipage}
\begin{minipage}{0.325\linewidth}
\begin{center}
\includegraphics[trim=0mm 0mm 0mm 0mm, clip, width=50mm]{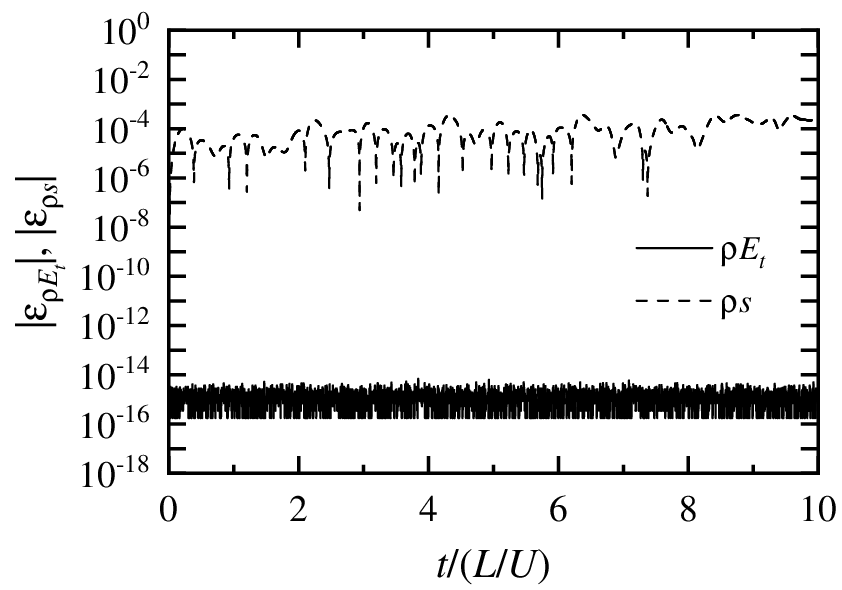} \\
{\small (c) $Ma = 0.5$}
\end{center}
\end{minipage}
\caption{Time variation of total energy error: uniform grid ($N = 81$).}
\label{vortex_et}
\end{figure}

Figure \ref{vortex_et} shows the relative errors, 
$\varepsilon_{\rho E_t} = (\langle \rho E_t \rangle - \langle \rho E_t \rangle_0)/\langle \rho E_t \rangle_0$ 
and $\varepsilon_{\rho s} = (\langle \rho s \rangle - \langle \rho s \rangle_0)/\langle \rho s \rangle_0$, 
in the total energy and entropy. 
The total energy error maintains the rounding error level, regardless of the Mach number. 
On the other hand, the entropy error is on the order of $10^{-6}$ for $Ma = 10^{-3}$ and 0.1 
and the order of $10^{-4}$ for $Ma = 0.5$. 
As mentioned in Subsection \ref{inviscid}, 
as the entropy equation cannot be derived discretely, 
the discrete conservation property deteriorates, but the error maintains a low level. 
We also confirmed that the total amount of magnetic helicity is zero.

\begin{figure}[!t]
\begin{minipage}{0.325\linewidth}
\begin{center}
\includegraphics[trim=0mm 0mm 0mm 0mm, clip, width=50mm]{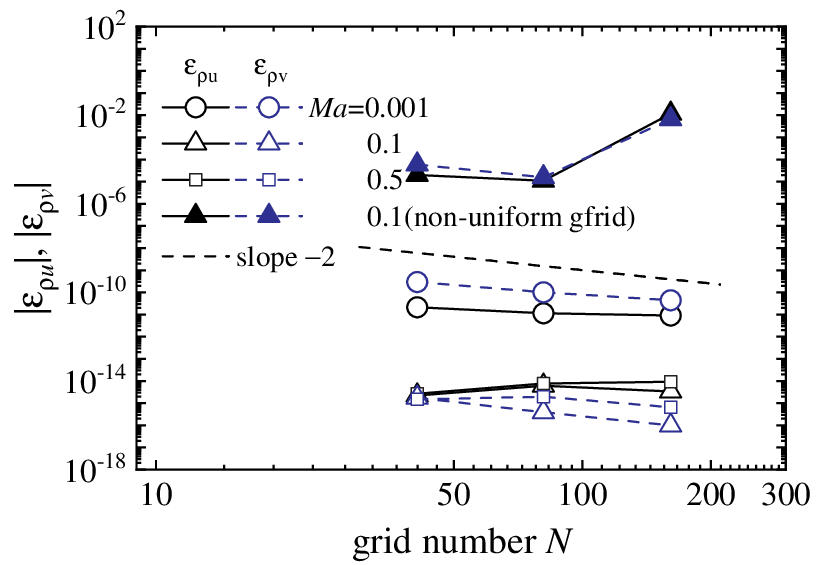} \\
{\small (a) $u$, $v$}
\end{center}
\end{minipage}
\begin{minipage}{0.325\linewidth}
\begin{center}
\includegraphics[trim=0mm 0mm 0mm 0mm, clip, width=50mm]{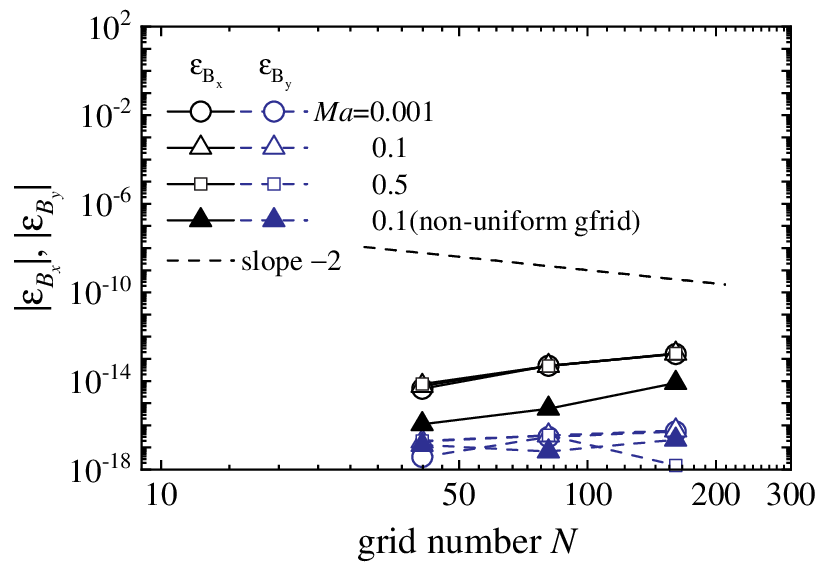} \\
{\small (b) $B_x$, $B_y$}
\end{center}
\end{minipage}
\begin{minipage}{0.325\linewidth}
\begin{center}
\includegraphics[trim=0mm 0mm 0mm 0mm, clip, width=50mm]{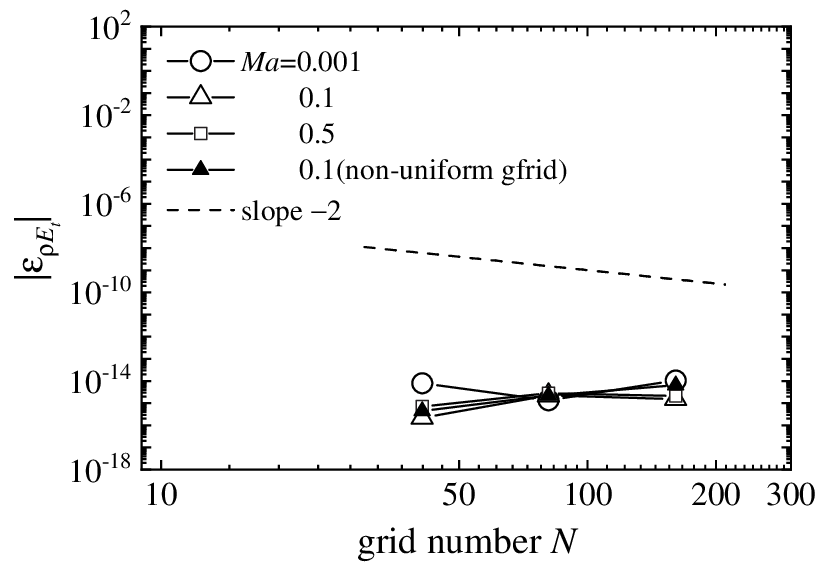} \\
{\small (c) $E_t$}
\end{center}
\end{minipage}
\caption{Errors of momentum, magnetic flux density, and total energy.}
\label{vortex_error}
\end{figure}

Figure \ref{vortex_error} shows the relationship between the relative errors, 
$\varepsilon_{\rho u}$, $\varepsilon_{\rho v}$, $\varepsilon_{B_x}$, 
and $\varepsilon_{\rho E_t}$, in the momentum, magnetic flux density, 
and total energy and the number of grid points $N$. 
An absolute error, $\varepsilon_{B_x}$, of $B_y$ is shown. 
At $Ma = 0.1$, the results using the nonuniform grid are also plotted. 
The momentum error becomes high at $Ma = 0.001$, but it is at a sufficiently low level. 
Also, the error remains constant regardless of $N$. 
The error at $Ma \ge 0.1$ is also a low value regardless of $N$. 
The conservation property deteriorates for the nonuniform grid. 
This tendency is similar to the case of incompressible flow \citep{Yanaoka_2023}. 
The magnetic flux density is also conserved discretely, regardless of $N$ and $Ma$. 
Regarding the total energy, the error is approximately $10^{-15} - 10^{-14}$, regardless of $N$ and $Ma$, 
and the energy conservation property is excellent.

The maximum error of the mass conservation law 
and the maximum divergence error of the magnetic flux density in this analysis 
are $4.83\times 10^{-12}$ and $5.52\times 10^{-11}$, respectively, for the uniform grid 
and $1.80\times 10^{-12}$ and $1.74\times 10^{-11}$, respectively, for the nonuniform grid.

\subsection{Taylor decaying vortex}
\label{decaying_vortex}

We confirm that this computational method can also be applied to incompressible flows. 
For high-Reynolds number flows with the decaying of total energy, 
the accuracy of the present numerical method is verified 
by comparing the calculation result with the exact solution. 
A Taylor decaying vortex analysis was performed to verify the accuracy and convergence. 
The solution to the Taylor decaying vortex problem \citep{Taylor_1923,Yanaoka_2023} 
under the magnetic field is given as
\begin{align}
   \Psi_z &= \frac{1}{k} \cos(k x) \cos(k y) e^{-\frac{2 k^2}{Re}t}, \\
   A_z &= \frac{1}{k} \cos(k x) \cos(k y) e^{-\frac{2 k^2}{Re_m}t}, \\
   p &= - \frac{1}{4} 
          \left[ \cos(2 k x) + \cos(2 k y) \right] e^{-\frac{4 k^2}{Re}t} 
        + \frac{1}{4 Al^2} 
          \left[ 4 \cos^2(k x) \cos^2(k y) - 1 \right] e^{-\frac{4 k^2}{Re_m}t},
\end{align}
where $k = 2\pi$. 
$\Psi_z$ and $A_z$ are the stream function and magnetic vector potential, respectively. 
The velocities are calculated as $u = \partial \Psi_z/\partial y$ 
and $v = -\partial \Psi_z/\partial x$. 
The magnetic flux densities are calculated 
as $B_x = \partial A_z/\partial y$ and $B_y = -\partial A_z/\partial x$. 
These equations are nondimensionalized by the maximum values, $U$ and $B$, 
of the velocity and magnetic flux density, respectively, 
and the wavelength, $L$, of the periodic vortex. 
From Eq. (\ref{vector_potential}), the electric potential is determined to be constant. 
The initial value of the density at the reference temperature $T_0$ is $\rho_0$ and is uniform. 
The initial value of internal energy is found from the equation of state.

The calculation area is $L \times L$, 
and the computational region in the $z$-direction is the grid spacing $\Delta x$. 
The exact solution is given as the initial condition, 
and the periodic boundary is set as the boundary condition. 
The reference values used in this calculation are as follows: 
the length is $l_\mathrm{ref} = L$, velocity is $u_\mathrm{ref} = U$, 
time is $t_\mathrm{ref} = L/U$, density is $\rho_\mathrm{ref} = \rho_0$, 
pressure us $p_\mathrm{ref} = (\kappa -1) \rho_0 c_v T_0$, 
temperature is $T_\mathrm{ref} = T_0$, 
internal energy is $e_\mathrm{ref} = c_v T_0$, 
magnetic flux density is $B_\mathrm{ref} = B$, 
magnetic vector potential is $A_\mathrm{ref} = B L$, 
and electric potential is $\psi_\mathrm{ref} = U L B$. 
In this calculation, the specific heat ratio is set as $\kappa = 5/3$, 
and the Mach number $Ma = U/c_0$ is changed as $Ma = 10^{-6}$ and $10^{-3}$. 
Here, $c_0$ is the sound speed at temperature $T_0$. 
Similarly to the existing research \citep{Liu&Wang_2001,Yanaoka_2023}, 
the Reynolds number $Re = 10^4$, the magnetic Reynolds number $Re_m = 50$, 
and the Alfv\'{e}n number $Al = 1$. 
To compare the decaying tendency of the vortex, 
we also calculate the conditions of $Re = 10^2$ and $Re_m = 1$. 
The Prandtl number is set as $Pr = 1$. 

Uniform and nonuniform grids with $N \times N \times 2$ grid points are used. 
$N$ is the number of grid points in the $x$- and $y$-directions, 
and $N = 11$, 21, 41, and 81. 
We investigated the convergence of the calculation results against the number of grid points. 
We use the same nonuniform grid as in the existing research \citep{Yanaoka_2023}. 
Rather than using nonuniform grids to capture the phenomena accurately, 
we use nonuniform grids to investigate changes in energy conservation properties with the grid. 
The time step is fixed at $\Delta t/(L/U) = 0.001$, 
and the calculation results at time $t/(L/U) = 0.5$, 
when the strength of the vortex is halved, are compared with the exact solution. 
We used the same time step as the previous study \citep{Liu&Wang_2001}. 
The Courant number is defined as $\mbox{CFL} = \Delta t U/\Delta x_\mathrm{min}$ 
using the reference velocity $U_\mathrm{ref} = U$ and the minimum grid width $\Delta x_\mathrm{min}$. 
Because the time step is fixed, the Courant number varies with the grid width. 
For the uniform grids, the Courant number changes as $\mbox{CFL} = 0.01-0.08$. 
For the nonuniform grids, it changes as $\mathrm{CFL} = 0.011-0.169$. 
The Courant number, $\mbox{CFL} = \Delta t (U+c_0)/\Delta x_\mathrm{min}$, 
defined with the sound speed changes as $\mbox{CFL} = 10-80$ for the uniform grids 
and $\mbox{CFL} = 11-169$ for nonuniform grids.

\begin{figure}[!t]
\begin{minipage}{0.48\linewidth}
\begin{center}
\includegraphics[trim=0mm 0mm 0mm 0mm, clip, width=70mm]{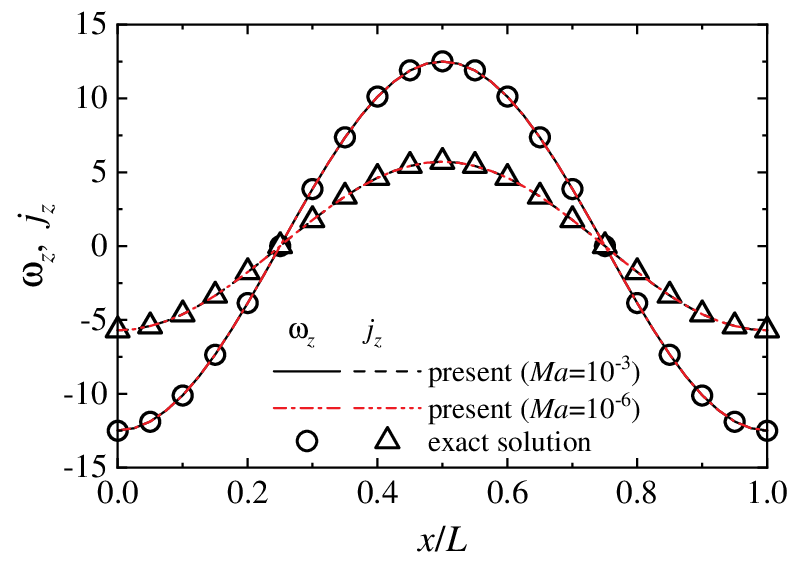} \\
{\small (a) Uniform grid}
\end{center}
\end{minipage}
\hspace{0.02\linewidth}
\begin{minipage}{0.48\linewidth}
\begin{center}
\includegraphics[trim=0mm 0mm 0mm 0mm, clip, width=70mm]{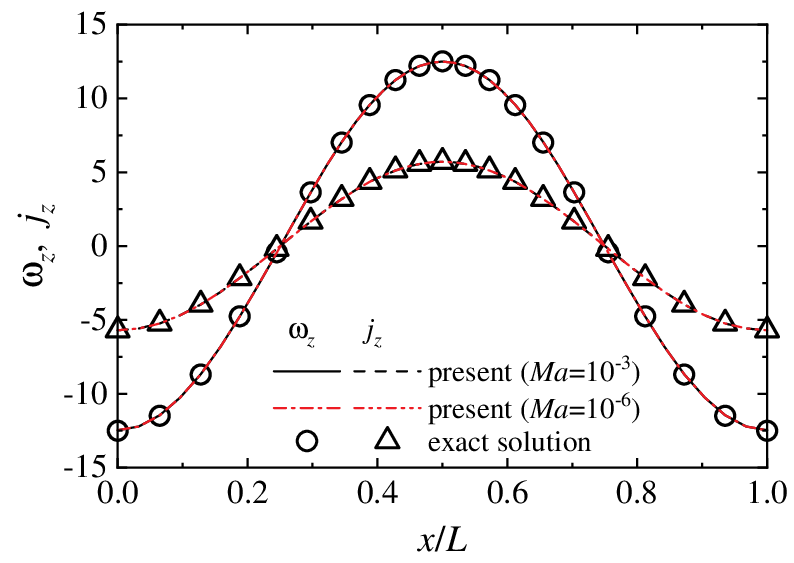} \\
{\small (b) Nonuniform grid}
\end{center}
\end{minipage}
\caption{Distributions of vorticity and current density: 
$N = 41$, $Re = 10^4$, $Re_m = 50$.}
\label{decay_wz_jz}
\end{figure}

We confirmed that the trends of the flow and magnetic fields at $t/(L/U) = 0.5$ are the same 
as in the previous study \citep{Yanaoka_2023}. 
Figure \ref{decay_wz_jz} shows the distributions for the vorticity $\omega_z$ 
and current density $j_z$ in the $z$-direction at $y/L = 0.5$ 
compared with the exact solution. 
For both the uniform and nonuniform grids, 
this calculation result agrees with the analytical solution, regardless of $Ma$. 
Even when $Ma = 10^{-6}$, the calculation does not become unstable, 
and a stable convergent solution is obtained.

\begin{figure}[!t]
\begin{minipage}{0.48\linewidth}
\begin{center}
\includegraphics[trim=0mm 0mm 0mm 0mm, clip, width=70mm]{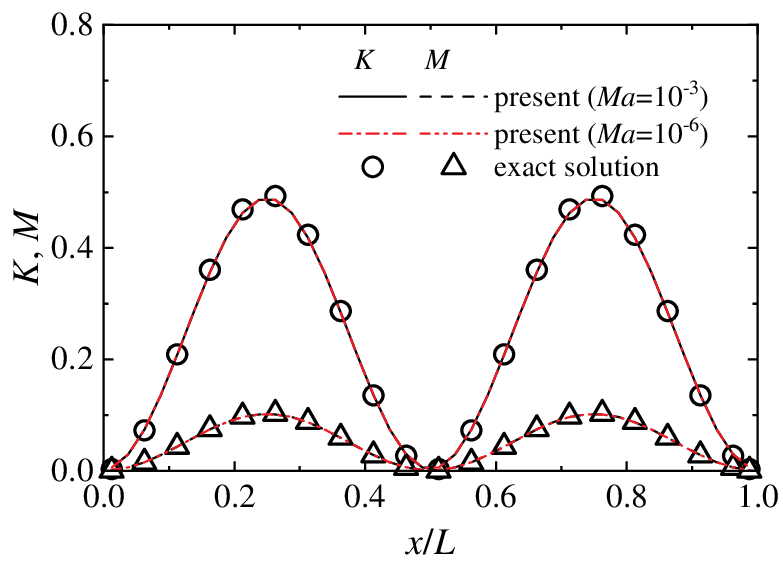} \\
{\small (a) Uniform grid}
\end{center}
\end{minipage}
\hspace{0.02\linewidth}
\begin{minipage}{0.48\linewidth}
\begin{center}
\includegraphics[trim=0mm 0mm 0mm 0mm, clip, width=70mm]{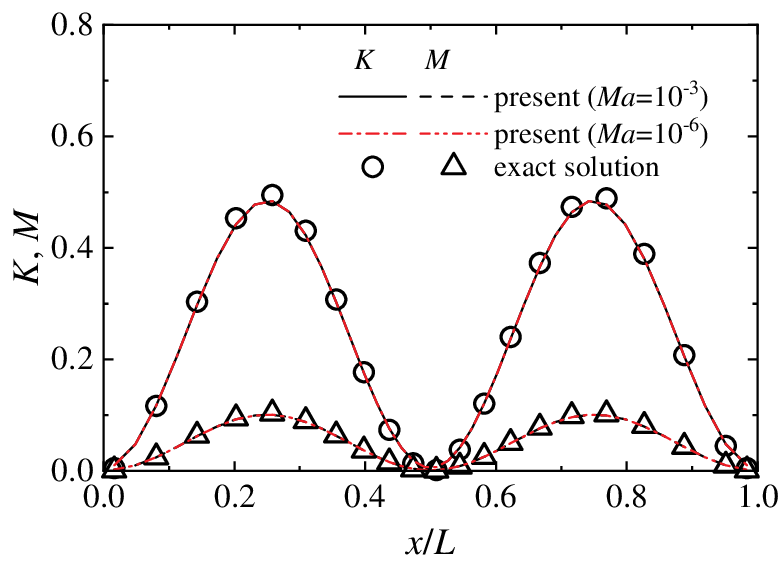} \\
{\small (b) Nonuniform grid}
\end{center}
\end{minipage}
\caption{Distributions of kinetic and magnetic energies: 
$N = 41$, $Re = 10^4$, $Re_m = 50$.}
\label{decay_k_m}
\end{figure}

Figure \ref{decay_k_m} compares the distributions of kinetic energy $K$ 
and magnetic energy $M$ at $y/L = 0.5$ with the exact solution. 
Here, $K$ and $M$ are dimensionless using the representative velocity, 
as in the case of incompressible MHD flow \citep{Yanaoka_2023}. 
Regardless of the lattice and $Ma$, this calculation result agrees with the exact solution.

\begin{figure}[!t]
\begin{minipage}{0.48\linewidth}
\begin{center}
\includegraphics[trim=0mm 0mm 0mm 0mm, clip, width=70mm]{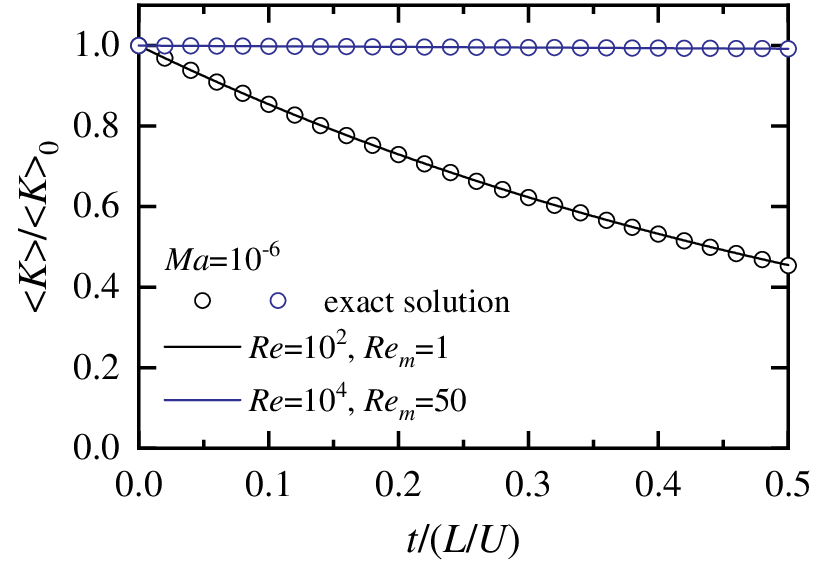} \\
{\small (a) $\langle K \rangle$}
\end{center}
\end{minipage}
\hspace{0.02\linewidth}
\begin{minipage}{0.48\linewidth}
\begin{center}
\includegraphics[trim=0mm 0mm 0mm 0mm, clip, width=70mm]{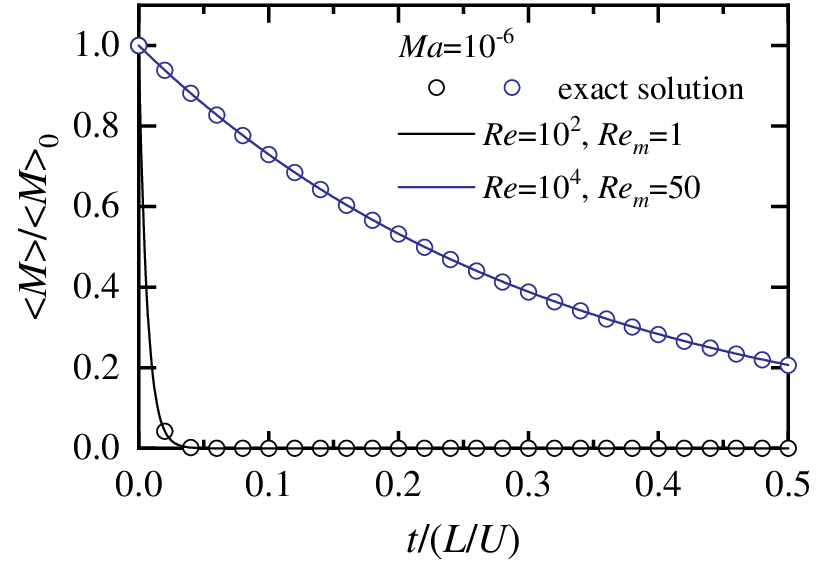} \\
{\small (b) $\langle M \rangle$}
\end{center}
\end{minipage}
\caption{Time variation of total amounts of kinetic and magnetic energies: 
uniform grid ($N = 41$).}
\label{decay_sum_k_m}
\end{figure}

To investigate the decaying trend of the vortex, 
Fig. \ref{decay_sum_k_m} shows the ratios, 
$\langle K \rangle$/$\langle K \rangle_0$ and $\langle M \rangle$/$\langle M \rangle_0$, 
of the total amount of energy to the initial value. 
This calculation used a uniform grid with $N = 41$. 
For all distributions, the calculation results agree well with the exact solution. 
We also confirmed that the results for the uniform and nonuniform grids were in complete agreement. 
At $Re = 10^4$ and $Re_m = 50$, 
the magnetic energy decreases with time, 
but the kinetic energy does not show a significant decay. 
Under the conditions of $Re = 10^2$ and $Re_m = 1$, 
the kinetic energy also decreases, and the magnetic energy decays rapidly. 
These results show that the present numerical method accurately captures 
energy attenuation in incompressible MHD flows.

\begin{figure}[!t]
\begin{minipage}{0.48\linewidth}
\begin{center}
\includegraphics[trim=0mm 0mm 0mm 0mm, clip, width=70mm]{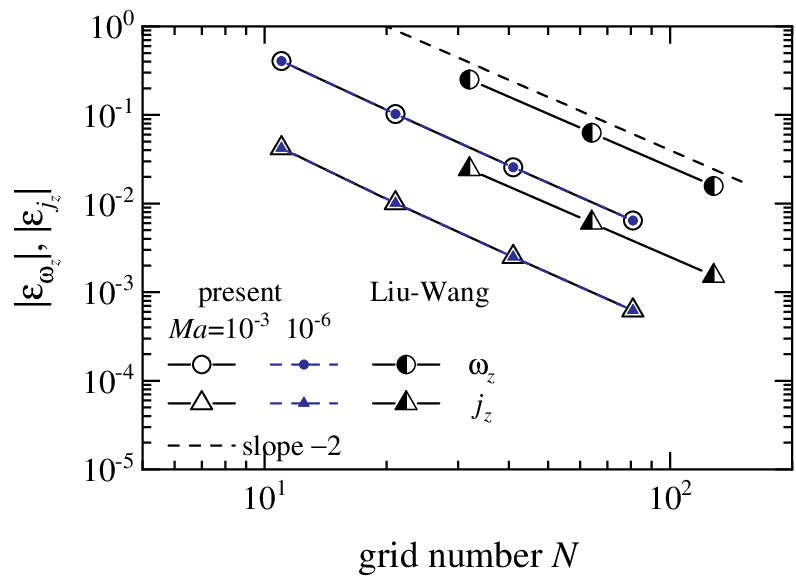} \\
{\small (a) Uniform grid}
\end{center}
\end{minipage}
\hspace{0.02\linewidth}
\begin{minipage}{0.48\linewidth}
\begin{center}
\includegraphics[trim=0mm 0mm 0mm 0mm, clip, width=70mm]{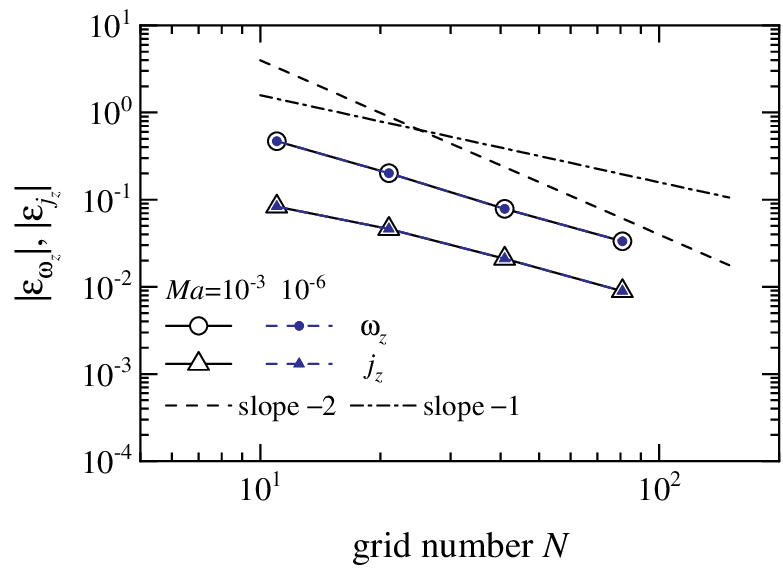} \\
{\small (b) Nonuniform grid}
\end{center}
\end{minipage}
\caption{Errors of vorticity and current density: 
uniform grid ($N = 41$), $Re = 10^4$, $Re_m = 50$.}
\label{decay_error_wz_jz}
\end{figure}

Figure \ref{decay_error_wz_jz} compares the $L^{\infty}$ errors, 
$|\varepsilon_{\omega_z}|$ and $|\varepsilon_{j_z}$, 
of the vorticity and current density with the previous result \citep{Liu&Wang_2001}. 
We can confirm that the calculation results are second-order accurate for the uniform grid. 
In addition, the error is lower than the existing result, 
and the calculation accuracy is good. 
For the nonuniform grid, quadratic convergence is not maintained. 
A slight decrease in the calculation accuracy of velocity and magnetic flux density worsens 
the convergence of vorticity and current density. 
This trend is similar to previous research \citep{Yanaoka_2023}. 
In addition, no change in the error depending on $Ma$ is observed.

\begin{figure}[!t]
\begin{minipage}{0.48\linewidth}
\begin{center}
\includegraphics[trim=0mm 0mm 0mm 0mm, clip, width=70mm]{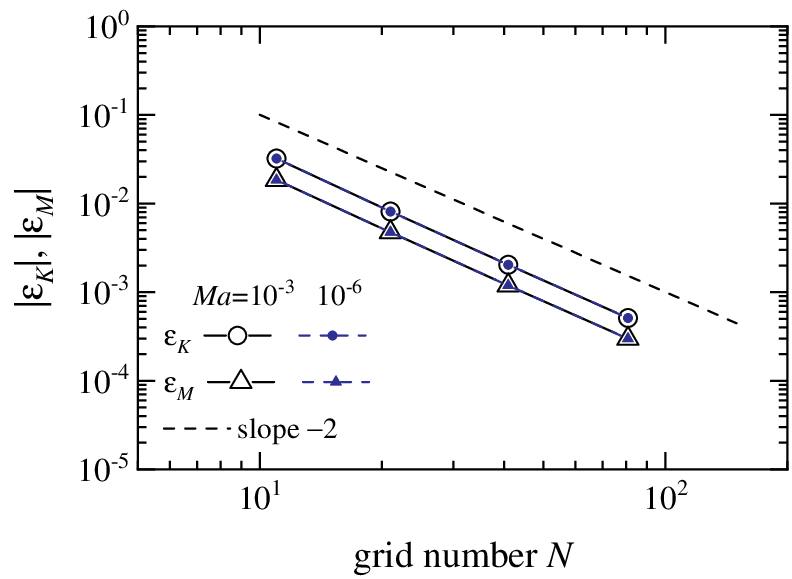} \\
{\small (a) Uniform grid}
\end{center}
\end{minipage}
\hspace{0.02\linewidth}
\begin{minipage}{0.48\linewidth}
\begin{center}
\includegraphics[trim=0mm 0mm 0mm 0mm, clip, width=70mm]{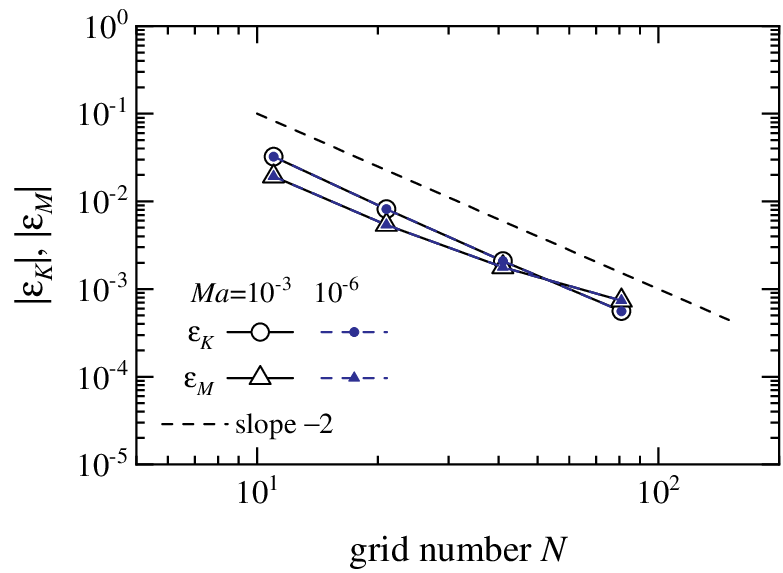} \\
{\small (b) Nonuniform grid}
\end{center}
\end{minipage}
\caption{Errors of kinetic and magnetic energies: 
uniform grid ($N = 41$), $Re = 10^4$, $Re_m = 50$.}
\label{decay_error_k_m}
\end{figure}

Figure \ref{decay_error_k_m} shows the changes in the relative errors of kinetic energy and magnetic energy, 
$|\varepsilon_K|$ and $|\varepsilon_M|$, with the grid resolution. 
The error decreases with a slope of $-2$ for both uniform and nonuniform grids, 
and the computational method is spatially second-order accurate. 
In addition, no change in the error with $Ma$ is observed.

The maximum error of the mass conservation law in this analysis 
is in the range $4.3 \times 10^{-13} - 6.7 \times 10^{-11}$ for the uniform grid 
and $4.3 \times 10^{-13} - 5.2 \times 10^{-8}$ for the nonuniform grid. 
The maximum divergence error of the magnetic flux density 
is in the range $1.0 \times 10^{-13} - 5.4 \times 10^{-11}$ for the uniform grid 
and $1.1 \times 10^{-13}- 2.3 \times 10^{-11}$ for the nonuniform grid. 
The Courant number increases for the nonuniform grid with $N = 81$, 
so the residuals of the mass conservation law and magnetic charge absence law increase.

In this analysis, the calculations were stable, 
and convergent solutions were obtained even at significantly low Mach numbers. 
A series of investigations revealed that the numerical method developed 
in this study can analyze flows ranging from incompressible flows to compressed flows 
at low Mach numbers.

\subsection{Orszag--Tang vortex}
\label{OT_vortex}

We investigate the influence of Mach number on the transition process to turbulent flow. 
The Orszag--Tang vortex \citep{Orszag&Tang_1979} is a model often used to verify numerical methods. 
Initial values for incompressible MHD flows are given in an existing study \citep{Orszag&Tang_1979}. 
The initial stream function, magnetic vector potential, and pressure are expressed, respectively, as
\begin{equation}
  \Psi_z = \frac{1}{k} \left[ \cos(k y) + \cos(k x) \right],
\end{equation}
\begin{equation}
  A_z = \frac{1}{k} \left[ \cos(k y) + \frac{1}{2} \cos(2 k x) \right],
\end{equation}
\begin{equation}
  p = - \cos(k x) \cos(k y),
\end{equation}
where $k = 2\pi$. 
The initial pressure is an exact solution for inviscid steady flow without an applied magnetic field. 
The initial value of the density at the reference temperature $T_0$ is $\rho_0$ and is uniform. 
The initial value of internal energy is found from the equation of state. 
The maximum values of velocity and magnetic flux density are $U$ and $B$, respectively, 
and the wavelength of the periodic vortex is $L$. 
The calculation area is $L \times L$, 
and the computational region in the $z$-direction is the grid spacing. 
The periodic boundary is set as the boundary condition.

The reference values used in this calculation are as follows: 
the length is $l_\mathrm{ref} = L$, velocity is $u_\mathrm {ref} = U$, 
time is $t_\mathrm{ref} = L/U$, density is $\rho_\mathrm{ref} = \rho_0$, 
pressure is $p_\mathrm{ref} = (\kappa -1) \rho_0 c_v T_0$, 
temperature is $T_\mathrm{ref} = T_0$, 
internal energy is $e_\mathrm{ref} = c_v T_\mathrm{ref}$, 
and magnetic flux density is $B_\mathrm{ref} = B$. 
The equations of the initial values are nondimensionalized using these reference values.

In this calculation, the specific heat ratio is $\kappa = 5/3$, 
and the Mach number is changed as $Ma = U/c_0 = 10^{-3}$, 0.1, 0.3, and 0.5. 
Here, $c_0$ is the sound speed at the initial temperature. 
Similarly to existing research \citep{Orszag&Tang_1979}, 
the Reynolds number $Re$ and the magnetic Reynolds number $Re_m$ were set to the same value. 
In this study, we vary $Re = Re_m = 100$, 200, 400, 1000, 
and set the Prandtl number as $Pr = 1$ and the Alfv\'{e}n number as $Al = 1$.

This analysis uses a uniform grid with $N \times N \times 2$. 
$N$ is the number of grid points in the $x$- and $y$-directions, 
and is changed to $N = 21$, 41, 81, 161. 
The time step is fixed at $\Delta t/(L/U) = 1.0\times10^{-3}$. 
The Courant number is defined as $\mathrm{CFL} = \Delta t U/\Delta x$ 
using the maximum velocity $U$ and the grid width $\Delta x$ and is $\mathrm{CFL} = 0.02 -0.16$. 
In addition, the Courant number considering the sound speed is $\mathrm{CFL} = 2-160$.

\begin{figure}[!t]
\begin{minipage}{0.48\linewidth}
\begin{center}
\includegraphics[trim=0mm 0mm 0mm 0mm, clip, width=70mm]{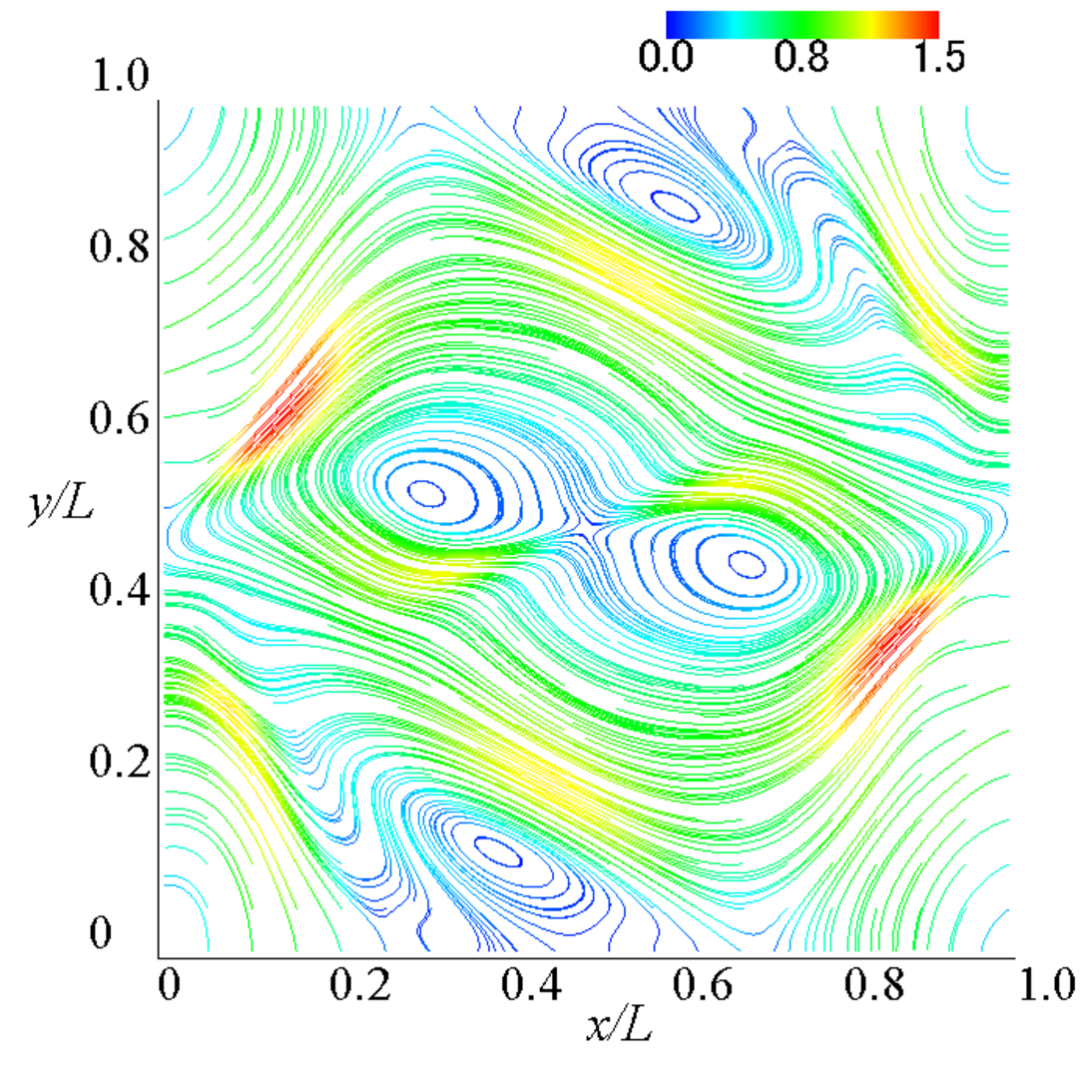} \\
\vspace*{-0.5\baselineskip}
{\small (a) $Ma = 10^{-3}$}
\end{center}
\end{minipage}
\hspace{0.02\linewidth}
\begin{minipage}{0.48\linewidth}
\begin{center}
\includegraphics[trim=0mm 0mm 0mm 0mm, clip, width=70mm]{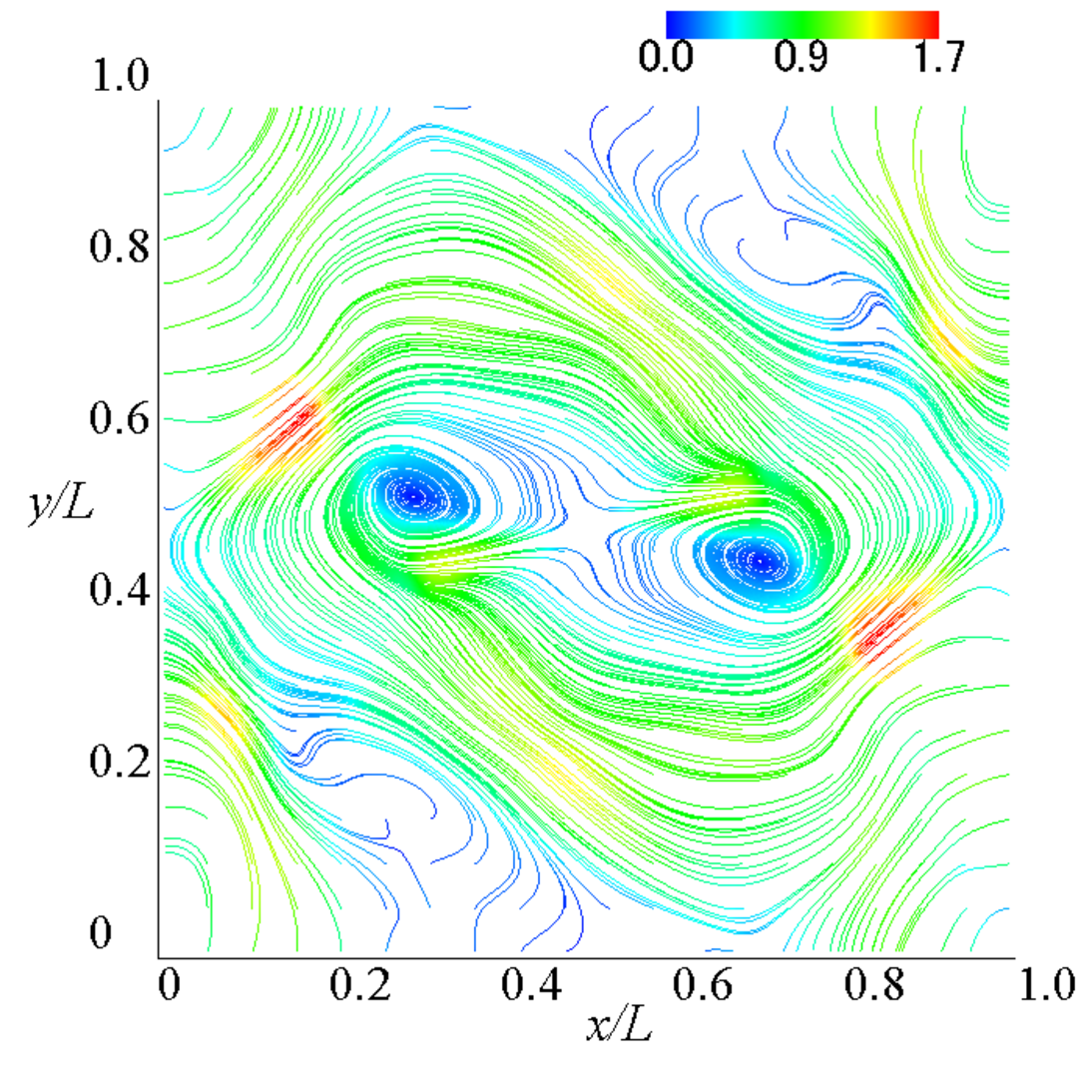} \\
\vspace*{-0.5\baselineskip}
{\small (b) $Ma = 0.5$}
\end{center}
\end{minipage}
\caption{Flow streamline: 
$N = 81$, $t/(L/U) = 0.3$, $Re = Re_m = 400$.}
\label{ot_streamline_Re400}
\end{figure}

\begin{figure}[!t]
\begin{minipage}{0.48\linewidth}
\begin{center}
\includegraphics[trim=0mm 0mm 0mm 0mm, clip, width=70mm]{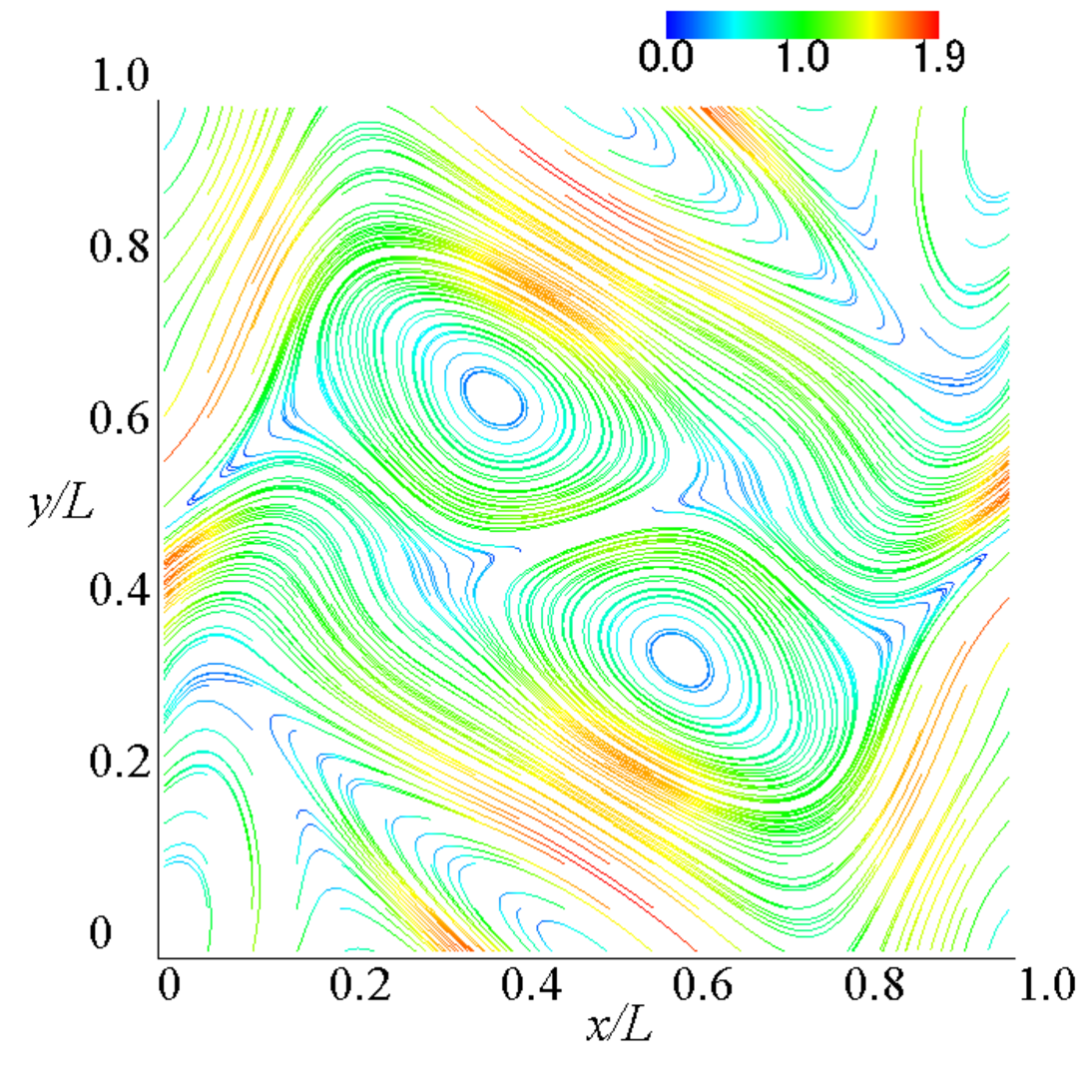} \\
\vspace*{-0.5\baselineskip}
{\small (a) $Ma = 10^{-3}$}
\end{center}
\end{minipage}
\hspace{0.02\linewidth}
\begin{minipage}{0.48\linewidth}
\begin{center}
\includegraphics[trim=0mm 0mm 0mm 0mm, clip, width=70mm]{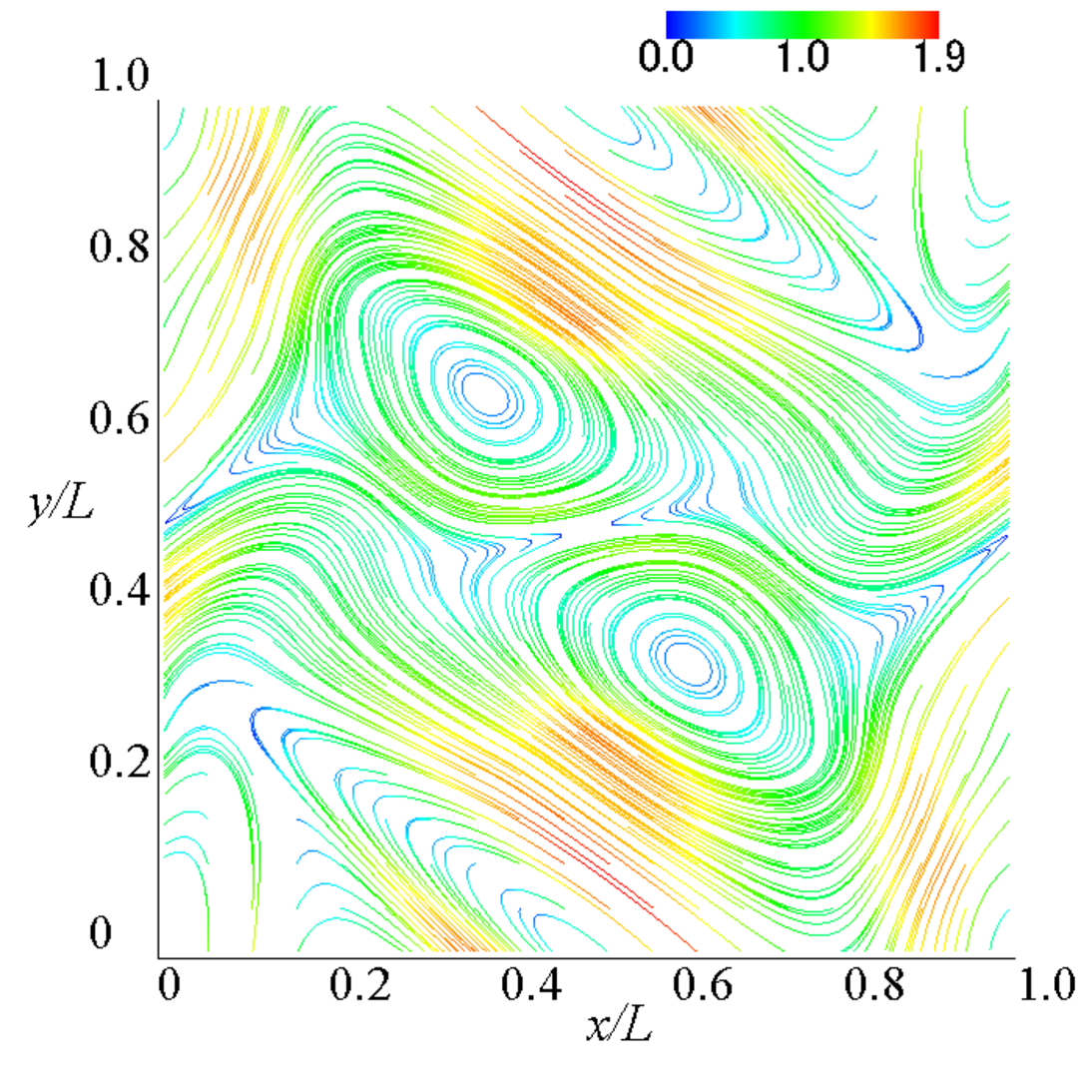} \\
\vspace*{-0.5\baselineskip}
{\small (b) $Ma = 0.5$}
\end{center}
\end{minipage}
\caption{Magnetic streamline: 
$N = 81$, $t/(L/U) = 0.3$, $Re = Re_m = 400$.}
\label{ot_magnetic_streamline_Re400}
\end{figure}

In Figs. \ref{ot_streamline_Re400} and \ref{ot_magnetic_streamline_Re400}, 
the flow streamlines and magnetic streamlines at time $t/(L/U) = 0.3$ are compared between $Ma=10^{-3}$ and 0.5. 
The current density enstrophy reaches a maximum around $t/(L/U) = 0.3$. 
The color level represents the magnitude of velocity and magnetic flux density. 
In the initial state, a circular vortex exists in the central region. 
Over time, a single vortex transforms into a vortex pair 
under the influence of the magnetic field. 
Dahlburg and Picone \citep{Dahlburg&Picone_1989} reported 
that in compressible flow, a finer flow structure appears than in incompressible flow, 
whereas no clear difference in the magnetic field depending on Mach number exists. 
This result is qualitatively consistent with existing results \citep{Dahlburg&Picone_1989,Warburton&Karniadakis_1999,Dumbser_et_al_2019, Fambri_2021}.

\begin{figure}[!t]
\begin{minipage}{0.48\linewidth}
\begin{center}
\includegraphics[trim=0mm 0mm 0mm 0mm, clip, width=70mm]{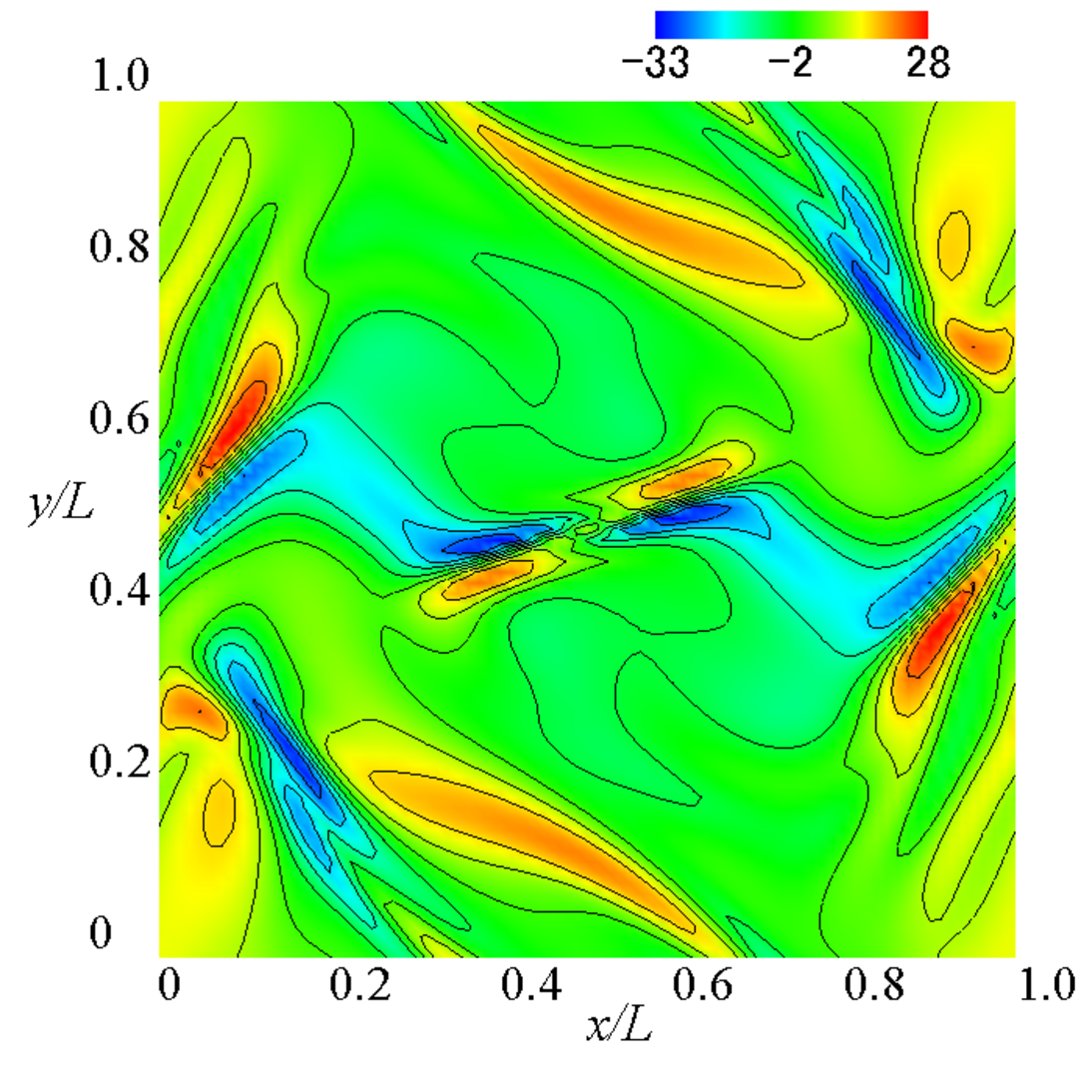} \\
\vspace*{-0.5\baselineskip}
{\small (a) $Ma = 10^{-3}$}
\end{center}
\end{minipage}
\hspace{0.02\linewidth}
\begin{minipage}{0.48\linewidth}
\begin{center}
\includegraphics[trim=0mm 0mm 0mm 0mm, clip, width=70mm]{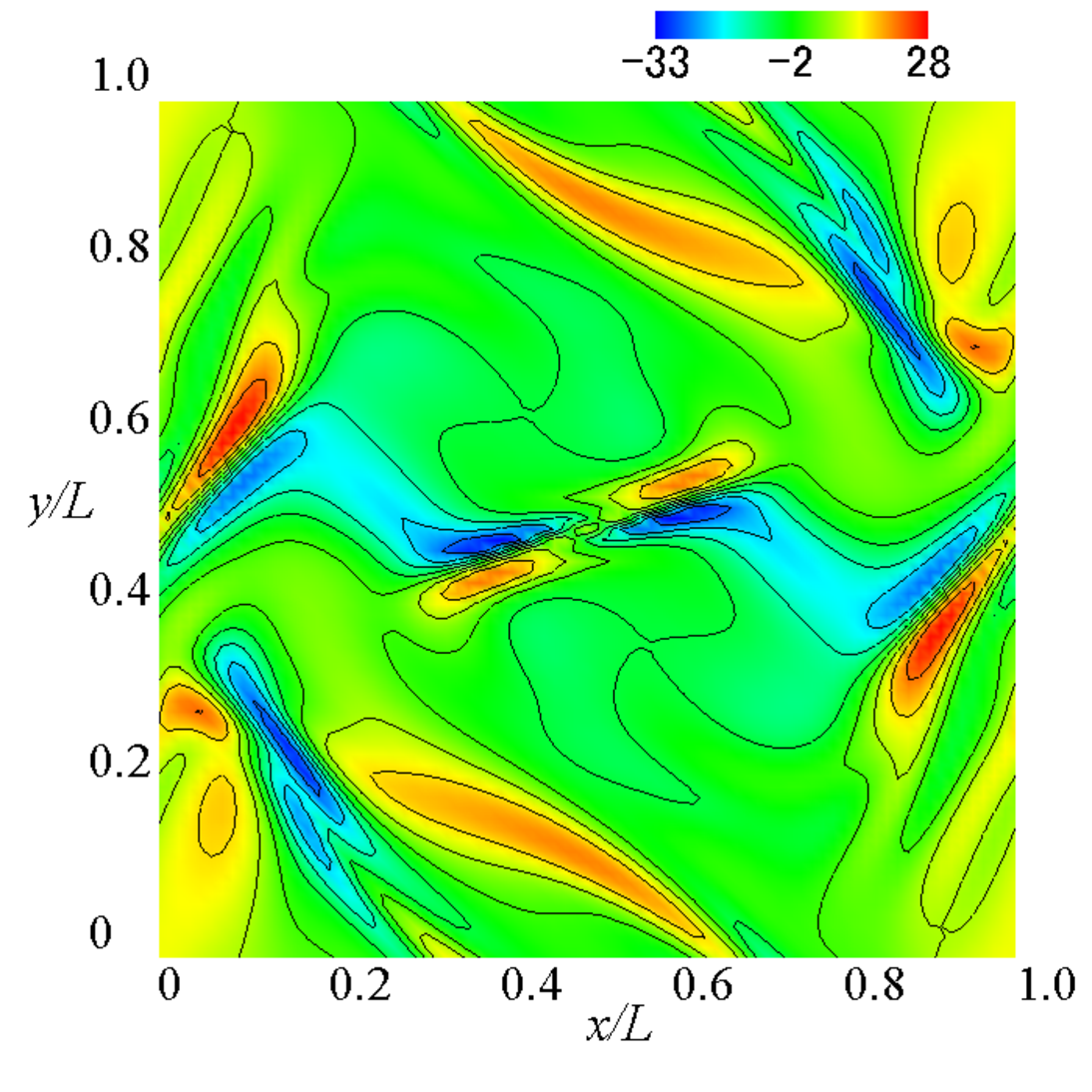} \\
\vspace*{-0.5\baselineskip}
{\small (b) $Ma = 0.1$}
\end{center}
\end{minipage}

\begin{minipage}{0.48\linewidth}
\begin{center}
\includegraphics[trim=0mm 0mm 0mm 0mm, clip, width=70mm]{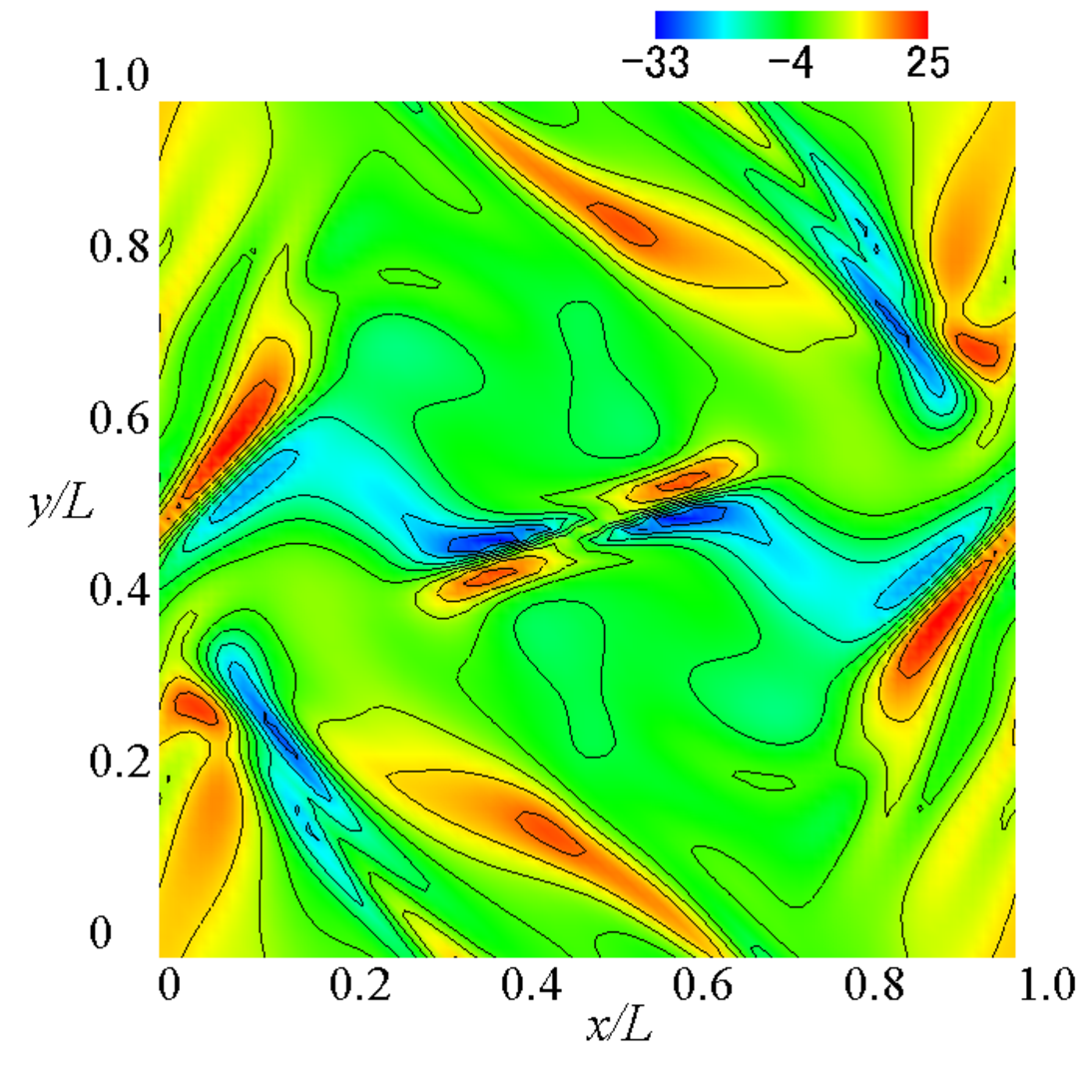} \\
\vspace*{-0.5\baselineskip}
{\small (c) $Ma = 0.3$}
\end{center}
\end{minipage}
\hspace{0.02\linewidth}
\begin{minipage}{0.48\linewidth}
\begin{center}
\includegraphics[trim=0mm 0mm 0mm 0mm, clip, width=70mm]{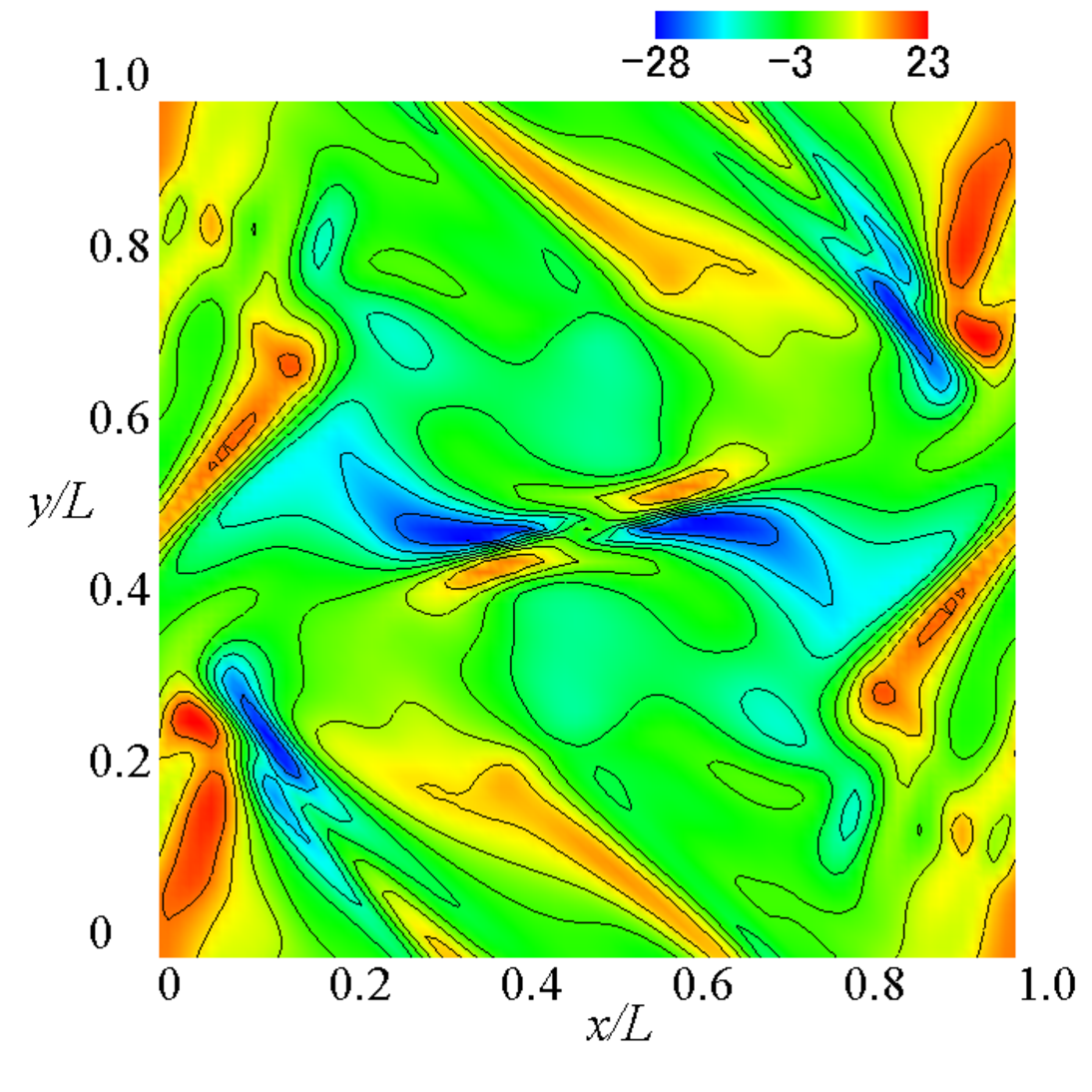} \\
\vspace*{-0.5\baselineskip}
{\small (d) $Ma = 0.5$}
\end{center}
\end{minipage}
\caption{Variation in vorticity contour with Mach number: 
$N = 81$, $t/(L/U) = 0.3$, $Re = Re_m = 400$.}
\label{ot_vortex_wz}
\end{figure}

\begin{figure}[!t]
\begin{minipage}{0.48\linewidth}
\begin{center}
\includegraphics[trim=0mm 0mm 0mm 0mm, clip, width=70mm]{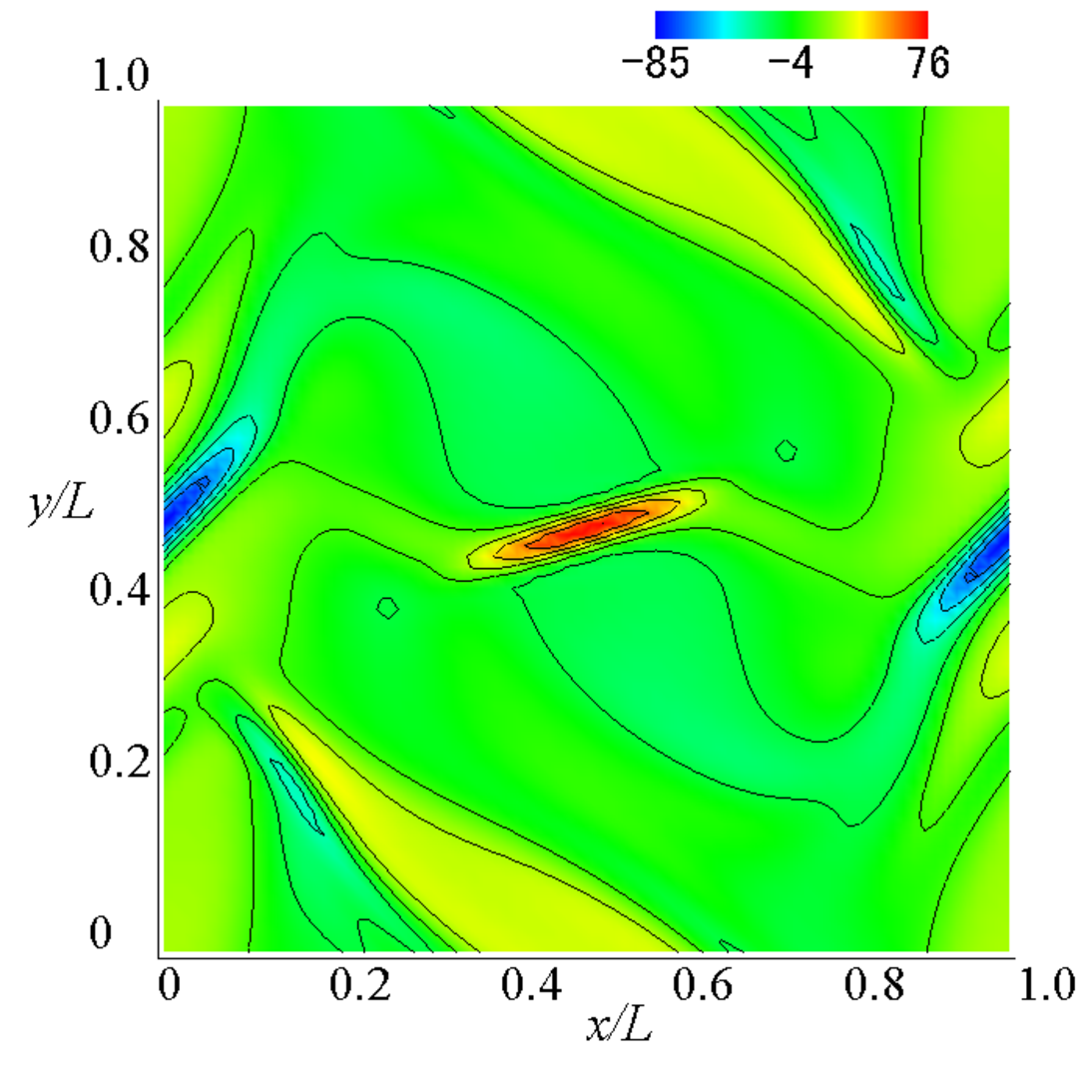} \\
\vspace*{-0.5\baselineskip}
{\small (a) $Ma = 10^{-3}$}
\end{center}
\end{minipage}
\hspace{0.02\linewidth}
\begin{minipage}{0.48\linewidth}
\begin{center}
\includegraphics[trim=0mm 0mm 0mm 0mm, clip, width=70mm]{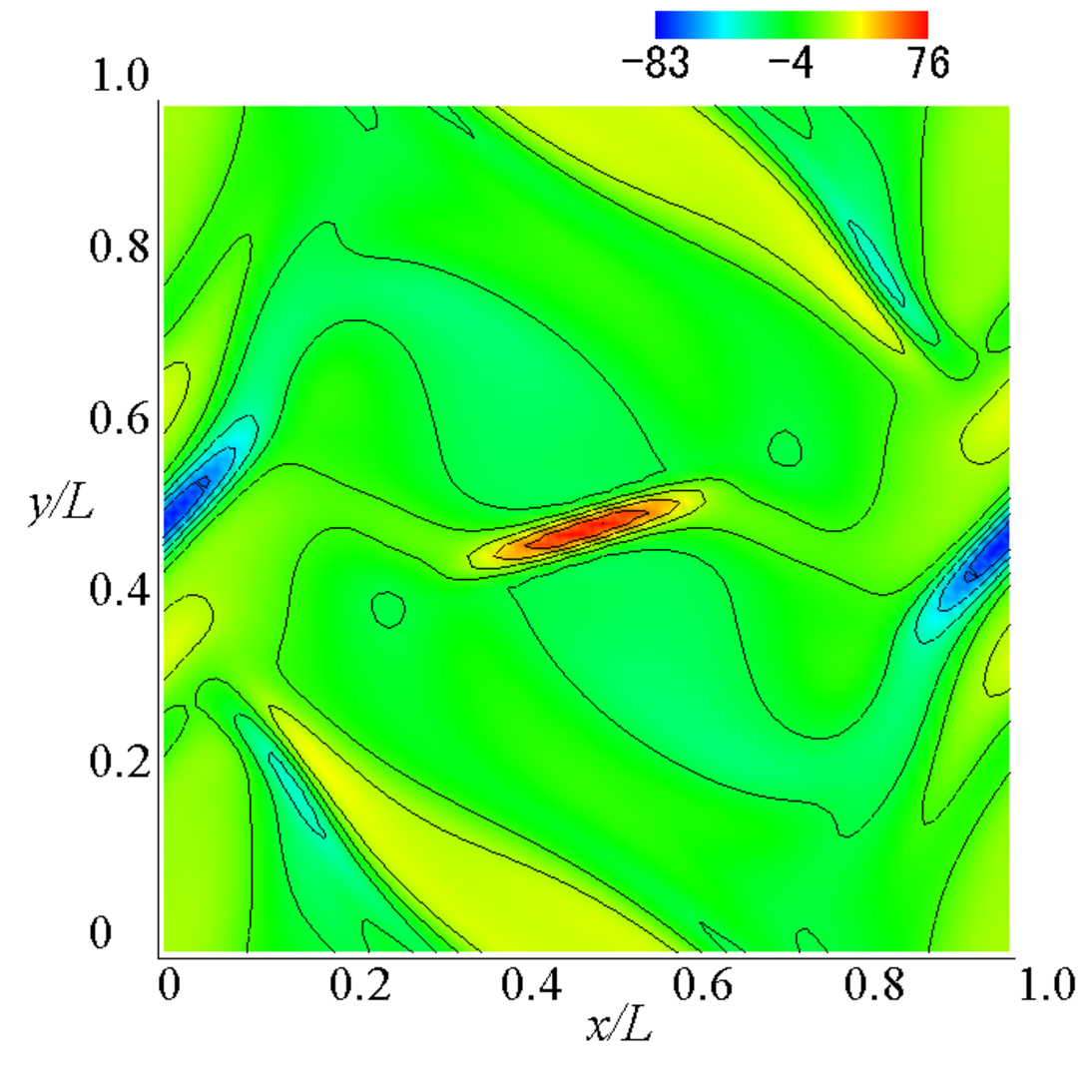} \\
\vspace*{-0.5\baselineskip}
{\small (b) $Ma = 0.1$}
\end{center}
\end{minipage}

\begin{minipage}{0.48\linewidth}
\begin{center}
\includegraphics[trim=0mm 0mm 0mm 0mm, clip, width=70mm]{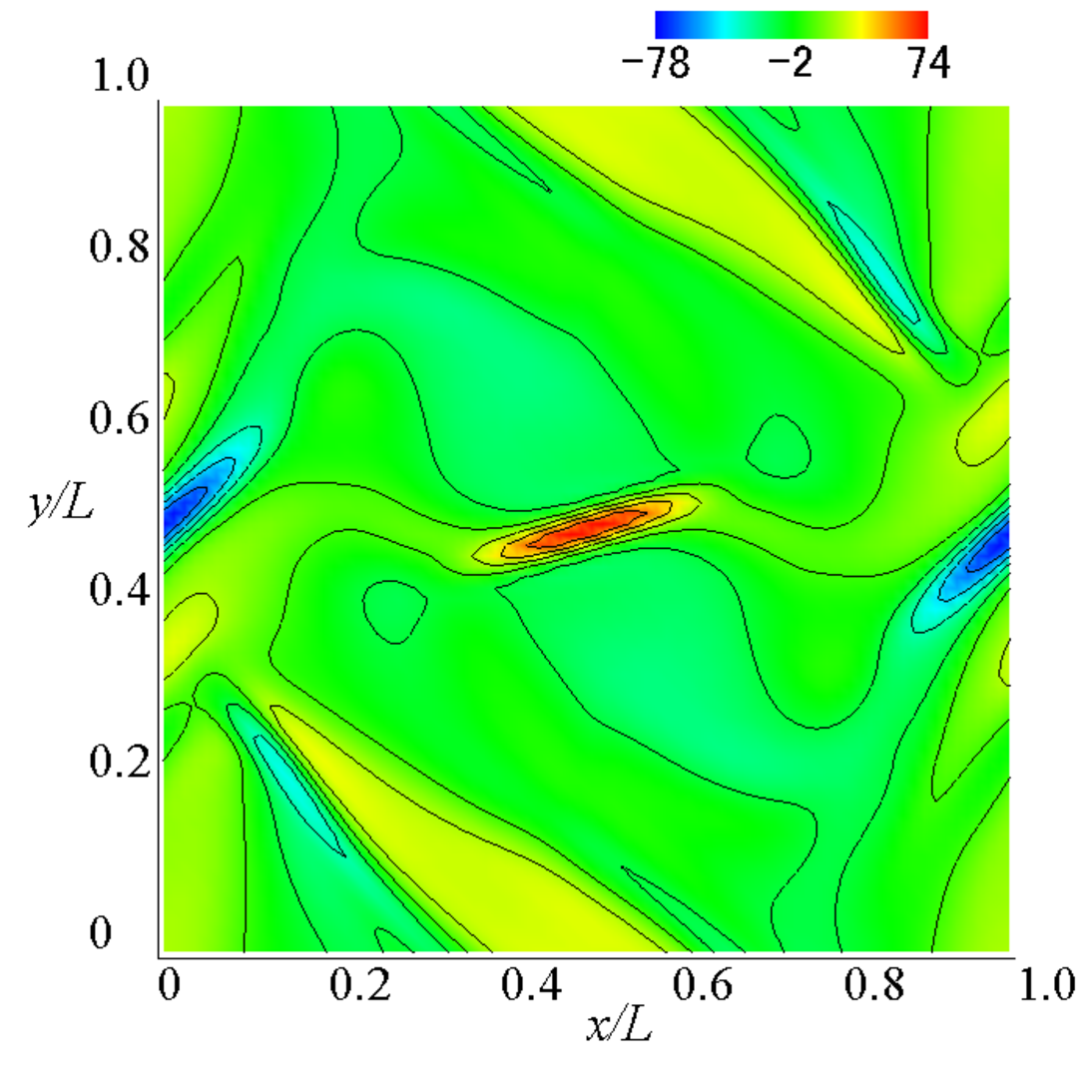} \\
\vspace*{-0.5\baselineskip}
{\small (c) $Ma = 0.3$}
\end{center}
\end{minipage}
\hspace{0.02\linewidth}
\begin{minipage}{0.48\linewidth}
\begin{center}
\includegraphics[trim=0mm 0mm 0mm 0mm, clip, width=70mm]{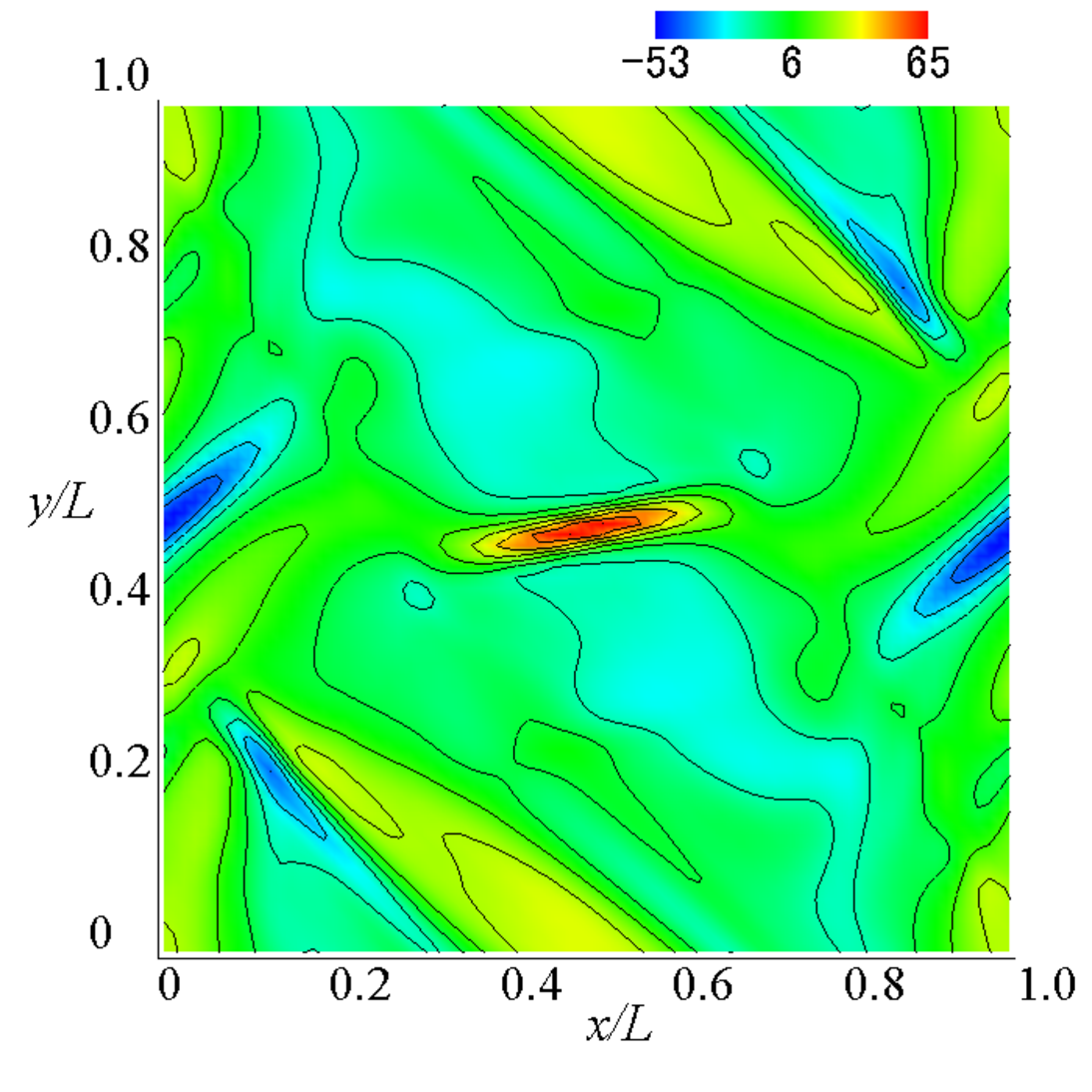} \\
\vspace*{-0.5\baselineskip}
{\small (d) $Ma = 0.5$}
\end{center}
\end{minipage}
\caption{Variation in current density contour with Mach number: 
$N = 81$, $t/(L/U) = 0.3$, $Re = Re_m = 400$.}
\label{ot_vortex_jz}
\end{figure}

Figures \ref{ot_vortex_wz} and \ref{ot_vortex_jz} show 
the contours of vorticity and current density for $Re = Re_m = 400$ at $t/(L/U) = 0.3$, respectively. 
At the region where the shear of flow is strong, 
the magnetic flux density increases, 
and the layers of high vorticity and current density occur. 
No significant difference is observed in the vorticity and current density distributions 
between $Ma = 10^{-3}$ and 0.1. 
As $Ma$ increases, the vortex and current layers parallel the $x$-direction. 
It is also seen that as $Ma$ increases, 
the absolute values of vorticity and current density decrease, 
and the dissipations of vorticity and current density become faster. 
At $Ma = 0.5$, small-scale structures with high vorticity occur. 
Dahlburg and Picone \citep{Dahlburg&Picone_1989} reported 
that compressibility generates more small-scale structures such as jets and breaks up the vortices. 
The tendency for small-scale structures to appear is similar to the existing study.

\begin{figure}[!t]
\begin{minipage}{0.48\linewidth}
\begin{center}
\includegraphics[trim=0mm 0mm 0mm 0mm, clip, width=70mm]{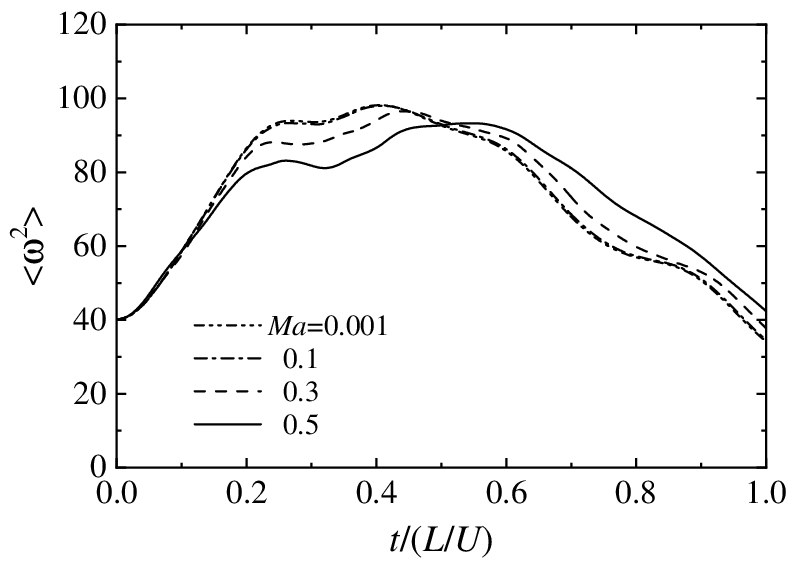} \\
{\small (a) $\omega_z$}
\end{center}
\end{minipage}
\hspace{0.02\linewidth}
\begin{minipage}{0.48\linewidth}
\begin{center}
\includegraphics[trim=0mm 0mm 0mm 0mm, clip, width=70mm]{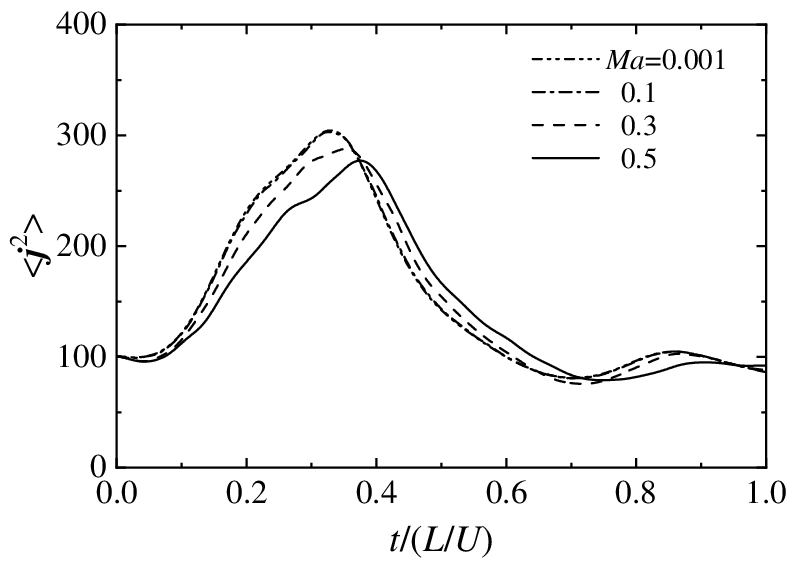} \\
{\small (b) $j_z$}
\end{center}
\end{minipage}
\caption{Variation in enstrophies of vorticity and current density with Mach number: 
$N = 81$, $t/(L/U) = 0.3$, $Re = Re_m = 400$.}
\label{ot_enstrophy_Re400}
\end{figure}

To quantitatively evaluate this dissipation, 
Fig.\ref{ot_enstrophy_Re400} shows the changes in the total amounts, 
$\langle \bm{\omega}^2 \rangle$ and $\langle \bm{j}^2 \rangle$, 
of vorticity and current density enstrophy depending on $Ma$. 
The total amount was calculated by area integration in the $x$-$y$ cross-section. 
The enstrophies for $Ma = 10^{-3}$ and 0.1 are almost the same level 
and show the same time variations. 
As $Ma$ increases, the enstrophy decreases, and the dissipation is accelerated. 
Additionally, the time when the dissipation reaches its maximum is delayed. 
Similarly to existing research \citep{Dahlburg&Picone_1989}, 
as $Ma$ increases, the fluctuation of $\langle \bm{\omega}^2 \rangle$ increases, 
and the local maximum value of $\langle \bm{j}^2 \rangle$ decreases. 
As $Ma$ increases, small-scale structures appear due to nonlinear effects, 
and the intensity of vorticity and current density decreases.

\begin{figure}[!t]
\begin{minipage}{0.325\linewidth}
\begin{center}
\includegraphics[trim=0mm 0mm 0mm 0mm, clip, width=50mm]{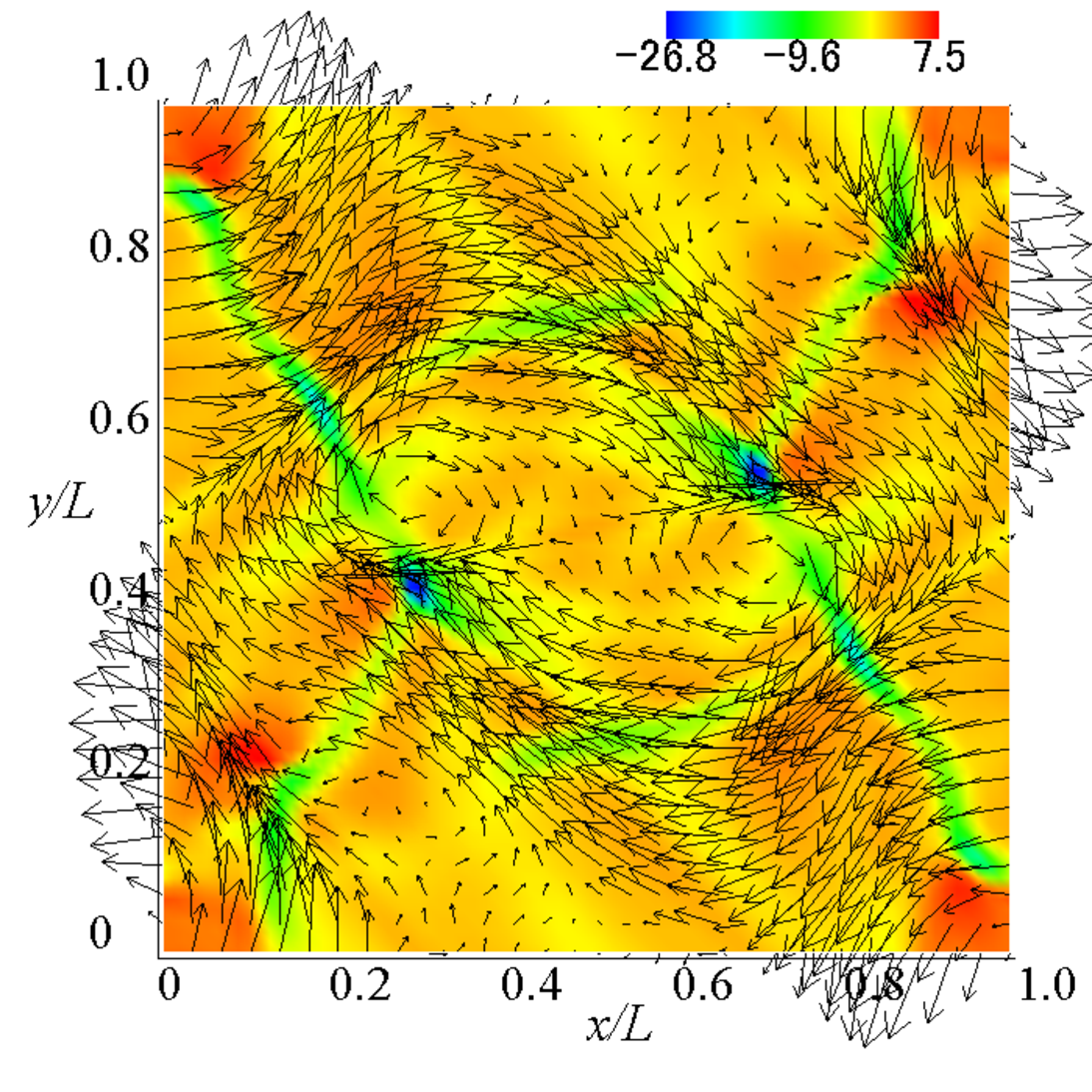} \\
{\small (a) Pressure work}
\end{center}
\end{minipage}
%
\begin{minipage}{0.325\linewidth}
\begin{center}
\includegraphics[trim=0mm 0mm 0mm 0mm, clip, width=50mm]{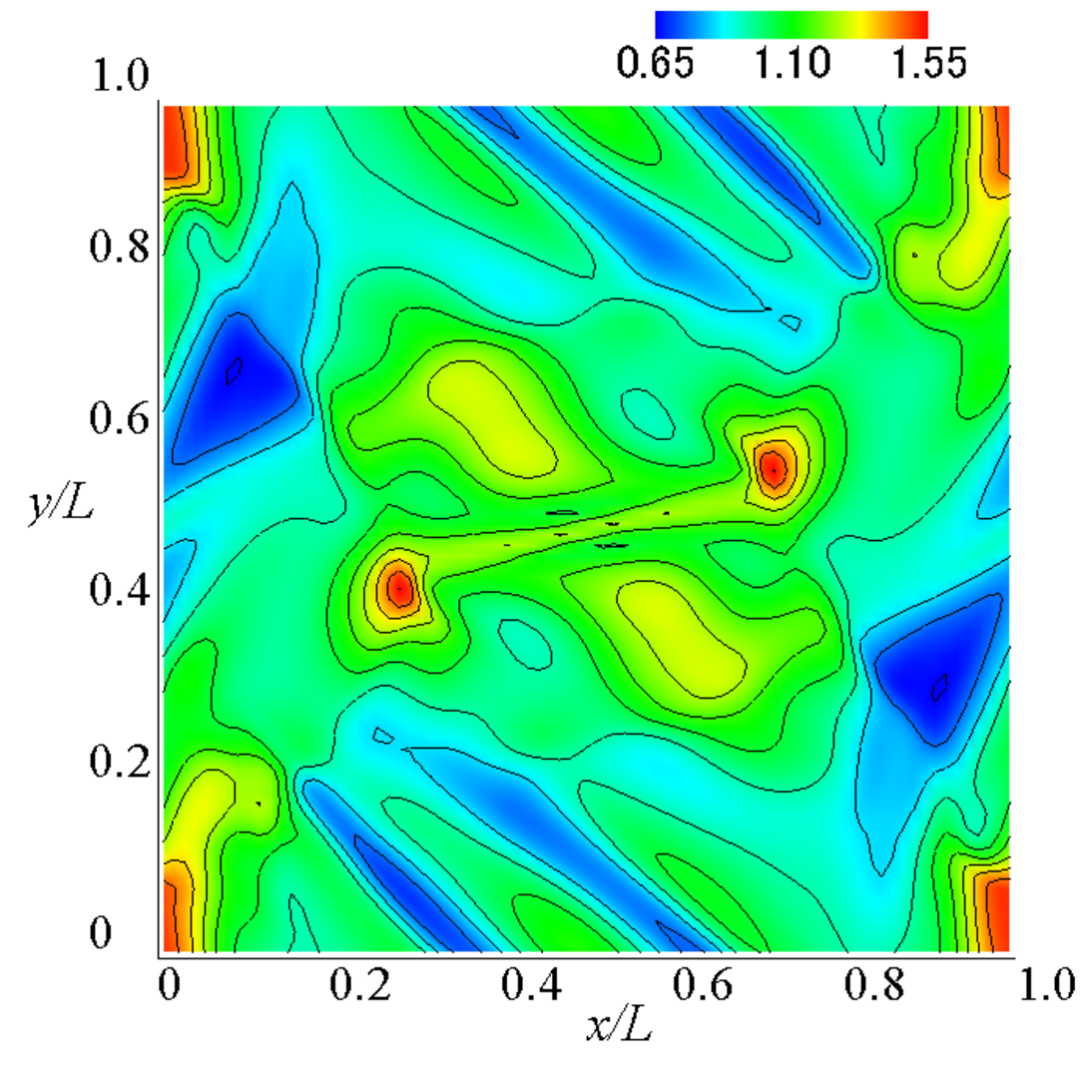} \\
{\small (b) Density}
\end{center}
\end{minipage}
\begin{minipage}{0.325\linewidth}
\begin{center}
\includegraphics[trim=0mm 0mm 0mm 0mm, clip, width=50mm]{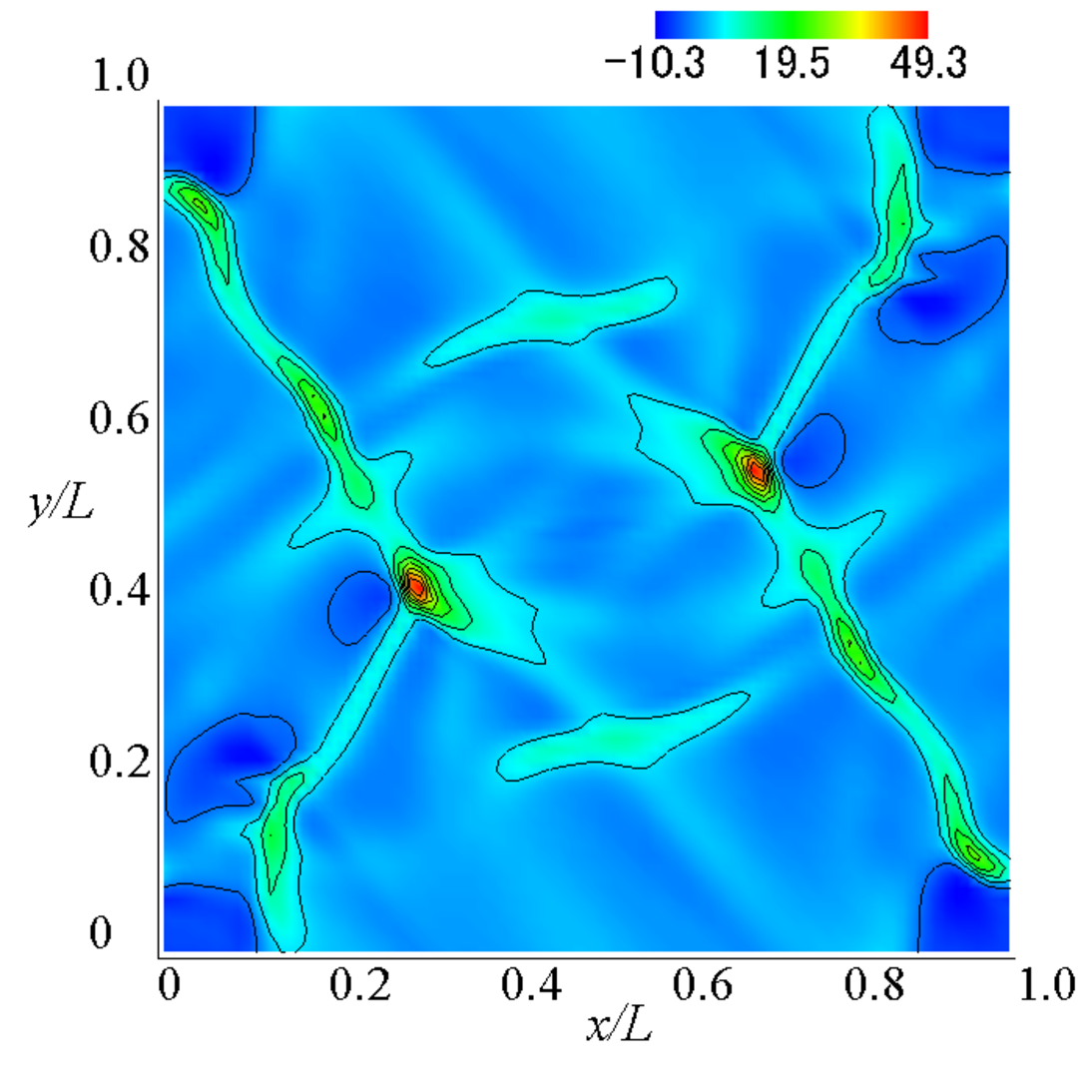} \\
{\small (c) Substantial derivative}
\end{center}
\end{minipage}
\caption{Pressure work, density contours, and substantial derivative of density: 
$N = 81$, $t/(L/U) = 0.3$, $Ma = 0.5$, $Re = Re_m = 400$.}
\label{ot_pwork_rho_Re400_Ma5E-1}
\end{figure}

The compressibility effect appears as $Ma$ increases. 
Figure \ref{ot_pwork_rho_Re400_Ma5E-1} shows 
the distributions of the dimensionless pressure work due to volume change, 
the velocity vector, and density for $Ma = 0.5$. 
The pressure work is defined as $Wp = (\kappa-1) (\kappa Ma^2 p+1) \nabla \cdot \bm{u}$. 
The gas expands in the region $Wp > 0$ and is compressed in the region $Wp < 0$. 
A compressed region appears as a thin layer around the vortex pair 
existing in the center of the computational domain. 
The density is high at both ends of the current layer, and the gas is compressed. 
There appears to be no clear correlation between the pressure work and density. 
Pressure work is expressed using the equation of state as $Wp = (\kappa-1) \rho e \nabla \cdot \bm{u}$. 
From the mass conservation equation, 
the substantial derivative $D \rho /D t$ of density can be expressed as $- \rho \nabla \cdot \bm{u}$. 
The substantial derivative of the density is shown in Fig. \ref{ot_pwork_rho_Re400_Ma5E-1}(c). 
A high compressive work $|Wp|$ appears in regions 
where the spatiotemporal variations in density are intensive. 
Therefore, it can be seen that there is a correlation 
between the spatiotemporal changes in density and pressure work rather than the density itself.

\begin{figure}[!t]
\begin{minipage}{0.48\linewidth}
\begin{center}
\includegraphics[trim=0mm 0mm 0mm 0mm, clip, width=70mm]{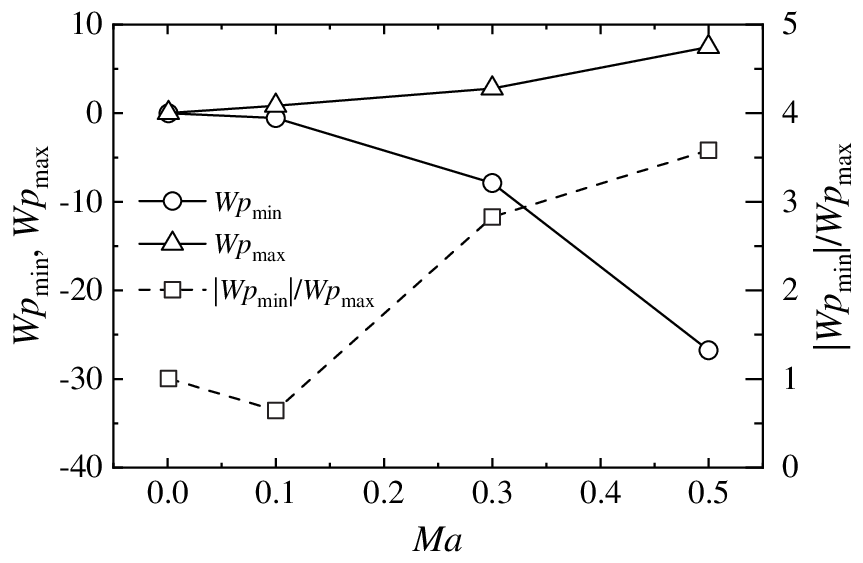} \\
{\small (a) Pressure work}
\end{center}
\end{minipage}
\hspace{0.02\linewidth}
\begin{minipage}{0.48\linewidth}
\begin{center}
\includegraphics[trim=0mm 0mm 0mm 0mm, clip, width=70mm]{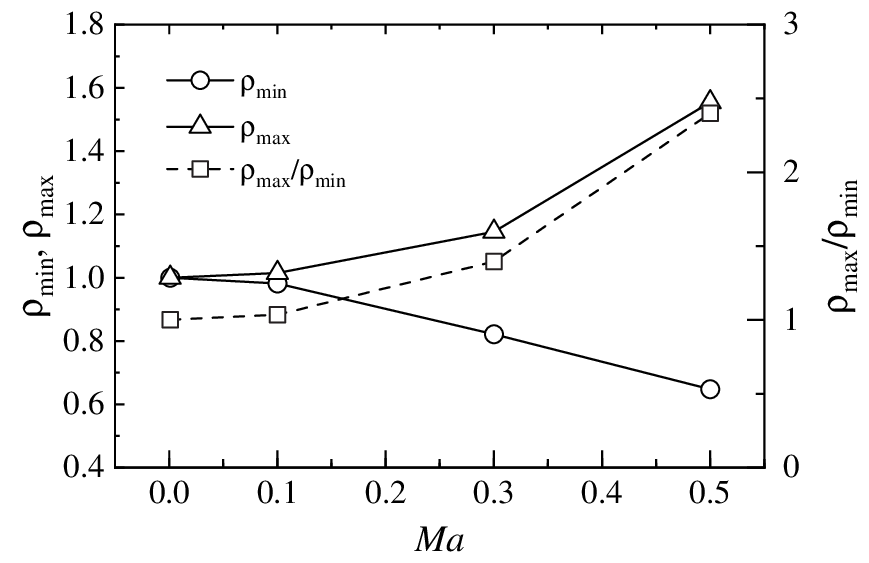} \\
{\small (b) Density}
\end{center}
\end{minipage}
\caption{Variations of maximum and minimum values of pressure work and density with Mach number: 
$N = 81$, $t/(L/U) = 0.3$, $Re = Re_m = 400$.}
\label{ot_pwork_rho_Re400}
\end{figure}

Figure \ref{ot_pwork_rho_Re400} shows the minimum value, $Wp_\mathrm{min}$, 
and maximum value, $Wp_\mathrm{max}$, of the pressure work, 
and their ratio, $|Wp_\mathrm{min}|/Wp_\mathrm{max}$. 
We also show the minimum value, $\rho_\mathrm{min}$, and maximum value, $\rho_\mathrm{max}$, 
of the density, and their ratio, $\rho_\mathrm{max}/\rho_\mathrm{min}$. 
At $Ma = 10^{-3}$, the pressure work is almost zero, 
and the density maintains the initial value. 
At $Ma = 0.1$, the compressibility effect begins to appear. 
As $Ma$ increases, the expansion work $Wp_\mathrm{max}$ 
and the compression work $|Wp_\mathrm{min}|$ increase, respectively, 
and at $Ma = 0.5$, the work ratio, $|Wp_\mathrm{min}|/Wp_\mathrm{max}$, reaches 3.6 times. 
In addition, the density ratio, $\rho_\mathrm{max}/\rho_\mathrm{min}$, reaches 2.4 times. 
The influence of compressibility is noticeable.

\begin{figure}[!t]
\begin{minipage}{0.48\linewidth}
\begin{center}
\includegraphics[trim=0mm 0mm 0mm 0mm, clip, width=70mm]{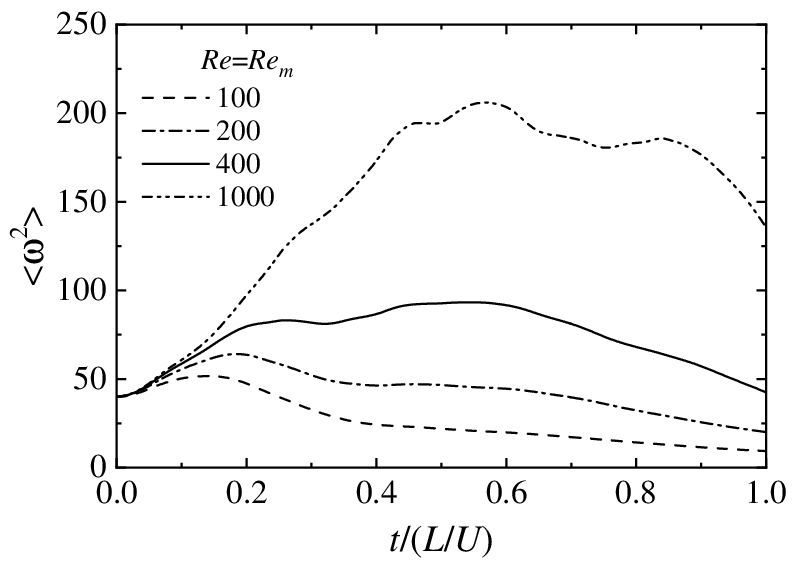} \\
{\small (a) $\omega_z$}
\end{center}
\end{minipage}
\hspace{0.02\linewidth}
\begin{minipage}{0.48\linewidth}
\begin{center}
\includegraphics[trim=0mm 0mm 0mm 0mm, clip, width=70mm]{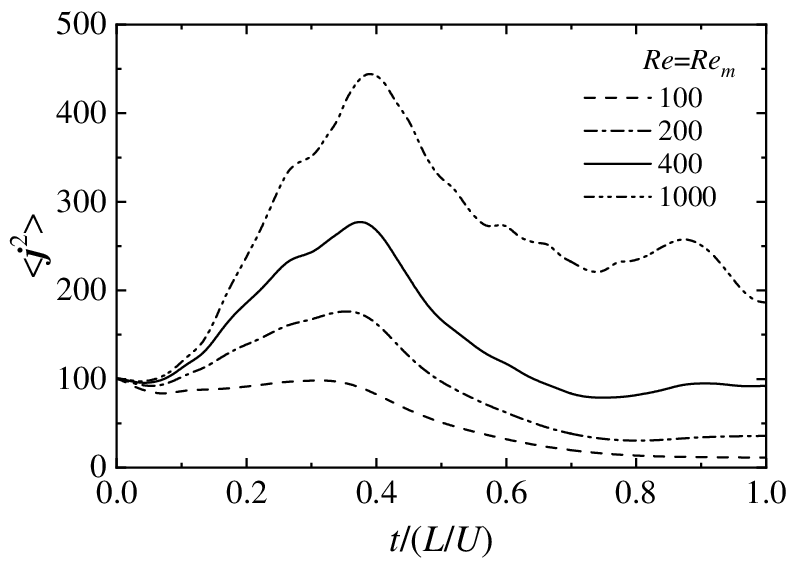} \\
{\small (b) $j_z$}
\end{center}
\end{minipage}
\caption{Time variations of enstrophies of vorticity and current density: 
$N = 81$, $Ma = 0.5$.}
\label{ot_enstrophy_wz_jz}
\end{figure}

Figure \ref{ot_enstrophy_wz_jz} shows the temporal changes in each total amount, 
$\langle \bm{\omega}^2 \rangle$ and $\langle \bm{j}^2 \rangle$, 
of vorticity and current density enstrophies for $Ma = 0.5$. 
For $Re = Re_m = 100$, 
$\langle \bm{\omega}^2 \rangle$ reaches its maximum 
when the total energy dissipation rate $\varepsilon_t$ reaches its maximum. 
As $Re$ and $Re_m$ increase, $\langle \bm{\omega}^2 \rangle$ reaches its maximum value 
later than when $\varepsilon_t$ reaches its maximum. 
The vorticity is strengthened even in the process of the total energy dissipation. 
The maximum value of $\langle \bm{j}^2 \rangle$ appears around $t/(L/U) = 0.30-0.39$, 
and as $Re$ and $Re_m$ increase, the time for the maximum to appear is slightly delayed. 
At $Re = Re_m = 400$ and 1000, a new local maximum appears 
in the distribution of $\langle \bm{j}^2 \rangle$ around $t/(L/U) = 0.8-1.0$, 
and $\varepsilon_t$ also becomes maximum. 
An interesting phenomenon occurs in which the current density is reintensified. 
As this analysis is a two-dimensional calculation, 
a three-dimensional analysis is necessary to elucidate such phenomena. 
This investigation is a future topic.

\begin{figure}[!t]
\begin{center}
\includegraphics[trim=0mm 0mm 0mm 0mm, clip, width=70mm]{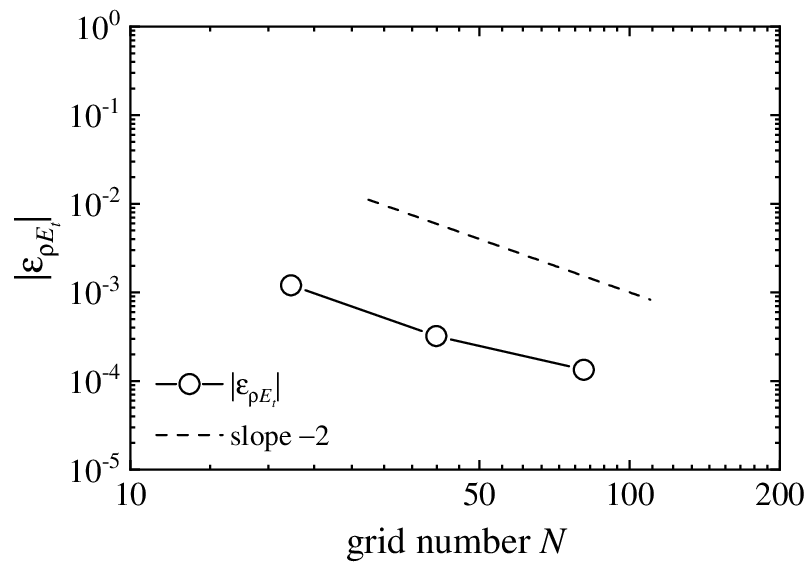}
\end{center}
\caption{Total energy error: 
$Ma = 0.5$, $Re = Re_m = 10^3$.}
\label{ot_error}
\end{figure}

Figure \ref{ot_error} shows the relative error of the total energy at $t/(L/U) = 1.0$. 
The error is defined as $\varepsilon_{\rho E_t} = (\langle \rho E_t \rangle - \langle \rho E_t \rangle_{161})/\langle \rho E_t \rangle_{161}$. 
Here, $\langle \rho E_t \rangle_{161}$ is the total amount obtained from the $N = 161$ grid. 
Regardless of the grid, the total energy error decreases with a slope of $-2$, 
indicating the second-order convergence.

For $Ma = 0.5$ and $Re = Re_m = 1000$ in this analysis, 
the maximum error of the mass conservation law is $9.71 \times 10^{-13}$, 
and the maximum divergence error of the magnetic flux density is $1.23 \times 10^{-10}$. 
Under different conditions, 
the maximum divergence error of the magnetic flux density is on the order of $10^{-13}$, 
and we found that the divergence error tends to increase as the Reynolds number increases. 
However, as the error is sufficiently low, 
we believe that the divergence-free condition of magnetic flux density is sufficiently satisfied 
even without correction of magnetic flux density.

\section{Conclusion}
\label{summary}

We constructed an energy-conserving finite difference method 
to analyze compressible MHD flows at low Mach numbers with the nonconservative Lorentz force. 
This analysis method discretizes the Lorentz force 
so that the transformation between conservative and nonconservative forms holds. 
Furthermore, the equations for total energy and magnetic helicity can be derived 
discretely from those for momentum, magnetic flux density, and magnetic vector potential. 
Even discretized equations satisfy the constraints of Gauss's law. 
This scheme simultaneously relaxes velocity, pressure, density, and internal energy, 
and stable convergence solutions can be obtained.

In this study, we analyzed four types of models 
and verified the accuracy and convergence of this numerical method. 
For the analysis of a three-dimensional ideal periodic inviscid MHD flow, 
momentum, magnetic flux density, and total energy are conserved 
in time in the case of a uniform grid. 
Using a nonuniform grid degrades the momentum conservation property, 
but the total energy is preserved discretely. 
Even without correction for the magnetic flux density, 
the divergence-free condition of the magnetic flux density is satisfied discretely.

Inviscid analysis of advective magnetic vortex can capture the behavior of vortex 
undergoing shear deformation in incompressible and low Mach number compressible flows. 
As the Mach number increases, the vortex and current layers become thin, 
and the magnitudes of vorticity and current density are locally enhanced. 
Even in a vortex advection problem, 
transport quantities such as total energy are conserved discretely.

To verify that this numerical method can also be applied to the analysis of incompressible flows, 
we analyzed a Taylor decaying vortex problem. 
A stable convergent solution is obtained even at a significantly low Mach number, 
and this calculation results agree with the exact solution. 
This numerical method can accurately predict the trend of energy attenuation.

In the analysis of the Orszag--Tang vortex, 
the influence of Mach number on the energy dissipation process was investigated. 
An increase in Mach number reduces the magnitude of vorticity and current density. 
In addition, compression work increases more than expansion work, 
and the influence of compressibility appears. 
An increase in Mach number slightly delays the transition to turbulent flow.

We verified that the numerical method proposed in this study can analyze 
flows ranging from incompressible to low Mach number flows. 
In the future, we plan to investigate flow control using nonuniform magnetic fields 
and nanoparticles using this analysis method.

\section*{Acknowledgment}

This research did not receive any specific grant from funding agencies 
in the public, commercial, or not-for-profit sectors. 
The author wishes to acknowledge the time and effort of everyone involved in this study.


\bibliography{jcp2024_low-mach_comp_mag_bibfile}
\end{document}